\documentclass[preprint]{elsarticle}
\usepackage{amssymb,amsmath}
\usepackage{graphicx}
\usepackage{feynmf}
\usepackage{array}
\usepackage{ifthen}

\usepackage{figwidth}

\newlength{\mywidth}
\newcommand{\slashed}[1]{\settowidth{\mywidth}{#1}\hspace*{0.5\mywidth}\makebox[0ex][c]{$#1$}\makebox[0ex][c]{$\slash$}\hspace*{0.5\mywidth}}

\newcommand{\eq}[1]{Eq.~(#1)}
\newcommand{\reffig}[1]{Fig.~\ref{#1}}
\newcommand{\reffigs}[1]{Figs.~\ref{#1}}
\newcommand{\reftable}[1]{Table~\ref{#1}}
\newcommand{\units}[1]{\, \mathrm{#1}}
\newcommand{\Tr}{\mathrm{Tr}}
\newcommand{\ansatz}[3]{\begin{array}{c}#1\\#2\\#3\end{array}}

\begin{document}

\ifthenelse{\equal{\Qclass}{revtex4}}{
\title{The topological susceptibility from grand canonical simulations in the interacting instanton liquid model: zero temperature calibrations and numerical framework.}

\author{Olivier Wantz}
\email[Electronic address: ]{O.Wantz@damtp.cam.ac.uk}
\affiliation{Department of Applied Mathematics and Theoretical Physics,
Centre for Mathematical Sciences,\\ University of Cambridge,
Wilberforce Road, Cambridge CB3 0WA, United Kingdom}
}{}
\ifthenelse{\equal{\Qclass}{elsarticle}}{
\title{The topological susceptibility from grand canonical simulations in the interacting instanton liquid model: zero temperature calibrations and numerical framework.}

\author[damtp]{Olivier Wantz}
\ead{O.Wantz@damtp.cam.ac.uk}
\address[damtp]{Department of Applied Mathematics and Theoretical Physics,
Centre for Mathematical Sciences,\\ University of Cambridge,
Wilberforce Road, Cambridge CB3 0WA, United Kingdom}
}{}

\bibliographystyle{plain}

\begin{abstract}
This is the first in a series of papers which ultimately aims on improving on the present estimates on the axion mass by modelling the topological non-perturbative QCD dynamics. Axions couple to instantons and their mass is set by the topological susceptibility whose temperature dependence we estimate with the interacting instanton liquid model (IILM). Since accurate finite temperature instanton calculations have problems and do not consider fluctuations in the topological charge, we develop an improved grand canonical version of the IILM to study topological fluctuations in the quark gluon plasma. In this first paper we will calibrate the model against the topological susceptibility at zero temperature, in the chiral regime of physical quark masses.
\end{abstract}

\maketitle

\section{Introduction}
\label{sec:introduction}

The strong interactions at finite temperature are believed to display a number of interesting, non-perturbative phenomena, among which are the confinement/deconfinement transition, spontaneous $P$ and $CP$ violation and chiral symmetry restoration. The latter is believed to have its origin in topological fluctuations. Lattice simulations, e.g.\ \cite{chu:grandy:huang:negele:instanton:lattice:evidence, wang:lu:wang:lattice:instanton:vacuum:dominance}, and phenomenology, e.g.\ \cite{diakonov:instanton:quarks,shuryak:instantons:liquid:I,shuryak:instantons:liquid:II,shuryak:instantons:liquid:III}, have shown that the chiral dynamics of QCD is well described by instanton models.

Another interesting question related to topological fluctuations is the determination of the axion mass. Axions couple to instantons and their mass is directly proportional to the topological susceptibility. The latter turns out to be a chiral property of QCD and can thus also be expected to be well described by the IILM. The main physical question that we want to address is the computation of the topological susceptibility and the systematic effects that pertain to its determination with instanton methods in the chiral regime of light, physical quark masses.

Instanton models are based on a combination of semi-classical expansion \cite{thooft:instanton:fluctuations} and variational approach \cite{diakonov:instanton:variational,diakonov:instanton:quarks}. Taking this variational path integral as a starting point, Shuryak investigated what has become known as the interacting instanton liquid model (IILM) \cite{shuryak:instantons:qcd:I,shuryak:instantons:qcd:II}. In \cite{schaefer:shuryak:iilm} many bulk properties were computed and seen to be consistent with the available lattice data and phenomenology. Some recent studies \cite{garcia:osborn:iilm:chiral,cristofotetti:foccioli:traini:negele:iilm} corroborate the earlier results that the IILM rather accurately describes the chiral properties of QCD, i.e.\ that instantons are the dominant degrees of freedom as far as the chiral regime of QCD is concerned. However, the IILM fails to reproduce confinement.

The topological susceptibility is a key parameter of QCD and has been investigated in many lattice studies. Comparatively few studies have addressed this quantity within the IILM \cite{shuryak:verbaarschot:screening}. One reason is that, so far, the IILM is based on a canonical ensemble. Although one can extract the topological susceptibility from the canonical ensemble through the decay of correlators, e.g.\ see \cite{shuryak:verbaarschot:screening} for IILM and \cite{Chiu:Aoki:JLQCD:TWQCD:topological:susceptibility:fixed_topology,Chiu:Aoki:JLQCD:TWQCD:topological:susceptibility:overlap} for lattice simulations, it is most natural to use the grand canonical ensemble to study the topological susceptibility. Recent investigations of the IILM in the grand canonical ensemble \cite{faccioli:guadagnoli:simula:neutron:dipole:moment,faccioli:strong:CP:theta} were based on canonical simulations and a fugacity expansion, while we will set up a grand canonical IILM that uses grand canonical Monte Carlo simulations; see also \cite{diakonov:polyakov:weiss:instanton:vacuum} for a `mean-field' study of the grand-canonical ensemble and \cite{munster:kamp:instanton:gas} for an exploratory investigation of grand canonical simulations in a simpler framework.

While developing the grand canonical IILM, we have found that the existing finite temperature IILM, \cite{shuryak:verbaarschot:interactions:finite:T}, displays an unphysical behaviour in that it does not allow for a thermodynamic limit. Specifically the instanton--instanton interaction, \eq{3.11} in \cite{shuryak:verbaarschot:interactions:finite:T}, contains a term that decays very slowly with instanton separation $R$,
\begin{equation}
 \ln\left(1+\frac{\beta}{R}\right) \xrightarrow{R \to \infty} \frac{\beta}{R}\,, \label{eq:dyon_dyon_interaction}
\end{equation}
and is not integrable. In their original paper the authors do discuss this long-range interaction and point out that they found the $O(1/R)$ dyon--dyon behaviour for a wide range of intermediate separations. It might well be that the interactions are still well described by this ansatz for the simulation boxes used in subsequent numerical investigations, e.g.\ \cite{schaefer:shuryak:iilm}, but for studying the large volume behaviour it is not appropriate.

We remedy this deficiency by re-deriving the interactions and setting up a numerical framework that avoids using parametric fitting formulas, such as (\ref{eq:dyon_dyon_interaction}); instead we will integrate the Lagrangian density for a pair of instantons exactly, i.e.\ numerically. To the best of our knowledge, the exact action density for a pair has not been published in the literature before; we will provide it for Harrington--Shepard calorons \cite{harrington:shepard:caloron}. The explicit form allows us to perform the numerical integration in an efficient way by exploiting the symmetries of the integrand, and an exact analytic computation for widely separated pairs is possible because of the localised nature of the integrands. Hence, the large separation interactions are under very good control. In particular, the large separation instanton--instanton interaction at finite temperature is not given by (\ref{eq:dyon_dyon_interaction}), as we will see more explicitly in the final paper of this series \cite{wantz:iilm:3}; in this paper, however, we will restrict ourselves to zero temperature. We hope that this framework can be build upon to include the non-trivial holonomy calorons \cite{kraan:baal:caloron:I,kraan:baal:caloron:II,kraan:baal:caloron:monopole}, which may play part in the confinement/deconfinement phase transition, and for which good fitting formulas will be even harder to come by because of their more complicated structure. As mentioned before, in this series we will restrict ourselves to the Harrington--Shepard calorons.

In section \ref{sec:saturate} we will review the standard strategy used to derive the partition function for an ensemble of background gauge fields, i.e.\ the semi-classical approximation. We will then re-derive the interactions for the so-called ratio ansatz, used to construct multi-instanton backgrounds from individual instantons, in section \ref{sec:interactions} and compare it with other available ans\"atze. In section \ref{sec:numerical:implementation} we present the numerical framework we have set up to deal with the simulations. Given that different ans\"atze are available, we will study their effect on some bulk properties in section \ref{sec:ansaetze} and we endeavour to get a handle on systematic uncertainties inherent in this approach. Finally we fix the free parameters of the model and summarise our results in section \ref{sec:parameters}. Finite temperature simulations will be dealt with in \cite{wantz:iilm:2} and \cite{wantz:iilm:3}.

\section{Saturating the path integral}
\label{sec:saturate}

The IILM path integral is an approximation to the fundamental path integral by saturating the latter with a given ansatz for the multi-instanton background. The functional measure consists of small fluctuations around that classical configuration. To make analytical progress, the action is expanded to quadratic order to define the `free' part that is used in perturbation theory. In general the background induces non-Gaussian fluctuations that need to be treated exactly. The directions of these zero modes can (sometimes) be integrated up, and correspond to the tangent space of a generically non-trivial manifold. The coordinates on this so-called moduli-space can be interpreted as those degrees of freedom whose quantum mechanics approximates the low-energy dynamics of the fundamental theory.

In order to discuss the approximations that are eventually used, we will now sketch the construction of the variational path integral, paying particular attention to the low lying modes. Details pertaining to the variational approach, gauge fixing and renormalisation can be found in the original papers \cite{diakonov:instanton:variational,diakonov:instanton:quarks}. We denote by $\phi$ the collection of bosonic fields. The classical action we write as $S_c$ and the classical interaction is defined as
\begin{equation}
 S_\mathrm{int} = S_c - N S_0\,,
\end{equation}
where $N$ is the number of instanton constituents of the background and $S_0$ the action of an individual instanton. Assume that the background has $N_\gamma$ (quasi) zero modes, which we denote collectively by $\gamma$. We can then write\footnote{This follows \cite{kleinert:path_integrals}.}, using the eigenfunctions of the free part of the action $\left.\delta^2 S/\delta \phi^2\right|_{\phi=\phi_c}$ at zero and finite $\gamma$,
\begin{equation}
\begin{split}
 \phi_c(x,0)+\phi(x) &= \phi_c(x,0)+\sum_{n=1}^{\infty} \zeta_n \eta_n(x,0)\,,\\
&= \phi_c(x,\gamma)+\sum_{n=N_\gamma+1}^{\infty} \bar{\zeta}_n \eta_n(x,\gamma) + O(\gamma^2)\,.
\end{split}
\end{equation}
This can be rearranged to (omitting the $x$-dependence for notational clarity)
\begin{equation}
\begin{split}
\phi(\{\gamma, \bar{\zeta}\}) &= \phi_c(\gamma)-\phi_c(0)+\sum_{n=N_\gamma+1}^{\infty} \bar{\zeta}_n \eta_n(\gamma)\,,\\
&= \sum_{m=1}^{\infty} \left[ \int d^n x \left(\phi_c(\gamma)-\phi_c(0)+\sum_{n=2}^{\infty} \bar{\zeta}_n \eta_n(\gamma)\right) \eta_m(0) \right] \eta_m(0)\,,
\end{split}
\end{equation}
where we used the fact that $\eta(x,0)$ forms a complete basis. Clearly $\phi(\gamma=0,\bar{\zeta}) \perp \eta_i$ with $i=1, \dots, N_\gamma$. The Jacobian for the variable change $\{\zeta_n\} \to \{\gamma,\bar{\zeta}_m\}$, follows from the following partial derivatives
\begin{eqnarray}
\left.\frac{\partial \zeta_n}{\partial \gamma_i}\right|_{\gamma=0} &=& \int \left( \partial_{\gamma_i} \phi_c(0) \,  \eta_n(0)- \phi (\bar{\zeta}) \partial_{\gamma_i} \eta_n(0)\right)\,,\\
\left.\frac{\partial \zeta_n}{\partial \bar{\zeta}_m}\right|_{\gamma=0} &=& \delta_{mn}\,.
\end{eqnarray}
Note the occurrence of the $\phi$ part. This will lead to new interactions which have no classical counterpart but are purely quantum mechanical; to 1-loop order, we are allowed to discard them. From this matrix structure it follows that the Jacobian is given by $\det ( \int \partial_{\gamma_n} \phi \, \eta_m )$.

Now, we do not know the set $\{\eta\}$ of exact low lying eigenfunctions. However, we can approximate it by constructing an orthonormal set of the known single particle zero modes that descend from the exact solutions used to build up the background field. With a slight abuse of notation, we substitute $\eta \to \bar{\eta} = O_B \eta$; $O_B$ is the matrix that generates an orthonormal basis from the original set $\{\eta\}$ of single particle zero modes. The Jacobian is then given by 
\begin{equation}
 \det \left( \int \partial_{\gamma_n} \phi \, \bar{\eta}_m \right)=\det \left( \int \partial_{\gamma_n} \phi \, \eta_m \right) \det O_B\,.
\end{equation}
It corresponds to the quantum mechanical gluonic interactions. The high-frequency eigenvalues, encoded in the determinant of the fluctuation operator $\delta^2 S/\delta \phi^2$ are assumed to be $N$-fold degenerate, and so the fluctuation determinant factorises.

In QCD we also need to introduce quarks and treat their interactions with the background field. In the case where the Dirac operator admits quasi-zero modes we can approximate the low frequency part in the same way as for the gluonic case; the high-frequency fluctuations will again be assumed to factorise. As for the Jacobian, we do not know the exact set of low-lying eigenfunctions for the superposition, but we approximate it by constructing an orthonormal set of the exact single particle zero modes $\xi_n$, i.e.\ $\bar{\xi}=O_F\xi$. The Dirac operator, truncated to that subspace, is then given by
\begin{equation}
 (\slashed{D}+m)_{\mathrm{low}} = \overline{(\slashed{D}+m)}_{ij} |\bar{\xi}_i\rangle \otimes \langle\bar{\xi}_j| = \left(\bar{\slashed{D}}_{ij}+m\delta_{ij}\right)|\bar{\xi}_i\rangle \otimes \langle\bar{\xi}_j|\,, \label{eq:quark:overlap}
\end{equation}
with $\bar{\slashed{D}}_{ij} = \langle \bar{\xi}_i|\slashed{D}|\bar{\xi}_j \rangle$. The matrix of overlaps is related to the single particle zero mode overlaps by
\begin{equation}
 \slashed{D}_\mathrm{low} = \bar{\slashed{D}} = O_F^\dagger \, \slashed{D} \, O_F\,,
\end{equation}
with $\slashed{D}_{ij} = \langle \xi_i|\slashed{D}|\xi_j \rangle$. Note that this is not a similarity transformation because $O_F$ is not unitary.

\section{Interactions in the IILM}
\label{sec:interactions}

We will now turn to instantons in QCD. In this paper, we will only discuss BPST instantons \cite{bpst:instanton}. In terms of the 't Hooft potential\footnote{Actually $1+\Pi$ is the 't Hooft potential.}
\begin{equation}
 \Pi(x,\{y,\rho\}) = \frac{\rho^2}{r^2}\,,
\end{equation}
with $r^2 = (x-y)^2$, the BPST instanton in singular gauge is given by
\begin{equation}
 A^a_\mu = - O_i^{ab} \zeta^b_{\mu\nu} \frac{\partial_\nu \Pi(x,\{y,\rho\})}{1+\Pi(x,\{y,\rho\})}\,,
\end{equation}
with $\zeta^b_{\mu\nu} = \bar{\eta}^b_{\mu\nu}$ for instantons, $\zeta^b_{\mu\nu} = \eta^b_{\mu\nu}$ for anti-instantons and $\eta$ the 't Hooft symbols. The collective coordinates are: $y$ the centre, $\rho$ the size and $O$ the colour orientation in the adjoint representation.

The simplest background configuration is the sum ansatz, as used for instance in \cite{diakonov:instanton:variational}. It was shown in \cite{shuryak:instantons:liquid:II}, that the sum ansatz produces an unphysical amount of repulsion; this is due to the fact that the field strength actually diverges at the individual centres, and is in sharp contrast to the individual singular gauge instanton whose field strength is finite at the centre\footnote{Note, however, that the field strength of the individual singular gauge instanton is not continuous at the centre and is only defined on the punctured Euclidean space. Incidentally, the winding about this singular point corresponds to the winding at infinity of the regular instanton.}. In this case, the author therefore proposed a different ansatz, inspired by 't Hooft's multi-instanton form, that stays finite at the centre of the instantons, and dubbed it the ratio ansatz. It is given by
\begin{equation}
 A^a_\mu = - \frac{\sum_i O_i^{ab} \zeta^b_{\mu\nu} \partial_\nu \Pi_i(x,\{y_i,\rho_i\})}{1+\sum_i \Pi_i(x,\{y_i,\rho_i\})}\,.
\end{equation}
In what follows we will refer to $R_E$ as the interactions or the ensemble generated by the ratio ansatz. We will compare the predictions from $R_E$ with those of the streamline ansatz $S$ \cite{verbaarschot:streamline} and another `hybrid' ratio-sum ansatz $R_H$ \cite{schaefer:shuryak:iilm}. This is summarised in \reftable{table:different:ansatz}.

\begin{table*}[tbp]
\begin{center}
 \begin{tabular}{c|m{0.8\textwidth}} 
Ansatz  & Description \\\hline\hline
$R_E$ & Interactions for ratio ansatz as derived in this paper. \\\hline
$R_H$ & Gluonic interactions are derived from the ratio ansatz whereas the quark overlaps use a sum ansatz, \cite{schaefer:shuryak:iilm}.\\\hline
$S$ & Interactions have been derived from the so-called streamline ansatz. These are only available at zero temperature, \cite{schaefer:shuryak:iilm} \cite{verbaarschot:streamline}.
 \end{tabular}
\end{center}
\caption{Several ans\"atze for the classical background field have been proposed. The following table summarises what they will be referred to throughout the rest of the paper.}\label{table:different:ansatz}
\end{table*}

Phenomenological considerations have lead to the conclusion that the QCD vacuum consists of a dilute ensemble of instantons; a fact corroborated by lattice studies and self-consistency checks within the IILM. Diluteness and the localised nature of instantons render negligible contributions other than two-body interactions\footnote{A note on terminology: whenever we use the word interaction, we mean a quantity `normalised' to the dilute gas, i.e.\ we subtract the dilute gas counterpart if the term naturally occurs in the exponential, as in the classical gluonic interactions, or we divide by the dilute gas counterpart if the interaction is a pre-exponential factor, as in the gluonic Jacobian or the Dirac determinant.}, given here for an instanton--anti-instanton pair,
\begin{equation}
 A^a_\mu = - \frac{\bar{\eta}^a_{\mu\nu}\partial_\nu \Pi_1(x,\{x_1,\rho_1\}) + O^{ab} \eta^b_{\mu\nu} \partial_\nu \Pi_2(x,\{x_2,\rho_2\})}{1+\Pi_1(x,\{x_1,\rho_1\})+\Pi_2(x,\{x_2,\rho_2\})}\,,\label{eq:ratio}
\end{equation}
with $O=O_1^t O_2$. The formulas for an like-charged pairs follow trivially from the above.

\subsection{Gluonic interactions}

The complete classical gluonic interaction is given by the sum over all the possible pairings. It is clear from the structure of (\ref{eq:ratio}) that the colour degrees of freedom can be completely factorised out. After some lengthy algebra the result for the squared field strength can be written in the form
\begin{multline}
 F^a_{\mu\nu} F^a_{\mu\nu} = I + (\Tr O^tO + (\bar{\eta} O \eta)_{\mu\nu\mu\nu}) J + (\bar{\eta} O \eta)_{\rho\mu\rho\nu} I_{\mu\nu} \\
+ (\bar{\eta} O \eta)_{\mu\rho\nu\sigma} I_{\mu\rho\nu\sigma} + (\eta O^tO \eta)_{\mu\rho\nu\sigma} J_{\mu\rho\nu\sigma} + (\bar{\eta} O \eta)_{\alpha\mu\alpha\rho} (\bar{\eta} O \eta)_{\beta\nu\beta\sigma} K_{\mu\rho\nu\sigma} \label{eq:interaction:gluonic:density}\,.
\end{multline}
The different contributions are given in appendix \ref{app:interaction:gluonic}.

\begin{figure}[tbp]
\begin{center}
 \includegraphics[width=\figwidth,clip=true,trim=0mm 0mm 15mm 10mm]{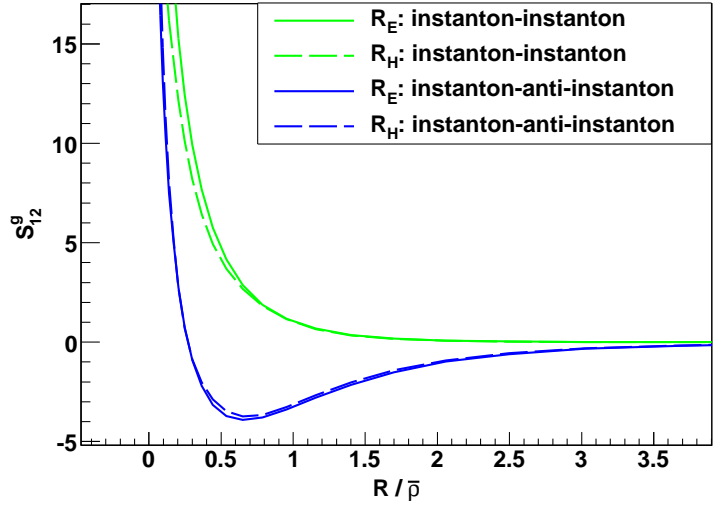}
\end{center}
 \caption{For instantons with equal sizes the interaction of $R_E$ agrees very well with $R_H$ for oppositely charged instantons. There is slight discrepancy for like-charged instantons, in that the repulsion is a bit steeper in the $R_E$ case. (We have set $\bar{\rho}=\sqrt{\rho^2_1+\rho^2_2}$.)}\label{fig:interaction:gluonic:equal}
\end{figure}

\begin{figure}[tbp]
\begin{center}
 \includegraphics[width=\figwidth,clip=true,trim=0mm 0mm 15mm 10mm]{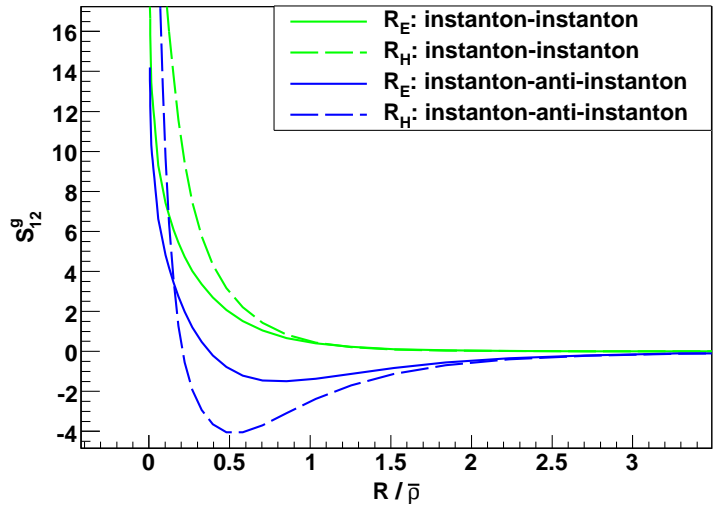}
\end{center}
 \caption{For unequal size parameters, e.g.\ $\rho_1/\rho_2=3$ in this case, large differences start to become apparent. The reason is that the dependence on the sizes is more complicated than the functional form $\sqrt{\rho_1\rho_2}$ used in $R_H$. Note that the attractive well is much deeper in the $R_H$ case which will eventually lead to a denser ensemble. (We have set $\bar{\rho}=\sqrt{\rho^2_1+\rho^2_2}$.)}\label{fig:interaction:gluonic:unequal}
\end{figure}

Factorising out the single instanton contributions and the coupling constant, the classical interaction between instantons is given by
\begin{eqnarray}
 S^g_{12}/S_0 &\equiv& V_{12} \equiv (S[A]/S_0 - 2)\,,\\
 S[A] &=& \frac{1}{4 g^2} \int F^a_{\mu\nu} F^a_{\mu\nu}\,,
\end{eqnarray}
where $S[A]$ is the action of the background gauge fields and $S_0=8\pi/g^2$ that of a single instanton. For equal sizes the agreement with $R_H$ is very good, see \reffig{fig:interaction:gluonic:equal}. However, for unequal sizes there are noticeable differences, see \reffig{fig:interaction:gluonic:unequal}. The discrepancy follows from the functional dependence on the size parameters being of the form $\sqrt{\rho_1\rho_2}$ in $R_H$. As can be seen from the asymptotic behaviours, see appendix \ref{app:interaction:gluonic:asymptotic}, the sizes enter rather in the combination $\sqrt{\rho^2_1+\rho^2_2}$, at least in the parameter regions of large and small separations; this is in agreement with \cite{diakonov:instanton:variational}\footnote{The authors of \cite{diakonov:instanton:variational} have considered the sum ansatz; for large separations, however, every ansatz is equivalent to the sum ansatz.}.

We will a adopt a couple of simplifications in the practical implementation that have been introduced in previous work. These are, on the one hand, the approximation of the high-frequency quantum interaction by an inverse running coupling constant evaluated at the scale of the mean instanton size and, on the other hand, the neglect of the Jacobian that introduces the collective coordinates and represents the low-frequency quantum interaction.

In the single instanton case the high frequency quantum fluctuations lead to charge renormalisation and the coupling constant is replaced by the running coupling at the scale given by the instanton size \cite{thooft:instanton:fluctuations}, $S_0(\rho)=8\pi/g^2(\rho)$. The same calculation has never been performed for a pair. In the original paper \cite{diakonov:instanton:variational}, the interaction part of high-frequency quantum fluctuations have been estimated to be subdominant to the classical interactions . In that paper the authors argue then that the quantum interaction can be estimated by modulating the (total) classical interaction with the inverse running coupling constant, a slowly varying function of the background field, at the scale provided by the mean instanton size $\bar{\rho}$. We will adopt the parametrisation put forth in \cite{shuryak:instantons:liquid:II,schaefer:shuryak:iilm} that estimates the scale of the running coupling constant on a pair-by-pair basis and uses the geometrical mean of the sizes to set this scale. The full gluonic interaction is then given by
\begin{equation}
 S^g_{12} = S_0(\sqrt{\rho_1 \rho_2}) V_{12}\,. \label{eq:interaction:gluonic}
\end{equation}

The Jacobian interaction is positive by definition and can therefore be interpreted as a repulsive (low-frequency) quantum interaction. A rough estimate of the large distance behaviour suggests that the asymptotic power-law decay is $O(1/R^6)$, with $R$ the separation between the pair. This is a faster decay than the well-known dipole--dipole interaction that follows from the classical action. For strong overlaps the Jacobian matrix will become approximately degenerate, and its determinant small, essentially because the matrix elements of the pair with the other instantons will be roughly equal. For complete degeneracy the singularity should be logarithmic because one singular value will tend to zero as the rank of the matrix decreases by one. The repulsion will thus be of the form
\begin{equation}
 \ln J_{12}^\mathrm{sing} \sim \ln \left(\frac{R^2}{\rho_1^2+\rho_2^2}\right)\,,
\end{equation}
with proportionality factor of order $O(1)-O(10)$ because, as we argued, the degeneracy is due to one overlapping pair and should not get contributions from other instantons. In (\ref{sec:numerical:implementation:interpolation:matching}) we will discuss the small separation asymptotic behaviour for the ratio ansatz; the analytical expressions are given in appendix \ref{app:interaction:gluonic:small}, and we note that the singular behaviour is also repulsive and logarithmic,
\begin{equation}
 I_{IA}^\mathrm{sing} \sim \ln \left(1+\frac{\rho_I^2+\rho_A^2}{R^2}\right)\,,
\end{equation}
with a proportionality factor that is again of order $O(1)-O(10)$. In the intermediate region it is harder to estimate the Jacobian interaction, but the logarithm should make its contribution subdominant. Also, the classical interaction is boosted by the quantum contribution through charge renormalisation. We conclude that the Jacobian interaction is probably negligible compared to the classical interactions.

Thus, the gluonic interactions we will use in this work will be given solely by the classical interaction.

\subsection{Quark Interactions}

The quark interaction arises from (\ref{eq:quark:overlap}), as is clear from our discussion in section \ref{sec:saturate}, and is purely quantum mechanical. As for the gluonic interaction, some further approximations have been used in the literature; we will adopt these, albeit rephrased sightly differently.

We will assume that the single instanton zero modes $\{\xi\}$ form a functional orthonormal basis, i.e.\ we neglect contributions arising from non-vanishing overlaps among the $\xi_i$. With this in mind, the finite dimensional low-frequency Dirac operator is then given by
\begin{equation}
 (\slashed{D}+m)_{ij}=\langle \xi_i|\slashed{D}+m|\xi_j \rangle=\slashed{D}_{ij}+m\delta_{ij}\,.
\end{equation}
To reiterate, we attribute the diagonal mass term to the requirement of orthonormality\footnote{Writing $H=H_{ij}|\psi_i\rangle \otimes \langle \psi_j|$ makes only sense if $\{\psi_i\}$ forms an orthonormal system, given the scalar product $\langle \cdot | \cdot \rangle$.\label{para:orthonormality}} rather than the degree of dilution of the instanton ensemble, e.g.\ \cite{nowak:rho:zahed:chiral_nuclear_dynamics}. On the practical level this is irrelevant in as far as we recover the same determinantal interaction as used in previous works.

The quark zero mode, in singular gauge and in the chiral representation, is given by \cite{grossman:dirac:zeromode}
\begin{eqnarray}
 \xi_I &=& \frac{1}{2\pi \rho_I} \sqrt{1+\Pi_I} \slashed{\mbox{$\partial$}} \frac{\Pi_I}{1+\Pi_I} 
\left(
\begin{array}{c}
 U_I \varphi\\
 0
\end{array}
\right)\,,\\
\xi_A &=& \frac{1}{2\pi \rho_A} \sqrt{1+\Pi_A} \slashed{\mbox{$\partial$}} \frac{\Pi_A}{1+\Pi_A} 
\left(
\begin{array}{c}
 0\\
 U_A \varphi
\end{array}
\right)\,,
\end{eqnarray}
with $\varphi_{\alpha a}=\epsilon_{\alpha a}$, normalised according to $\epsilon_{12}=1$. Finally, $U_i$ is the $3 \times 3$ colour matrix describing the collective coordinates for the colour embedding; it is related to the adjoint representation by $O^{ab}=1/2 \Tr(U \tau^a U^\dagger \tau^b)$, with $U=U_I^\dagger U_A$.

The Dirac operator, as defined above, is anti-hermitian. Eventually we need to diagonalise it, but, since readily available routines work with hermitian matrices, we display here the matrix elements of $i\slashed{D}$. Within the ratio ansatz, the matrix elements $T_{IA}= \int \xi_I^\dagger i \gamma_\mu \slashed{D}_\mu \xi_A$ are as follows
\begin{equation}
T_{IA} = \int d^4x \frac{1}{4\pi^2\rho_I \rho_A} \frac{1}{2}\Tr (U \tau^{+}_{\beta}) I_\beta \,. \label{eq:quark_interaction}
\end{equation}
The concrete realisation of $I_\beta$ is given in appendix \ref{app:interaction:quark}.

\begin{figure}[tbp]
\begin{center}
 \includegraphics[width=\figwidth,clip=true,trim=0mm 0mm 15mm 10mm]{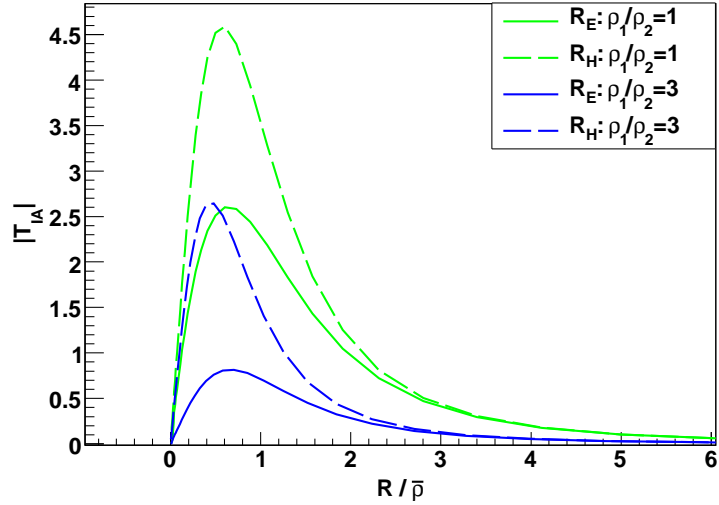}
\end{center}
 \caption{The relatively large discrepancy is due to the fact that $R_E$ uses the full ratio ansatz in the Dirac operator whereas $R_H$ uses the sum ansatz. (We have set $\bar{\rho}=\sqrt{\rho^2_1+\rho^2_2}$.)}\label{fig:overlap:quark}
\end{figure}

The rather large difference between $R_E$ and $R_H$, see top of \reffig{fig:overlap:quark}, is due to the fact that the latter use a sum ansatz. The ratio ansatz was introduced to remove the unphysical divergence in the field strength; no such problem afflicts the overlap matrix elements. On top of that the quark determinant is a pre-exponential factor and as an effective interaction the extra logarithmic factor should make it rather insensitive to its exact form, see \cite{shuryak:verbaarschot:interactions:finite:T}. Within our numerical framework, the full ratio ansatz does not produce any additional overhead and has the merit to be more consistent with the gluonic interactions. We have checked that upon neglect of the contributions special to the ratio ansatz, i.e.\ simplifying the overlaps so as to recover the sum ansatz, our results agree very well with those of $R_H$, apart form the aforementioned discrepancy in the instanton size parametrisation. As for the gluonic interactions, the colour matrices could again be completely factorised out.

\begin{figure}[tbp]
\begin{center}
 \includegraphics[width=\figwidth,clip=true,trim=0mm 0mm 15mm 10mm]{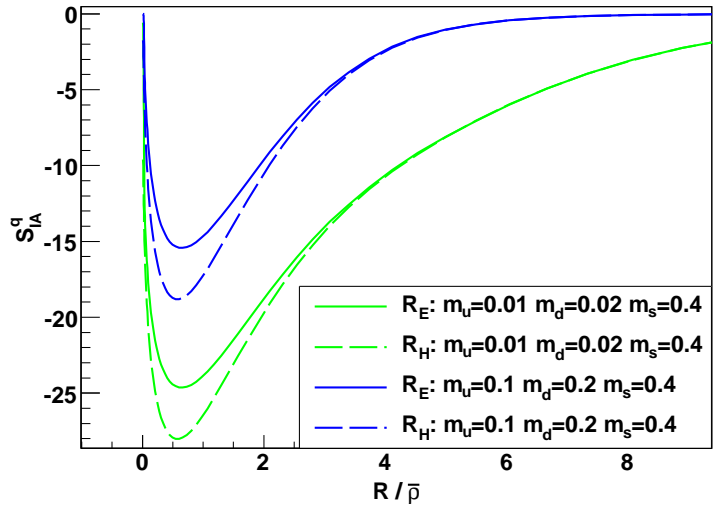}
\end{center}
 \caption{On the level of the effective interaction, the difference between $R_E$ and $R_H$ is not as pronounced, i.e.\ the relative difference has decreased substantially. We can clearly see that light quark masses lead to a stronger attractive interaction between instantons and anti-instantons. Note that the relative difference between the ans\"atze $R_E$ and $R_H$ does not seem to depend strongly on the quark masses. The instantons have been set up with equal sizes. (We have set $\bar{\rho}=\sqrt{\rho^2_1+\rho^2_2}$.)}\label{fig:interaction:quark}
\end{figure}

Note that the Dirac operator only connects instantons to anti-instantons due to the extra $\gamma$-matrix factor as compared to the mass operator, which vanishes between instantons and anti-instantons. Therefore the quark fluctuation operator has the following form
\begin{equation}
m \mathbb{I} - i 
\left(
 \begin{array}{cc}
  0 & T\\
  T^\dagger & 0
 \end{array}
\right)\,,
\end{equation}
with $T$ the $N_I \times N_A$ matrix of overlaps $T_{IA}$, and $N_I$ ($N_A$) is the number of instantons (anti-instantons); the $0$-matrices are $N_I \times N_I$ and $N_A \times N_A$ dimensional, respectively; finally, $\mathbb{I}$ is the identity operator on the quasi-zero mode space of dimension $(N_I+N_A) \times (N_I+N_A)$. To diagonalise $i\slashed{D}$, it suffices to know the singular-value-decomposition of $T$. The left and right singular vectors, $\psi^L$ and $\psi^R$, are defined by
\begin{eqnarray}
 T \psi^R_n &=& \lambda_n \psi^L_n\,,\\
 T^\dagger \psi^L_n &=& \lambda_n \psi^R_n\,.
\end{eqnarray}
The singular eigenvalues $\lambda_n$ are always positive. The kernel of the Dirac operator is spanned by the $\lambda=0$ singular eigenvectors, $\psi^K$, of either $T$ or $T^\dagger$, depending on whether $N_I<N_A$ or $N_I>N_A$. We can then construct the eigenvalue decomposition of the Dirac operator. The non-zero eigenvalue part has the following eigensystem
\begin{equation}
 \left\{
 \left.
 \left[
 \lambda_n,
 \left(
 \begin{array}{c}
  \psi^L_n\\
  \psi^R_n
 \end{array}
 \right)
 \right],
 \left[
 -\lambda_n,
 \left(
 \begin{array}{c}
  -\psi^L_n\\
  \psi^R_n
 \end{array}
 \right)
 \right]
 \right|
 n \in \{1,\dots,\min(N_I,N_A)\}
 \right\}\,.
\end{equation}
Finally, the kernel is spanned by the eigensystem
\begin{equation}
 \left\{
 \begin{array}{cl}
  
 \left\{
 \left.
 \left[
 0,
 \left(
 \begin{array}{c}
  \psi^K_n\\
  0
 \end{array}
 \right)
 \right]
 \right|
 n \in \{1,\dots,N_A-N_I\}
 \right\},
 & N_I<N_A\,, \\[3ex]

 \left\{
 \left.
 \left[
 0,
 \left(
 \begin{array}{c}
  0\\
  \psi^K_n
 \end{array}
 \right)
 \right]
 \right|
 n \in \{1,\dots,N_I-N_A\}
 \right\},
 & N_I>N_A\,.

 \end{array}
 \right.
\end{equation}

Note that the non-zero eigenvalues come in pairs. Together with the zero eigenvalues, the determinant of the Dirac operator can be written as
\begin{equation}
 \det(i\slashed{D}) = m^{|Q|} \prod_n^{\min(N_I,N_A)} (m^2+\lambda^2_n)\,,
\end{equation}
with $Q=N_I-N_A$ the topological charge. If we are only interested in the determinant, and not so much in the eigensystem, this can be put in the equivalent form
\begin{equation}
 m^{|Q|}
 \left\{
 \begin{array}{cl}
 \det(TT^\dagger + m^2), & Q<0 \\
 \det(T^\dagger T + m^2), & Q>0
 \end{array}
 \right. \,.\label{eq:interaction:quark:determinant}
\end{equation}

Upon placing this term into the exponential, the normalised determinant of quark zero mode overlaps leads to an effective interaction. The normalisation consists of dividing (\ref{eq:interaction:quark:determinant}) by $m^{N_I+N_A}$. The quark interaction is thus given by
\begin{equation}
 S^q_{N_f} = - \sum_{n=1}^{N_f}
 \left\{
 \begin{array}{cl}
 \ln\det(TT^\dagger + m^2_n) - N_I \ln m^2_n, & Q<0 \\
 \ln \det(T^\dagger T + m^2_n) - N_A \ln m^2_n, & Q>0
 \end{array}
 \right. \,,\label{eq:interaction:quark}
\end{equation}
with $N_f$ the number of active quark flavours. Note that the quark interaction is always attractive. This follows from the fact that we can write the overlap matrix for each flavour as $\mathbb {I} + \frac{T^2}{m^2_n}$, and this form makes it explicit that the determinant is bounded from below by unity because the smallest eigenvalue is easily seen to satisfy $\lambda_{\mathrm{min}} \ge 1$.

This exhausts the interactions in the IILM because the fluctuation operator of the ghost part is positive definite and its lack of zero modes prevents the construction of the low frequency part of the spectrum within the moduli-space approximation. We are thus left with the high-frequency part which, as in the other cases, is assumed to factorise and cannot lead to interactions.

\section{Numerical Implementation}
\label{sec:numerical:implementation}

\subsection{Interpolation and asymptotic matching}
\label{sec:numerical:implementation:interpolation:matching}

The decoupling of the colour degrees of freedom is a computational benefit: by using global $SO(4)$ transformations, without loss of generality, we place the first instanton at the origin and the second along the $z$-direction. The initial orientational dependence is then factored out of the integrand and combines with the colour matrices as in \cite{diakonov:instanton:variational}. These integrations are too time consuming to perform during actual simulations; instead, they are computed beforehand to fill interpolation tables that are, in turn, used during the simulations. The interpolation grid is three-dimensional, and depends on $\rho_1$, $\rho_2$ and $R=|x_1-x_2|$. For numerical stability we choose to use simple linear interpolation.

A uniform grid can, of course, only extend over a finite region and we must decide which portion of the parameter space to cover. We took the single instanton moduli-space measure as a guide for the size grid because, suitably normalised, it can be interpreted as a probability density. We choose the lower limit, $\rho_\mathrm{min}\approx \frac{2}{30}  \Lambda$, to be a fairly small quantile\footnote{It corresponds to less than the millionth quantile.}. Here, $\Lambda$ is the scale at which QCD starts to become strongly coupled. The upper limit is set to $\rho_\mathrm{max}=\Lambda$. Larger instantons cannot be treated consistently in the IILM because it uses perturbation theory, which breaks down below $\Lambda$.

We believe that these choices cover the relevant parameter space, and we sample the sizes from the interval $[\rho_\mathrm{min},\rho_\mathrm{max}]$. As a consistency check we monitored the actual size distribution and did not find any evidence for a significant weight at the edges of the sample interval. We therefore conclude that this procedure is well-defined.

The classical gluonic interaction in the ratio ansatz suffers from gauge singularities that prevent us from extending the grid down to vanishingly small instanton separations, $R \to 0$. The opposite limit, $R \to \infty$, cannot be covered either unless we use a non-uniform measure on $\mathbb{R}_+$. In principle this would seem like the most elegant approach, however, it is not feasible practically because the numerical integration becomes inaccurate at larger separations; the only remedy would be to set very small error tolerances for the numerical integrations, but that is computationally prohibitive. Therefore, we decided to use matching formulas for both the large and small separation regimes.

The rationale is not to derive accurate formulas in absolute terms but to get the absolute value from the interpolation results at a matching point $R_m$. The matching formulas are thus to be understood as accurate in a relative sense, i.e.\ the asymptotic interactions $f_\mathrm{asy}$ should behave, asymptotically, like the exact numerical interactions $f_\mathrm{ex}$. This ensures that we reproduce the correct fall-off or singularity behaviour. Thus, we compute the interactions according to
\begin{equation}
f(R) = f_\mathrm{asy}(R) \frac{f_\mathrm{ex}(R_m)}{f_\mathrm{asy}(R_m)}\,,
\label{eq:asymptotic:matching}
\end{equation}
whenever they fall out of the grid. Since the localisation of the instantons is set by the sizes, it is natural for the matching point to be proportional to the former. Eventually, the exact proportionality factor follows from an `optimisation' procedure, given that we aim for the interpolated interactions to be correct at the one percent level.

The full gluonic interaction consists of different pieces that are added together, (\ref{eq:interaction:gluonic:density}). We could use (\ref{eq:asymptotic:matching}) for these subinteractions term by term but it turns out that such a matching is numerically rather unstable. Thus, even though we are only interested in asymptotic relations, we need a systematic procedure that insures that the different asymptotic subinteractions are added up with the correct magnitude relative to each other.

\begin{figure}[tbp]
\begin{center}
 \includegraphics[width=0.8\figwidth]{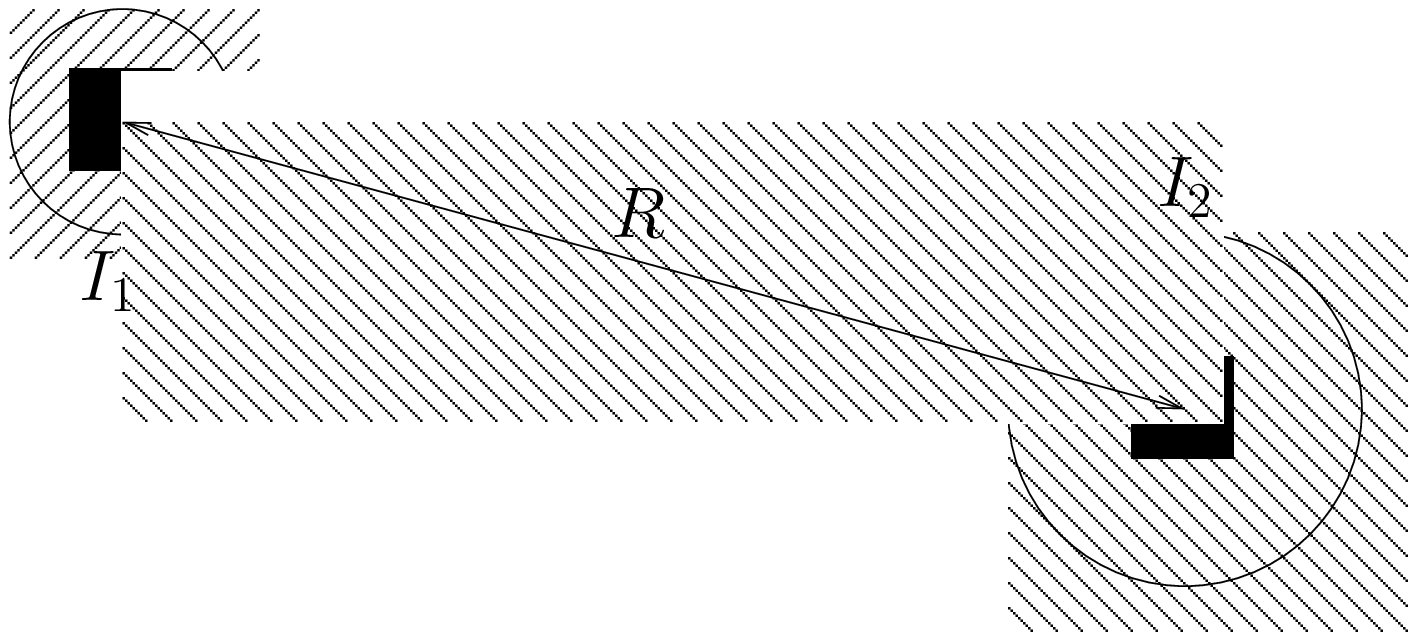}
\end{center}
 \caption{The instantons $I_1$ and $I_2$ are so far apart that, within the shaded region that give the dominant contribution to the field strength of each, the partner's field strength is roughly constant and fixed at $x_\mu-R_\mu \approx -R_\mu$. We can then safely extend the integration region to be all of $\mathbb{R}^4$, with a negligible error due to the rather strong localisation of the individual instantons. }\label{fig:interaction:separation:large}
\end{figure}

For the large separation case we want the instantons to be so far apart from each other that within the region in which the field strength for $I_1$ is strong the field strength of $I_2$ hardly changes: we can approximate $x_\mu-R_\mu \approx -R_\mu$\footnote{We use a translation to place $I_1$ at the origin.}. Since the field strength is negligible at and beyond $R_\mu$, we can safely extend the integration region to cover all of $\mathbb{R}^4$. The field strength of $I_2$ behaves as a constant, and we can use the rather simple rational expression for the interaction in terms of the 't Hooft potential to find exact results. We add to this the analogous contribution from $I_1 \leftrightarrow I_2$. The configuration is illustrated in \reffig{fig:interaction:separation:large}.

\begin{figure}[tbp]
\begin{center}
 \includegraphics[width=\figwidth]{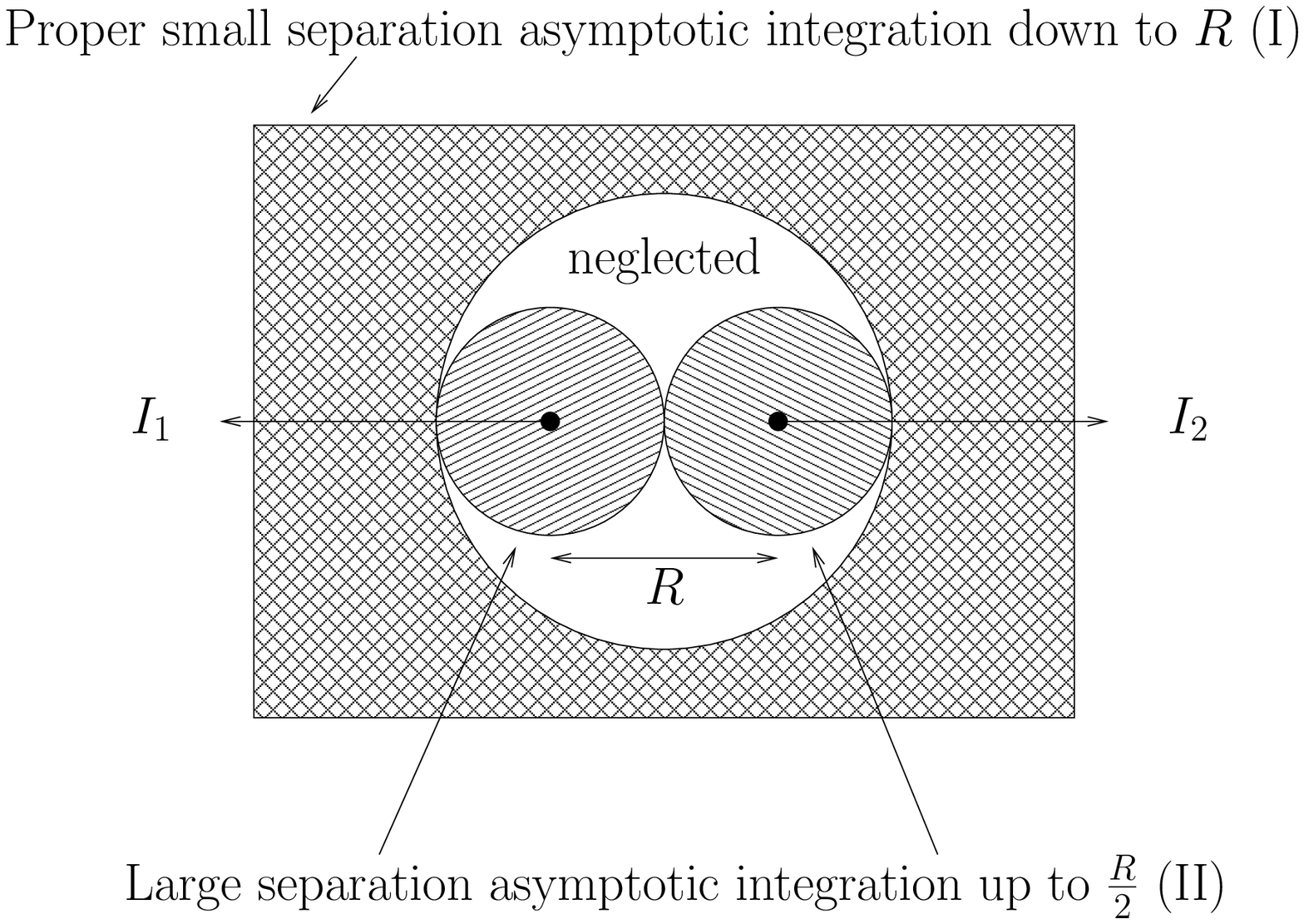}
\end{center}
 \caption{The instantons $I_1$ and $I_2$ are strongly overlapping. We approximate the integral by, first, integrating over $I_1$ keeping $I_2$ fixed at $R_\mu$, as in the large separation case but with upper limit $R/2$; to this we add the analogous contribution from $I_2$. Secondly, the, possibly, singular behaviour is picked up by integrating from infinity down to $R$, and approximating the arguments to be $x_\mu-R_\mu/2 \approx x_\mu$ and $x_\mu+R_\mu/2 \approx x_\mu$ respectively.}\label{fig:interaction:separation:small}
\end{figure}

We shall call this the zeroth order approximation, and it is clear that, to this order, terms odd in derivatives of $I_i$ will vanish due to $O(4)$ symmetry. However, it turns out numerically that, upon combining all the different terms from (\ref{eq:interaction:gluonic:density}), the zeroth order terms are sufficient. In particular, no  non-integrable terms are present; as we will demonstrate in \cite{wantz:iilm:3}\footnote{At finite temperature, the present framework remains virtually unchanged. The only modifications are that the 't Hooft potential, $\Pi$, changes and that the integration region becomes $S^1 \times \mathbb{R}^3$.}, the finite temperature interactions do not include terms such as (\ref{eq:dyon_dyon_interaction}) that prohibit the thermodynamic limit. Our formulas are given in appendix \ref{app:interaction:gluonic:large}.

In principle, we can compute the neglected terms by going to first order, i.e.\ $g(x-R) \approx g(-R) + x_\mu \partial_\mu g(-R)$, or beyond. Such higher order contributions will typically no longer converge on $\mathbb{R}^4$. It seems natural to cut them off at $R$, and this will generally lead to logarithms, $\ln(1+R^2/(\rho_1^2+\rho_2^2))$, together with rational functions. However, in contrast to the fitting formulas of \cite{schaefer:shuryak:iilm}, the Taylor expansion of our asymptotic formulas produce only power-law like decays for large separations; in addition they fall off more strongly than the zeroth order terms and thus will not produce non-integrable interactions either. We have thus achieved our goal of deriving interactions that allow us to study the thermodynamic limit of the IILM.

We now turn to the case of asymptotically small separations. A typical situation is depicted in \reffig{fig:interaction:separation:small}. The rationale is to split the integration into 2 regions.

\begin{enumerate}
 \item[I] The far-field region beyond both centres, placed symmetrically around the origin; we approximate the arguments by $x_\mu \pm R_\mu/2 \approx \pm x_\mu$.
\item[II] The region around each instanton up to $R/2$, with $R$ the pair separation. We integrate around $\pm R_\mu/2$ keeping the arguments of the partner instanton fixed at $x_\mu \mp R_\mu/2 \approx \mp R_\mu/2$. This is similar to the large separation case, but here we only integrate up to $R/2$.
\end{enumerate}

Region I accounts for possible singularities. After adding up all the different subinteractions, the singularities from region I dominate the total interaction. Since we need the region II approximations anyway in the large separation case, it does not represent any extra overhead to use them as well in the small separation limit.

\begin{figure}[tbp]
\begin{center}
 \includegraphics[width=\figwidth,clip=true,trim=0mm 0mm 15mm 10mm]{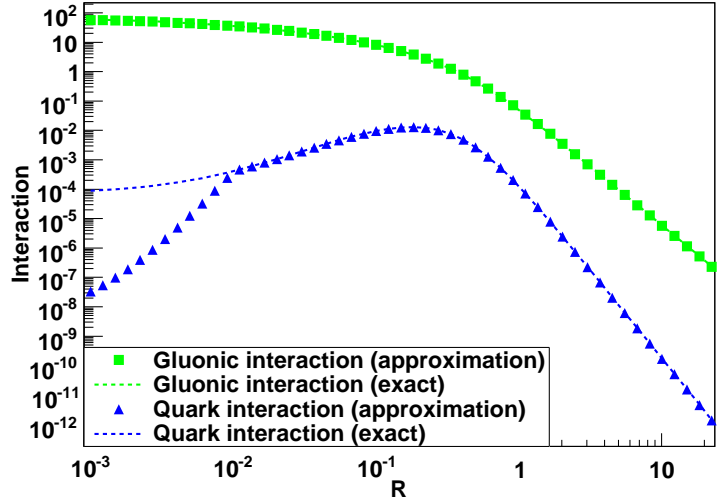}
\end{center}
 \caption{The gluonic interaction is well approximated by the combination of interpolation and asymptotic matching. The quark overlap is very poorly approximated by the zeroth order small separation asymptotic formula; it tends to zero with too high a power as compared to the exact result. The correct behaviour can in principle be obtained from higher orders, and we've estimated that the second order contribution will suffice. In practice, the quark interaction in this region is completely irrelevant as compared to the gluonic interaction.}\label{fig:interaction:approximation:gluonic:quark}
\end{figure}

\begin{figure}[tbp]
\begin{center}
 \includegraphics[width=\figwidth,clip=true,trim=0mm 0mm 15mm 10mm]{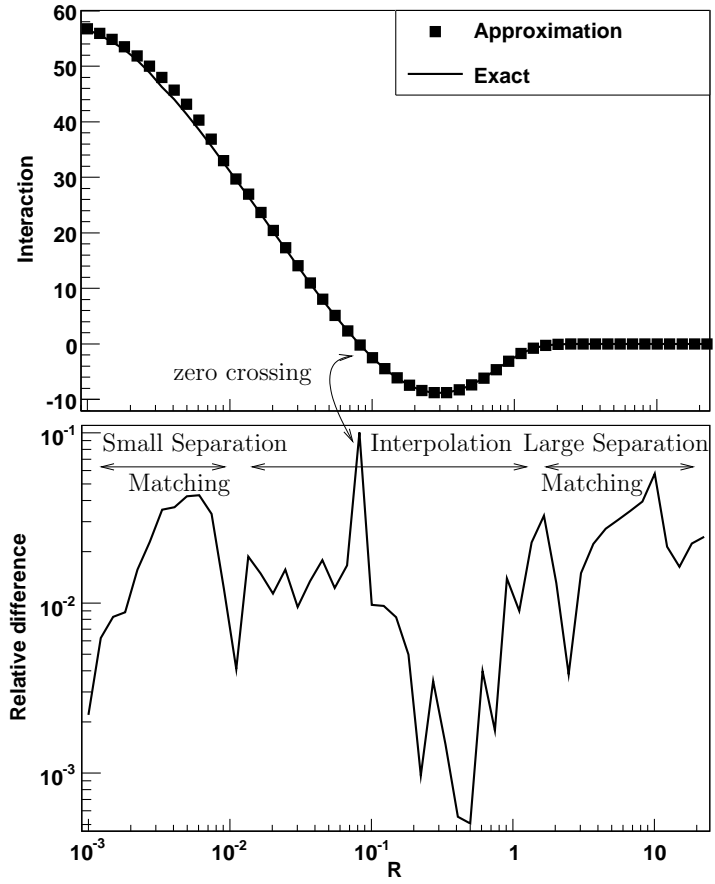}
\end{center}
 \caption{The total pair interaction is accurate on the one percent level. Note that the error in region I from the quark interaction is completely negligible. The spike in the bottom plot is due to the sensitivity to zero crossings.}\label{fig:interaction:approximation:total}
\end{figure}

In \reffigs{fig:interaction:approximation:gluonic:quark} and \ref{fig:interaction:approximation:total} we plot the exact and approximate result for the gluonic, quark and total interaction. Note that some subinteractions in the gluonic sector are poorly approximated by the zeroth order asymptotic matching formulas in region I. However, those terms that do exhibit singularities completely dominate, and the total gluonic interaction is well approximated for all separations. The quark overlap consists of just one term, which is not well described in region I. We estimated that the correct asymptotic behaviour can be obtained at second order. However, this is unnecessary since the quark interaction is bounded in that region and the gluonic interaction completely dominates. Combining the gluonic and fermionic interactions, we see that the total pair interaction is accurate on the one percent level.

Note that we use interpolation up to very strong overlaps; also, instantons rarely enter region I during simulations because it makes up a very small part of the total volume box. In any case, region I overlaps will almost certainly be rejected in Monte Carlo moves, and therefore quantities that are computed solely from the quark interaction, as for instance the quark condensate, are very insensitive to large errors in region I. Ultimately, this is the reason why we chose the interpolation grid to cover such strong overlaps.

\subsection{Monte Carlo}

Previous studies, lattice results and phenomenology indicate that the instanton ensemble is fairly dilute. Therefore, we organise the partition function into a dilute gas measure times the exponential of interactions,
\begin{eqnarray}
 Z &=& \sum_{N_I, N_A}^\infty \frac{1}{N_I !} \frac{1}{N_A !} \prod_i^{N_I} d(\rho_i) \prod_j^{N_A} d(\rho_j) \exp\left(-(S^g+S^q)\right) \label{eq:iilm:partition_function}\,,\\
 &\equiv& \sum_{N_I, N_A}^\infty \frac{1}{N_I !} \frac{1}{N_A !} Z_{N_I,N_A}\,,\\
 S^g &=& \sum_{ij} S^g_{ij}\,,
\end{eqnarray}
where $S^g_{ij}$ is given in (\ref{eq:interaction:gluonic}) and $S^q$ is given in (\ref{eq:interaction:quark}). We follow \cite{schaefer:shuryak:iilm} and use one-loop accuracy for the charge renormalisation factor that modulates the classical gluonic interaction, i.e.\ $S_0(\sqrt{\rho_1 \rho_2}) \to \beta_1(\sqrt{\rho_1 \rho_2})$, with $\beta_1$ given below in (\ref{eq:running:coupling:one-loop}). Although not really consistent, the single instanton density is given at two-loop in order to replace the pre-exponential bare by running coupling constants \cite{diakonov:instanton:variational}; the former were induced by the transformation to collective coordinates. The two-loop single instanton measure is then given by
\ifthenelse{\equal{\Qclass}{revtex4}}{
\begin{align}
 d(\rho) &= d_0^g(\rho) \, d_0^q(\rho)^{N_f}\,, \label{eq:size_distribution} \\
 d_0^g(\rho) &= C_{N_c} \rho^{-5} \beta_1(\rho)^{2N_c} \exp\left[-\beta_2(\rho) +\left(2N_c-\frac{b'}{2b} \right) \frac{b'}{2b}\frac{\ln\beta_1(\rho)}{\beta_1(\rho)} \right]\,,\\
d_0^q(\rho) &=m \rho \exp \left(-\frac{1}{3} \ln m\rho + \frac{\frac{1}{3}\ln m\rho + 2\alpha -(6 \alpha + 2 \beta ) (m\rho)^2 + 2 A_1 (m\rho)^4 - 2 A_2 (m\rho)^6}{1-3(m\rho)^2 + B_1 (m\rho)^4 + B_2 (m\rho)^6 + B_3 (m\rho)^8} \right)\,.
\end{align}
}{}
\ifthenelse{\equal{\Qclass}{elsarticle}}{
\begin{align}
 d(\rho) &= d_0^g(\rho) \, d_0^q(\rho)^{N_f}\,, \label{eq:size_distribution} \\
 d_0^g(\rho) &= C_{N_c} \rho^{-5} \beta_1(\rho)^{2N_c} \exp\left[-\beta_2(\rho) +\left(2N_c-\frac{b'}{2b} \right) \frac{b'}{2b}\frac{\ln\beta_1(\rho)}{\beta_1(\rho)} \right]\,,\\
d_0^q(\rho) &=m \rho \exp \left(-\frac{1}{3} \ln m\rho \right.\\
& + \left. \frac{\frac{1}{3}\ln m\rho + 2\alpha -(6 \alpha + 2 \beta ) (m\rho)^2 + 2 A_1 (m\rho)^4 - 2 A_2 (m\rho)^6}{1-3(m\rho)^2 + B_1 (m\rho)^4 + B_2 (m\rho)^6 + B_3 (m\rho)^8} \right)\,.
\end{align}
}{}
For the quark term, $d_0^q$, we use the generalisation of 't Hooft's \cite{thooft:instanton:fluctuations} result valid for arbitrary mass \cite{dunne:hur:lee:min:instanton:determinant:mass}. The different terms in $d_0^g$ are given by
\begin{align}
 C_{N_c} &= \frac{0.466 \,e^{-1.679 N_c}}{(N_c-1)!(N_c-2)!}\,,\\
 \beta_1(\rho) &= -b \ln (\rho \Lambda)\,,&  b &= \frac{11}{3} N_c - \frac{2}{3} N_f\, ,\label{eq:running:coupling:one-loop}\\
 \beta_2(\rho) &= \beta_1(\rho) + \frac{b'}{2b}\ln \left( \frac{2}{b} \beta_1(\rho) \right)\,,& b' &= \frac{34}{3} N_c^2 - \frac{13}{3} N_c N_f + \frac{N_f}{N_c}\,.
\end{align}
Note that the above has been derived in Pauli-Villars regularisation.

Being an interacting many-body system, the partition function cannot be evaluated analytically, and we choose Monte Carlo methods to cope with it numerically. More precisely, we will use the Metropolis algorithm to sample the important integration regions of the partition function. This is, of course, all well known, but it seems appropriate to introduce the, possibly less known, Monte Carlo moves corresponding to insertion and deletion of instantons needed for grand canonical simulations.

Following the usual strategy of imposing detailed balance, the simplest insertion/deletion algorithm consists of randomly placing an instanton in the box and randomly selecting an instanton to be removed. Imposing detailed balance and considering the case of an instanton, we arrive at
\begin{equation}
 \frac{1}{V} p^\mathrm{eq}_{N_I,N_A} \mathcal{A}_{N_I,N_I+1} = \frac{1}{N_I+1} p^\mathrm{eq}_{N_I+1,N_A} \mathcal{A}_{N_I+1,N_I}\,. \label{eq:detailed:balance:ins:del}
\end{equation}
As usual, $p^\mathrm{eq}_{N_I,N_A}=Z_{N_I,N_A}/Z$\footnote{Note that we neglect the factorial terms in the definition of the equilibrium probability density $p^\mathrm{eq}$ because they are an artifact as far as the measure is concerned. They have been introduced to render the integration volume simple, i.e.\ the product of the single instanton moduli-spaces $M^{N_I}$. During the integration process all the permutations of a given set of coordinates are generated, but, since the instantons are indistinguishable, they really correspond to only one configuration. To correct for this overcounting, we then have to divide by a factor of $N_I!$. The important point is that for the transition probabilities these factorial factors are irrelevant.} is the probability to be in the state $\{N_I,N_A\}$. The acceptance probability $\mathcal{A}_{ij}$ is implicitly defined through (\ref{eq:detailed:balance:ins:del}), and the Metropolis algorithm defines it to have the following form, \cite{frenkel:smit:molecular:simulation},
\begin{eqnarray}
 \mathcal{A}_{N_I,N_I+1} &=& \min(1,\mathcal{A})\,,\\
 \mathcal{A}_{N_I+1,N_I} &=& \min(1,\mathcal{A}^{-1})\,.
\end{eqnarray}
Plugging this into (\ref{eq:detailed:balance:ins:del}) we finally arrive at
\begin{equation}
 \mathcal{A} = \frac{V}{N_I+1} \frac{p^\mathrm{eq}_{N_I+1,N_A}}{p^\mathrm{eq}_{N_I,N_A}}\,.  \label{eq:detailed:balance:ins:del:acceptance}
\end{equation}

The difference to ordinary Monte Carlo moves, as used in the canonical ensemble\footnote{That is, updates for the positions, sizes and colour orientations}, is that the proposal probabilities do not cancel and the transition matrix is not symmetric. In this specific case, the proposal probability for an insertion is $\mathcal{P}^{\mathrm{ins}}_P=1/V$, corresponding to the probability to place the instanton randomly within the box, whereas the proposal probability for a deletion is $\mathcal{P}^{\mathrm{del}}_P=1/(N_I+1)$, corresponding to the probability to select an instanton among the $N_I+1$ available.

When we perform the standard updates, it is easy to monitor the acceptance rates and tune the the proposal probabilities to achieve good rates, i.e.\ $50\%$ say. For the move described by (\ref{eq:detailed:balance:ins:del:acceptance}) we do not have a parameter to tune though. At $T=0$, this is not really a big issue because it turns out that the acceptance rate is $\approx 0.4$, even for the rather small quark masses that we will use. This is still acceptable and does not really justify the overhead of more sophisticated update algorithms.

\begin{table*}[tbp]
\begin{center}
 \begin{tabular}{c|c|c|c|c|c|c|c}
  $\delta_{C}$ & $\delta_{GC}$ & $\langle N \rangle$ & $\xi_N$ & $\langle Q^2 \rangle$ & $\xi_{Q^2}$ & $\langle S_{\mathrm{int}} \rangle$ & $\xi_\mathrm{int}$ \\\hline
  $0.5$ & $0.5$ & $101.4(6)$ & $2500$ & $2.4(1)$ & $200$ & $-5.005(5)$ & $1200$ \\\hline
  $0.6$ & $0.4$ & $100.8(6)$ & $2100$ & $2.30(5)$ & $130$ & $-5.0010(5)$ & $1400$ \\\hline
  $0.8$ & $0.2$ & $102.6(6)$ & $4000$ & $2.35(4)$ & $130$ & $-5.010(2)$ & $2400$ \\\hline
  $0.9$ & $0.1$ & $102.1(6)$ & $5000$ & $2.5(1)$ & $270$ & $-5.016(3)$ & $3700$ \\\hline
 \end{tabular}
\end{center}
\caption{The sample size is roughly equivalent for each set, with $200$ independent configurations generated according to the autocorrelation time $\xi_N$. Considering some bulk properties, we see that the sampling does not really depend on the a-priori-probabilities $\delta_i$. Even though the autocorrelation times are only rough estimates, we will take the data at face value and choose $\delta_{C}=0.6$ and $\delta_{GC}=0.4$ for the remaining simulations.}\label{table:mc:ins:del}
\end{table*}

Also, note that such an insertion/deletion step is a fairly large change as compared to the normal coordinate updates, and so these grand canonical moves actually help to sweep through phase space more quickly.

Finally, we need to decide how many grand canonical moves we perform per coordinate update, that is we need to fix the a-priori-probabilities $\delta_C$ and $\delta_{GC}$. We found that, for $T=0$, the ensemble is not sensitive at all to this parameter, see \reftable{table:mc:ins:del}. Since we will ultimately be interested in computing the topological susceptibility, we will aim to achieve low autocorrelation times for the instanton number, $N=N_I+N_A$, and topological charge, $Q=N_I-N_A$, i.e.\ we will perform rather more insertion/deletion moves than less. In practice we perform canonical moves only $60\%$ of the time. 

\subsection{Fermionic determinant}
\label{sec:fermion_determinant}

As mentioned in the introduction, we want to study the IILM for 'physical' quark masses. In that case, we must make sure that the simulation box is large to be insensitive to finite size effects. In the lattice community it is common practice to use a box length that corresponds to $4-5$ times the wavelength of the lightest propagating degree of freedom, which is the pion. In practice, we want to circumvent the need for extremely large boxes by studying the thermodynamic limit, $V \to \infty$.

As compared to fitting formulas, our combination of interpolation and asymptotic matching results in a rather substantial computational overhead. This is particularly so in the quenched case. For unquenched simulations the situation is less drastic as the computationally most demanding part is the evaluation of the determinant and/or the determination of the eigensystem of the Dirac operator. Increasing the simulation box, i.e.\ increasing the number of instantons, this becomes the bottleneck to large volume simulations.

The Monte Carlo changes are, however, of a rather simple form, changing only one column of the overlap matrix $T$ at a time. We can therefore use decomposition update techniques to reduce the complexity from $O(N^3)$ to $O(N^2)$.

For the update step we only need to evaluate the determinant (\ref{eq:interaction:quark:determinant}). Given the fact that $m^2 + TT^\dagger$, respectively $m^2 + T^\dagger T$, is a positive definite hermitian matrix, the fastest evaluation will be achieve by using the Cholesky decomposition. An added bonus is that the Cholesky decomposition and its algorithm are known to be very stable.

Focusing on $M^2 = m^2 + TT^\dagger = L D L^\dagger$, an update $T' = T + \Delta T$ can be written as two rank 1 updates for $M^2$, of the form
\begin{equation}
 M'^2 = M^2 + \Phi\Phi^\dagger - \Psi\Psi^\dagger\,, \label{eq:mass:matrix:update}
\end{equation}
with $\Phi$, $\Psi$ vectors. Details are given in appendix \ref{app:cholesky:update}, where we also discuss more efficient ways to deal with adding and removing instantons, and the corresponding updates. The Cholesky decomposition can be updated efficiently when it only changes by rank 1 matrices, that is transformations of the form
\begin{equation}
 M'^2 = L' D' L^{'\dagger} = M^2 + \alpha z z^\dagger = L ( D + \alpha w w^\dagger) L^\dagger\,,
\end{equation}
where $Lw=z$. The algorithms then compute the decomposition of $D + \alpha w w^\dagger = \tilde{L} \tilde{D} \tilde{L}^\dagger$, which can be achieved in $O(N^2)$ because $D$ is diagonal. Furthermore, the matrix $\tilde{L}$ has a special form which allows an efficient matrix multiplication, $L' = L \tilde{L}$, in $O(N^2)$. Details can be found in \cite{gill:golub:murray:saunders:modify:factorization}. The algorithm we use in practice is known to be unstable for downgrading, $\alpha<0$, unless the resulting matrix, $M'^{2}$, is known to be positive definite. Since upgrading, $\alpha>0$, is always stable, it is important to perform the two consecutive updates in the order given by (\ref{eq:mass:matrix:update}).

In general we will be performing grand canonical simulations, and need to keep track of two decompositions, one for $m^2 + TT^\dagger$ and one for $m^2 + T^\dagger T$. Furthermore, we deal with 3 active quarks so that each Monte Carlo update entails $2 \cdot 2 \cdot 3 = 12$ rank 1 updates. We find that for an ensemble with $100$ instantons and $100$ anti-instantons we achieve a computational gain of a factor of $2$ as compared to the full Cholesky decomposition.

\section{Different Ensembles}
\label{sec:ansaetze}

To be predictive, the IILM should not depend too sensitively on the chosen ansatz. Given the fact that, for instance, the streamline and the ratio ansatz have quite different functional forms for overlapping pairs, the insensitivity of the model to specific background ans\"atze can only be determined a posteriori. On a heuristic, level we expect insensitivity to emerge if the ensemble stabilises in a rather dilute form so that the precise functional form of the repulsion is irrelevant. The large separation limit is a priori unproblematic because all ans\"atze are constructed such that, asymptotically, they approach the simple sum ansatz, i.e.\ $A=A_1+A_2$, with $A$ the gauge field.

\begin{figure}[tbp]
\begin{center}
 \includegraphics[width=\figwidth,clip=true,trim=0mm 0mm 15mm 10mm]{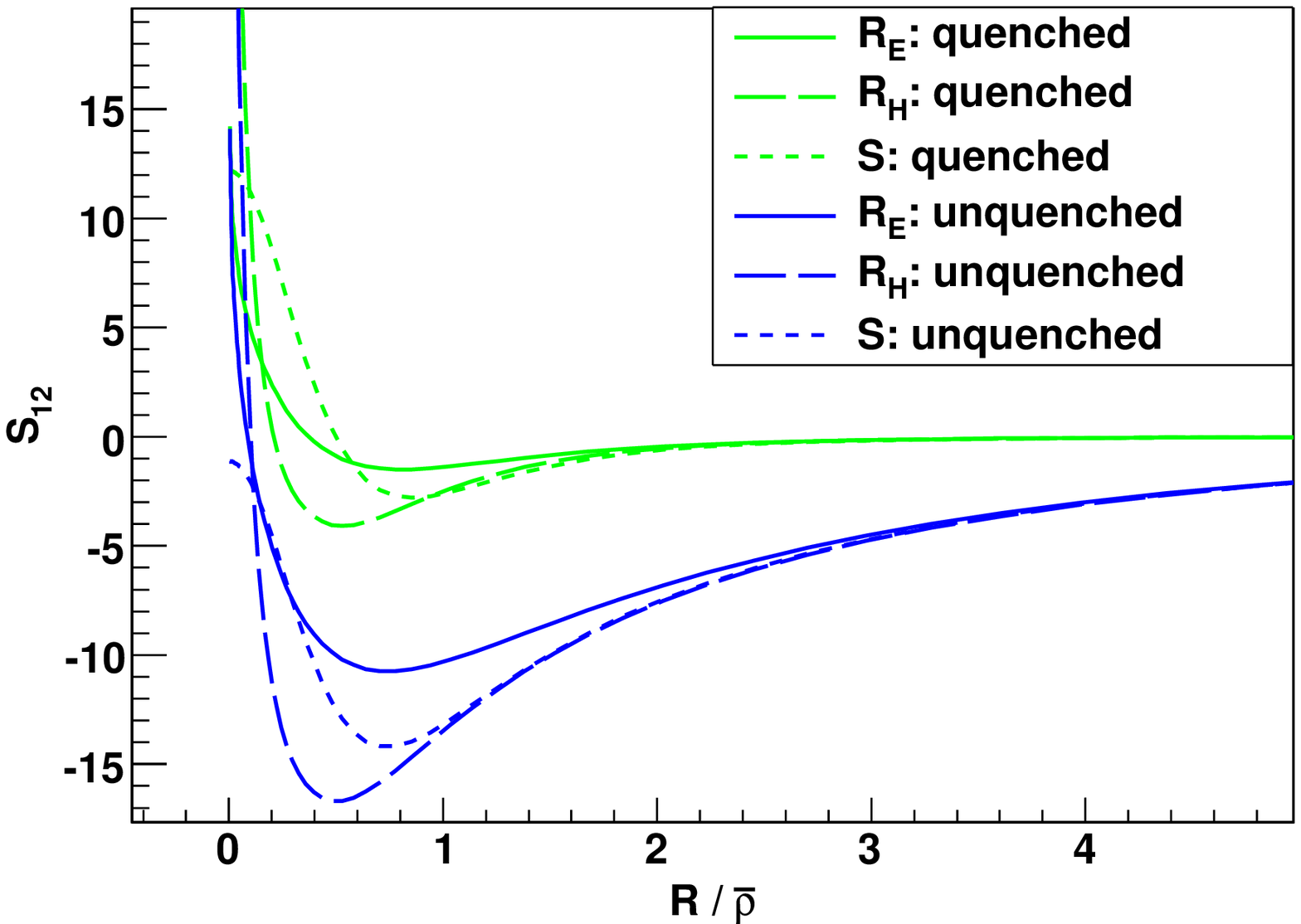}
 \includegraphics[width=\figwidth,clip=true,trim=0mm 0mm 15mm 10mm]{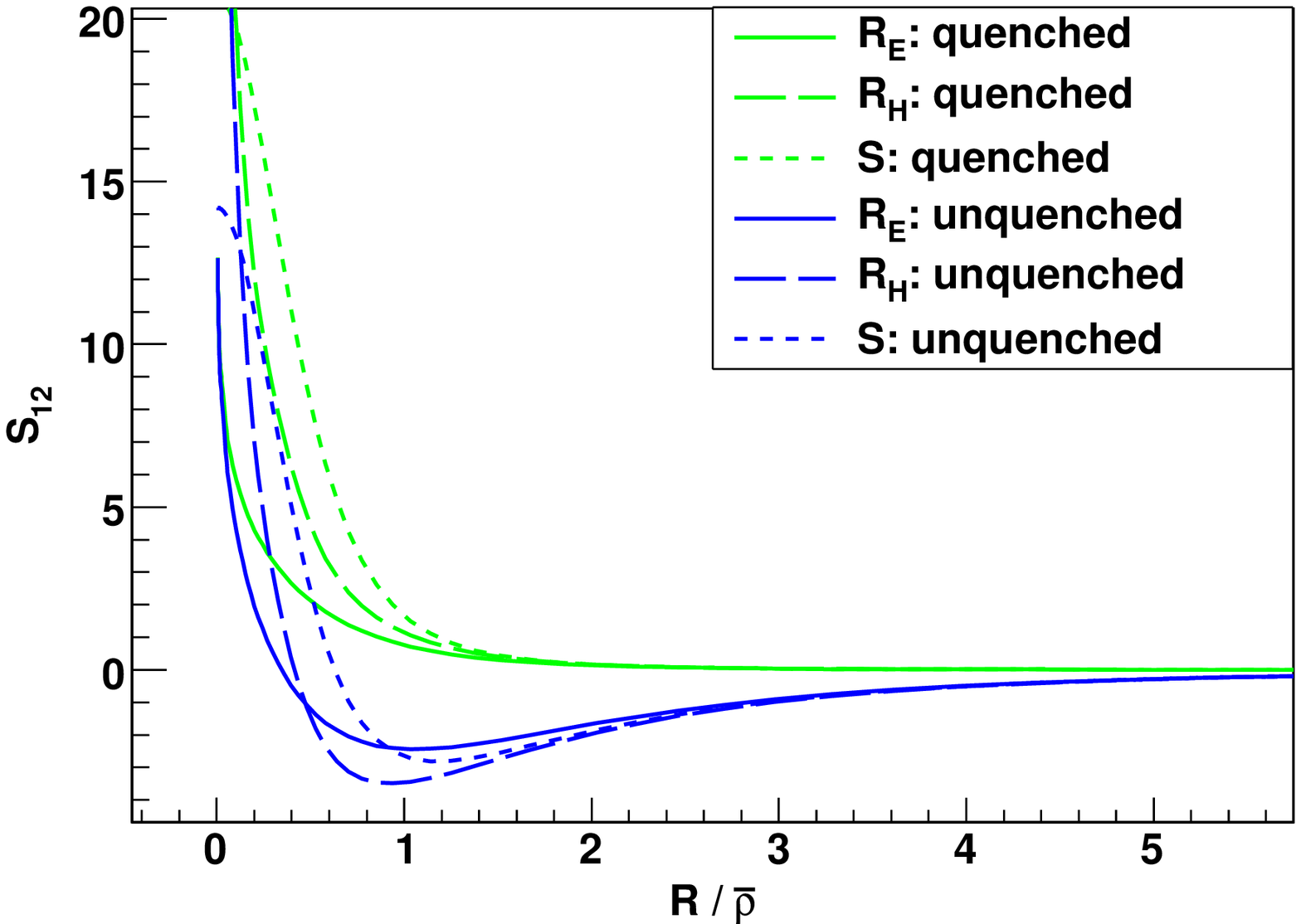}
\end{center}
 \caption{\textit{Top:} Most attractive colour orientation. \textit{Bottom:} Random colour orientation. For both graphs the instantons have different size parameters. We expect the $R_H$ ensemble to be denser than the $R_E$ ensemble because the attraction well is deeper, whereas the excluded volume due to repulsion is not that different. Along the same lines, the $S$ ansatz should lead to a rather more dilute system in the quenched case. For unquenched simulations, the deeper attraction well, and the steeper and broader repulsion, of the $S$ interactions might lead to an ensemble roughly equivalent to the $R_E$ one.}\label{fig:interaction:total}
\end{figure}

First, we will frame our discussion on the pair interactions. The total effective interaction of a pair of oppositely charged partners is given by
\begin{equation}
 S_{12} = S_0(\sqrt{\rho_1 \rho_2}) V_{12} - \sum_{n=1}^{N_f}\ln \left( \frac{|T_{12}|^2 + m_n^2}{m_n^2} \right)\,.\label{eq:interaction:pair}
\end{equation}
Identically charged pairs only feel the gluonic interaction as $T_{II}=T_{AA}=0$. As expected, the ratio and streamline ansatz are markedly different only for strongly overlapping pairs, see \reffig{fig:interaction:total}, where strongly overlapping pairs are characterised by $R \leq \sqrt{\rho_1^2+\rho_2^2}$.

In the quenched case, we notice that the $R_E$ ansatz has a higher absolute minimum as compared to the $R_H$ ansatz, occurring roughly for the same separation. So we expect the ensemble to become slightly more dilute because it will not be as favourable, energetically, for instantons to come close. For unequal sizes, however, the repulsion is weaker in the $R_E$ case which would favour a denser ensemble, as less volume is excluded. The streamline ansatz will lead to a substantially more dilute system because the core repulsion is broader, excluding more volume for the instantons to move through.

In the unquenched case, the difference in the absolute interaction strength is much more pronounced between $R_E$ and $R_H$. We therefore expect that the $R_E$ ansatz should be quite a bit more dilute as compared to the $R_H$ ansatz. Considering that the streamline ansatz has a deeper minimum than the $R_E$ ansatz, the former will favour instantons to come closer. However, it has more excluded volume. Both trends work in opposite directions, and there is a possibility that they lead to roughly identical ensembles, at least on the level of the instanton density.

\begin{figure}[tbp]
\begin{center}
\includegraphics[width=\figwidth,clip=true,trim=0mm 0mm 15mm 10mm]{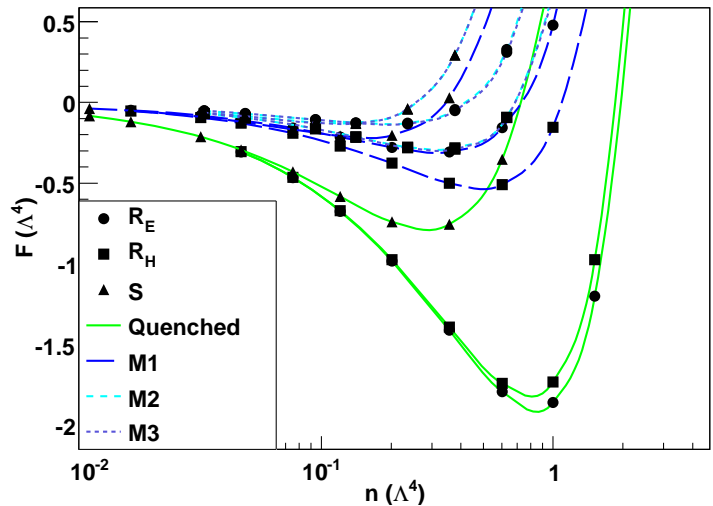}
\end{center}
\caption{The different quark masses are summarised in \reftable{table:adiabatic:unquenched:gc:M}. The simulations were performed in the topologically trivial sector, i.e.\ $N_I=N_A$. For the quenched and the $M_1$ case we fixed $N=64$ as in \cite{schaefer:shuryak:iilm}. The other two unquenched simulations have $N=200$. We clearly see that small quark masses suppress instanton contributions to the QCD vacuum, but also that there exists a finite limit for the instanton density as the quark masses vanish; this is in contrast to the dilute gas approximation which suppresses instanton contributions completely for zero quark masses. For unquenched simulations the free energy for the $R_E$ ensemble roughly agrees with that of the $S$ ensemble, although the equilibrium densities are rather different. Still, the approximate equality between the free energies might be interpreted as evidence of an approximate equivalence between both ensembles for bulk properties, e.g.\ equivalent pressure, since it is directly related to the free energy. However, the $R_E$ liquid does not seem to converge towards the $S$ ensemble as we lower the quark masses.}\label{fig:adiabatic}
\end{figure}

We will address these issues in more detail by performing canonical simulations and minimising the free energy. This follows closely \cite{schaefer:shuryak:iilm}, and also serves to validate our code against their results. Note that the simulations are performed in the topologically trivial sector, for which $N_I=N_A$, i.e.\ the topological charge $Q=0$. In \reffig{fig:adiabatic} we plot the free energy $F=-\ln Z /V$ against the instanton density $n=(N_I + N_A)/V$. As expected from our considerations of the pair interactions, in the quenched case the $R_E$ and $R_H$ ans\"atze are only slightly different, with $R_E$ leading to a slightly denser ensemble. Also, the interactions stored in that ensemble\footnote{The difference in the free energies is directly related to the difference of the interaction per instanton} are a bit lower, again as could be anticipated from the pair interactions. The $S$ ansatz leads to a much more dilute instanton ensemble, and our data reproduces well that of \cite{schaefer:shuryak:iilm}.

Ultimately, we will be interested in smaller quark masses. It is clear from (\ref{eq:interaction:pair}) and \reffig{fig:interaction:quark} that smaller masses increase the quark interaction strength as compared to the gluonic counterpart, which stays constant and is responsible for the core repulsion. From a purely energetic point of view, smaller quark masses should then lead to denser ensembles. However, we clearly see in \reffig{fig:adiabatic} that the ensembles become more dilute. The reason is that the small quark masses enter the instanton size distribution; in turn, the density, in the dilute gas limit, is entirely set by the size distribution, i.e.\ $n=2\int d\rho d(\rho)$. Bringing it into the action, we can interpret the size distribution as the energy cost needed to insert an instanton into the box. This is a well-known fact, namely that small quark masses suppress instanton contributions to the QCD vacuum because the different topological vacua become equivalent in the limit of vanishing quark masses; phrased differently, the energy barrier has disappeared, and only field configurations with topological charge $Q=0$ survive. In the dilute gas approximation this leads to the disappearance of instantons altogether. As \reffig{fig:adiabatic} shows, this is not true for an interacting instanton ensemble, where the instanton density converges to a finite limit as the quark mass is lowered\footnote{Remember that the simulations take place in the topologically trivial sector, i.e.\ $N_I=N_A$ or $Q=0$.}. The results from \reffig{fig:adiabatic} also show that the $R_E$ ansatz generates an ensemble that differs more and more from the $R_H$ ansatz, as was anticipated from our considerations of the pair interactions. We can also clearly see that the $R_E$ ensemble does not converge to the $S$ ensemble.

\begin{table}[tbp]
\begin{center}
 \begin{tabular}{c|c|c|c}
 & $m_u$ & $m_d$ & $m_s$ \\\hline
$M_1$ & $0.1$ & $0.1$ & $0.7$ \\\hline
$M_2$ & $0.05$ & $0.05$ & $0.3$ \\\hline
$M_3$ & $0.012$ & $0.022$ & $0.44$ \\\hline
 \end{tabular}
\end{center}
\caption{We use three different sets of quark masses and investigate how the instanton liquid depends on them.}\label{table:adiabatic:unquenched:gc:M}
\end{table}

\begin{figure}[tbp]
\begin{center}
\includegraphics[width=\figwidth,clip=true,trim=0mm 0mm 15mm 10mm]{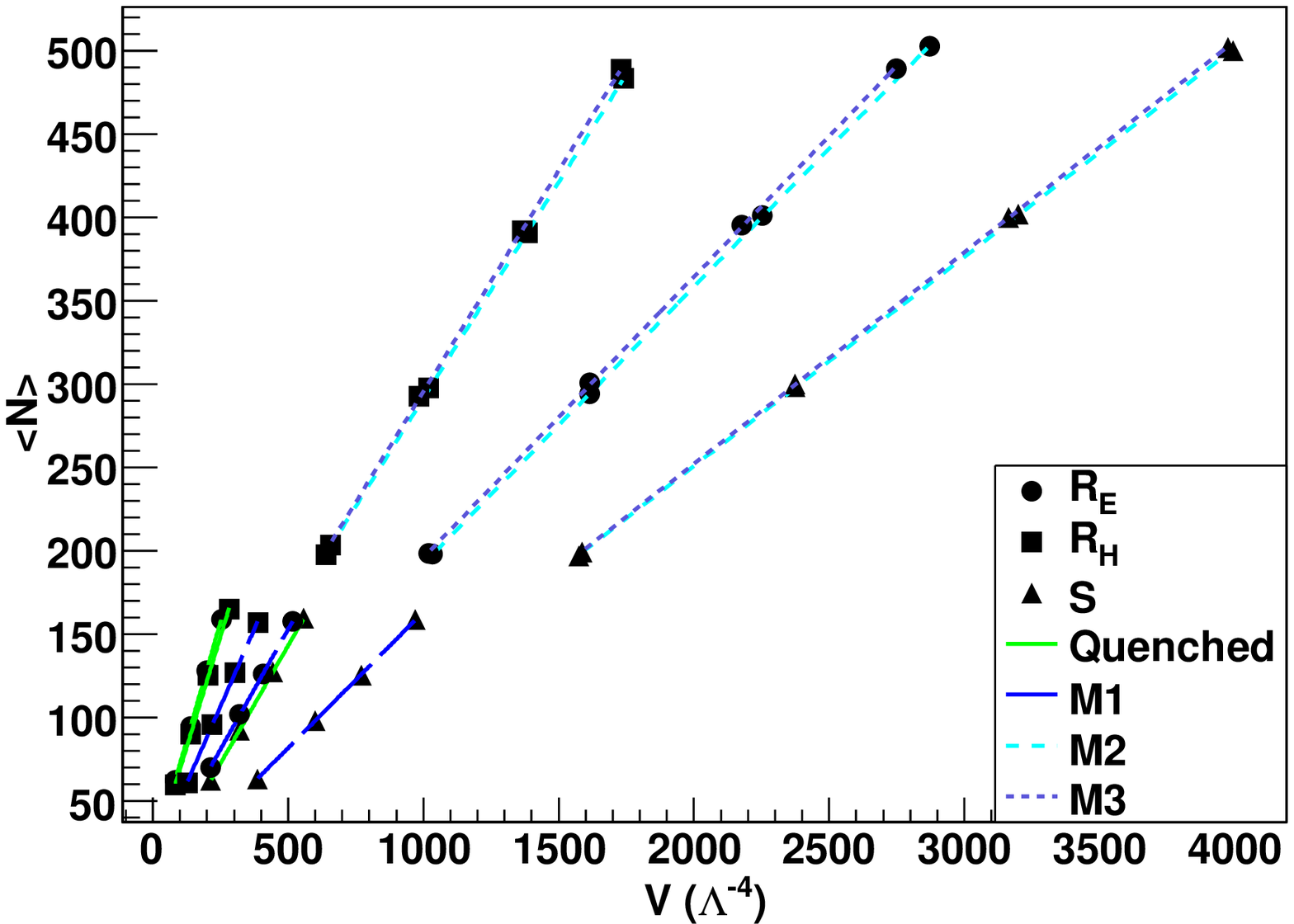}
\end{center}
 \caption{As anticipated from the canonical study, the instanton number for both ratio ans\"atze is very similar in the quenched case. There seems to exist a finite limit for the instanton density as the quark masses vanish, and instantons will be present in the QCD vacuum even in the chiral limit; this is in sharp contrast to dilute gas approximations.}\label{fig:adiabatic:gc:N}
\end{figure}

\begin{figure}[tbp]
\begin{center}
\includegraphics[width=\figwidth,clip=true,trim=0mm 0mm 15mm 10mm]{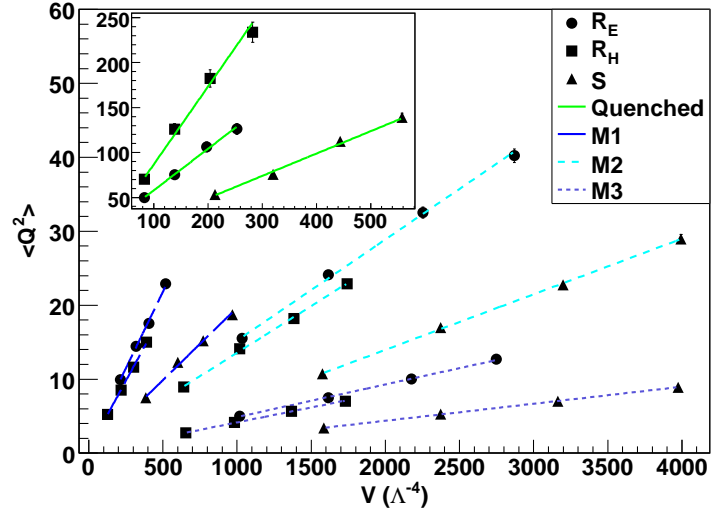}
\end{center}
 \caption{The topological susceptibility, the slope of the graphs, is very sensitive to the quark masses. It is screened by small quark masses and will vanish in the chiral limit. This is expected as QCD with massless quarks does not have topologically inequivalent vacua; in this case the so-called $\theta$ parameter is not physical and can be rotated away by a chiral rotation of the quark fields. See also \cite{shuryak:verbaarschot:screening} for another work on the topological susceptibility in the IILM.}\label{fig:adiabatic:gc:Q2}
\end{figure}

So far we have framed the discussion essentially in terms of the instanton density. To investigate the similarities and differences in more detail, we will look at a few bulk properties and their dependence on the different ans\"atze in the thermodynamic limit and the grand canonical ensemble. We clearly see how the density decreases with the quark masses, but approaches a finite limit, \reffig{fig:adiabatic:gc:N}. As we have discussed before, the quark masses will suppress fluctuations to inequivalent topological sectors and in the limit of vanishing masses only the trivial $Q=0$ sector will survive. The fluctuations between topological sectors are encoded in the topological susceptibility $\chi=\langle Q^2 \rangle /V$, which vanishes with the quark masses, see \reffig{fig:adiabatic:gc:Q2}. Both the instanton number and the topological charge fluctuations exhibit a nice scaling with the volume.

\begin{figure}[tbp]
\begin{center}
\includegraphics[width=\figwidth,clip=true,trim=0mm 0mm 15mm 10mm]{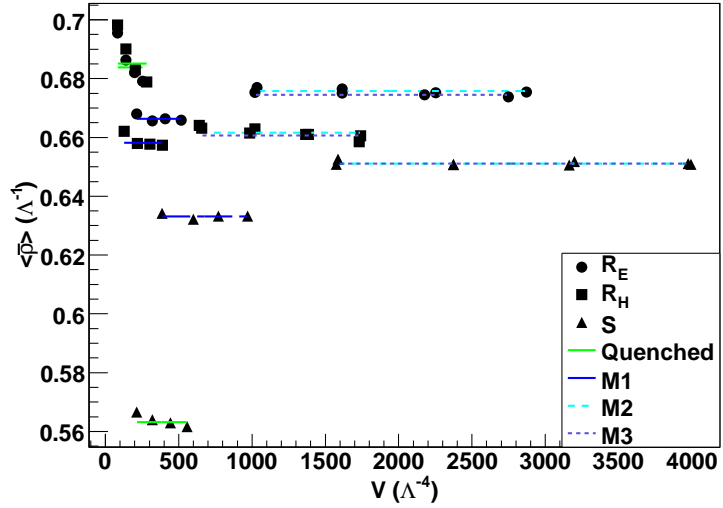}
\end{center}
 \caption{For the quenched and the $M_1$ simulations, which have been fixed to $N \approx 64$ for the smallest volume, the simulation boxes are still a too small, as can be seen by the systematic drift. For the other two unquenched simulations ($N \approx 200$ for the smallest volume) we are much closer to the thermodynamic limit, although there are still systematic deviations for the $R_H$ ansatz. In any case, the different ans\"atze give rather similar results. Also the mean instanton size approaches a unique limit for small quark masses.
}\label{fig:adiabatic:gc:rho}
\end{figure}

\begin{figure}[tbp]
\begin{center}
\includegraphics[width=\figwidth,clip=true,trim=0mm 0mm 15mm 10mm]{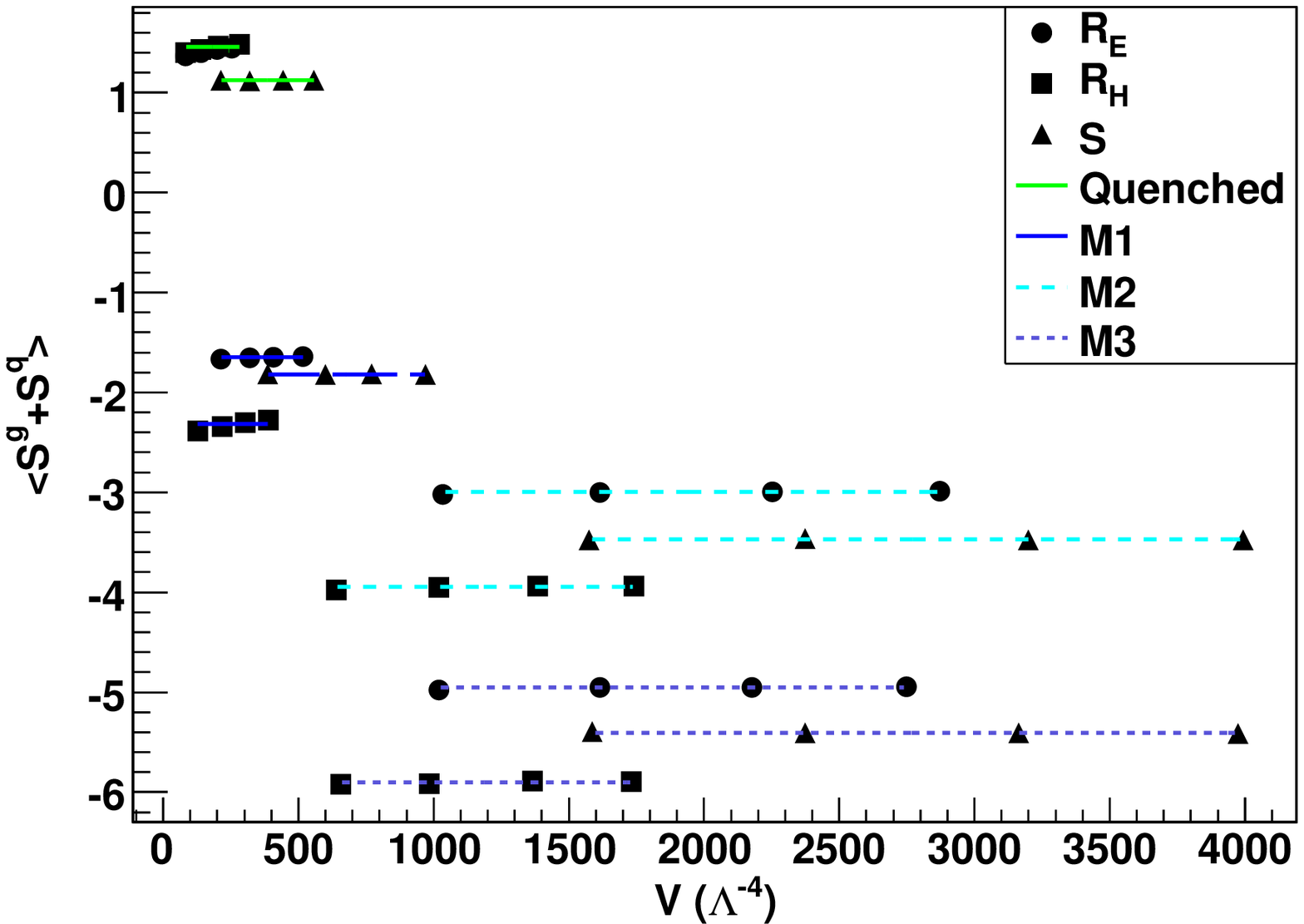}
\end{center}
 \caption{As anticipated from the pair interaction considerations, the interactions are very similar in the quenched sector for $R_E$ and $R_H$. For full simulations, the differences between the ans\"atze stay constant as the quark masses vary, with $R_H$ leading to the strongest attractive interactions, as was expected. Note that the pair interactions are less sensitive to finite size effects compared to the mean instanton size.}\label{fig:adiabatic:gc:interaction}
\end{figure}

We will now turn to an intensive quantities, the mean instanton size $\bar{\rho}$ \reffig{fig:adiabatic:gc:rho}. For the quenched and the first unquenched, $M_1$, simulations, which have been calibrated to achieve $N \approx 64$ for the smallest volume as in the canonical simulations, we see that the simulation boxes are not large enough, even though the density and the charge fluctuations seem to suggest otherwise, i.e.\ display good scaling with $V$. For the other two unquenched simulations, tuned to $N \approx 200$ for the smallest volume, we have reached volume sizes large enough to perform a thermodynamic limit. It is worth noticing that the mean instanton size is a rather robust quantity, and does neither depend strongly on the ansatz nor on the quark masses. This makes it a good quantity to use when comparing data from different ensembles, e.g.\ \cite{cristofotetti:foccioli:traini:negele:iilm} where the authors establish that the scale at which the IILM is operating is given by the inverse of the mean instanton size. Another intensive quantity, the interaction per instanton, is less sensitive to finite size effects, see \reffig{fig:adiabatic:gc:interaction}. The data shows that the weaker repulsion of the $R_E$ ansatz as compared to the $R_H$ ansatz dominates over the deeper attractive well of the latter. Therefore, the total interaction in the $R_E$ ensemble is slightly lower, leading to a denser system. The stronger repulsion for the $S$ interactions leads to more excluded volume; this, in turn, leads to lower interactions and a more dilute ensemble. These conclusions are in agreement with the direct measurement of the instanton density \reffig{fig:adiabatic:gc:N}.

\begin{table*}[tbp]
\begin{center}
 \begin{tabular}{c|c|c|c|c}
  & $n$ & $\chi$ & $\langle \rho \rangle$ & $\langle S^g + S^q \rangle/V$ \\\hline\hline
 Quenched & $\ansatz{0.567(1)\;[R_E]}{0.532(1)\;[R_H]}{0.282(1)\;[S]}$ & $\ansatz{0.46(3)}{0.86(5)}{0.24(1)}$ & $\ansatz{0.6837(2)}{0.6850(1)}{0.5631(1)}$ & $\ansatz{1.420(1)}{1.455(1)}{1.122(1)}$ \\\hline
 $M_1$ & $\ansatz{0.288(2)}{0.370(2)}{0.163(1)}$ & $\ansatz{0.041(1)}{0.0369(8)}{0.0196(6)}$ & $\ansatz{0.6662(2)}{0.6581(3)}{0.6331(3)}$ & $\ansatz{-1.648(1)}{-2.311(2)}{-1.818(3)}$\\\hline
 $M_2$ & $\ansatz{0.1660(7)}{0.259(1)}{0.1255(8)}$ & $\ansatz{0.0136(4)}{0.0126(2)}{0.0075(2)}$ & $\ansatz{0.6757(2)}{0.6615(2)}{0.6510(2)}$ & $\ansatz{-2.996(1)}{-3.945(2)}{-3.472(2)}$\\\hline
 $M_3$ & $\ansatz{0.1686(8)}{0.265(1)}{0.1269(5)}$ & $\ansatz{0.00440(7)}{0.00403(7)}{0.00230(5)}$ & $\ansatz{0.6744(2)}{0.6606(2)}{0.6511(2)}$ & $\ansatz{-4.954(1)}{-5.902(2)}{-5.407(2)}$\\\hline
 \end{tabular}
\end{center}
\caption{Thermodynamic extrapolations for the instanton density, the topological susceptibility, the mean instanton size and the mean interaction. The data has been obtained from \reffigs{fig:adiabatic:gc:N}, \ref{fig:adiabatic:gc:Q2}, \ref{fig:adiabatic:gc:rho} and \ref{fig:adiabatic:gc:interaction} respectively.}\label{table:adiabatic:gc}
\end{table*}

We are mostly interested in unquenched results, and the following comments relate this sector. From the data of \reftable{table:adiabatic:gc}, we can infer that results for $\chi^{1/4}$ have a $12\%$ systematic ansatz dependence. The topological susceptibility is surprisingly similar for the $R_E$ and $R_H$ ansatz in the unquenched sector. The mean instanton size is indeed a rather robust quantity and only affected on the $3\%$ level by these systematics. Also, note that the mass dependence on $\langle \rho \rangle$ is rather small, with differences not larger than $5\%$. The instanton interactions and the $n^{1/4}$ agree within $20\%$, and the latter converges to a fixed limit as the quark masses vanish.

\section{Fixing parameters}
\label{sec:parameters}

\subsection{Quenched case}
\label{sec:parameters:quenched}

\begin{figure}[tbp]
\begin{center}
\includegraphics[width=\figwidth,clip=true,trim=0mm 0mm 15mm 10mm]{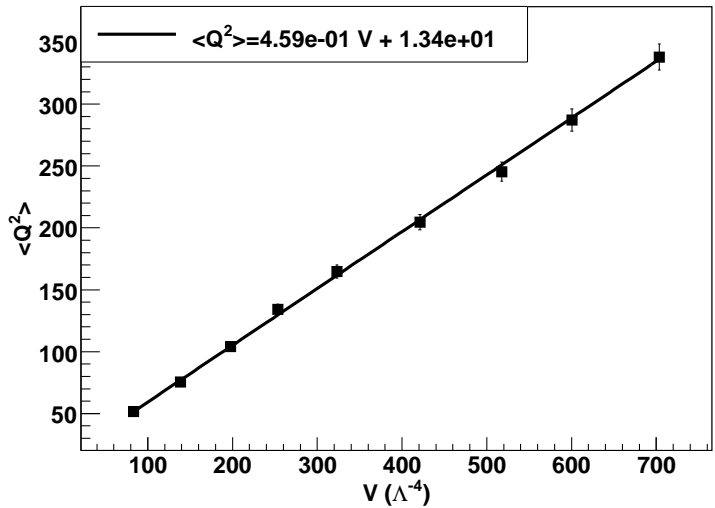}
\end{center}
 \caption{The fluctuations of the topological charge $\langle Q^2 \rangle$ show a nice linear dependence with the volume box $V$, as it should be for an extensive quantity. From this we infer the topological susceptibility $\chi_\mathrm{top}=\lim_{V \to \infty} \frac{\langle Q^2 \rangle}{V}$.}\label{fig:quenched:chi}
\end{figure}

In the quenched case, the IILM has only one freely adjustable constant, the lambda parameter $\Lambda$. We need one observable, from the lattice say, to fix it. Different approaches can be chosen. In the early works, $\Lambda$ was determined by fixing the instanton density to $1 \units{fm}^{-4}$ at $T=0$. To compare this with the lattice is not straightforward as the classical instanton content is convoluted with the quantum mechanical fluctuations. With the discovery of the KvBLL calorons, there is a renewed interest in studying the topological structures on the lattice, see for instance \cite{bruckmann:gattringer:ilgenfritz:al:filtering}. Since the topological susceptibility is well measured on the lattice and is easily accessible within the IILM, it is a natural candidate. The lambda parameter is then given by
\begin{equation}
 \Lambda = \sqrt[4]{\frac{\chi_\mathrm{lat}}{\chi_\mathrm{IILM}}}\,.
\end{equation}
We will use $\chi^{1/4}_\mathrm{lat} = 193 \units{MeV}$, \cite{durr:fodor:hoebling:kurth:su3:topological:susceptibility}. The topological susceptibility in the IILM is extracted from \reffig{fig:quenched:chi} by using the definition
\begin{equation}
 \chi_\mathrm{top}=\lim_{V \to \infty} \frac{\langle Q^2 \rangle}{V}\,.
\end{equation}
This yields $\Lambda=234(1) \units{MeV}$. The error is purely statistical. The instanton density turns out to be $n=0.543 \, \Lambda^4 = 1.02(2) \units{fm}^{-4}$, fairly close to the usually quoted phenomenological value of $n=1 \units{fm}^{-4}$. We find that even for these larger volumes the mean instanton size is still evolving towards lower values, as in \reffig{fig:adiabatic:gc:rho}. The largest volume then leads to the upper bound $\bar{\rho} < 0.57 \units{fm}$. Using a simple fit to $\bar{\rho} = \bar{\rho}_{\infty} + \alpha V^{-0.25}$ to extrapolate to the asymptotic value, we find $\bar{\rho}_\infty \approx 0.53 \units{fm}$; this is rather large compared to the phenomenological value of $\bar{\rho} \approx 0.33 \units{fm}$.

To estimate the systematic error due to the dependence on the ansatz, we will use the data from \reftable{table:adiabatic:gc}. The fact that we take a fourth root reduces the rather large differences in $\chi_\mathrm{IILM}$ to about $15\%$ for $\Lambda$, i.e.\ $\Lambda=234(35) \units{MeV}$. Our value has been obtain through simulations in $\mathrm{PV}$ regularisation. To compare it with lattice data, we will convert it to the $\overline{\mathrm{MS}}$ scheme, \cite{hasenfratz:hasenfratz:scales}, $\Lambda_{\overline{\mathrm{MS}}}/\Lambda_\mathrm{PV} = \exp(-1/22)$. This gives $\Lambda_{\overline{\mathrm{MS}}}=224(33)$ and compares well with the lattice result $\Lambda_{\overline{\mathrm{MS}}}=259(20)$ \cite{goeckeler:horsley:irving:pleiter:rakow:schierholz:stueben:lambda:parameter}.

\subsection{Unquenched case}
\label{sec:parameters:unquenched}

We want to use realistic quark masses. These are fairly small, and one must worry whether such light degrees of freedom will fit into the simulation box. The usual approach, used in the lattice community and also in work on the IILM \cite{cristofotetti:foccioli:traini:negele:iilm}, is to compute the pion mass from a set of unphysical quarks and to fix the volume box such that $Lm_{\pi} >5$; chiral perturbation theory can then be used to extrapolate to physical masses. Ultimately, the lattice wants to test the predictions of chiral perturbation theory as well, and in recent years, the computing power and, most importantly, the algorithms have improved to such an extent that physical quark mass simulations are becoming feasible; however, these are still immensely costly simulations, and $2+1$ flavour simulations were rare until recently.

\begin{figure}[tbp]
\begin{center}
 \includegraphics[width=\figwidth,clip=true,trim=0mm 0mm 15mm 10mm]{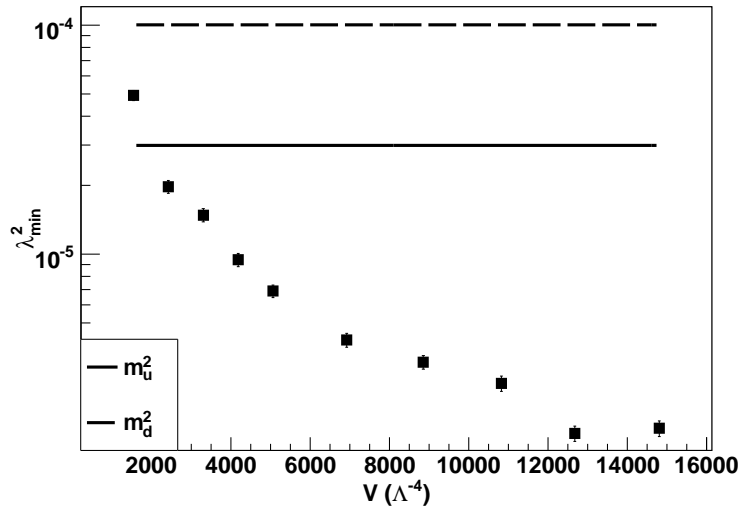}
\end{center}
 \caption{The simulations are performed for different simulation boxes. The average of the smallest Dirac eigenvalue, $\langle \lambda_\mathrm{min} \rangle$, is smaller than the quark masses for all but the smallest simulation box.}\label{fig:lambda_min}
\end{figure}

We follow a rather more modest rationale by simply demanding that the quark mass be at least so small as to be comparable to the lowest eigenvalue of the Dirac operator, $\langle \lambda_\mathrm{min} \rangle$, see \reffig{fig:lambda_min}. This sets the smallest box we use in our simulations. We then use ever larger volumes and extrapolate to the thermodynamic limit.

In \cite{cristofotetti:foccioli:traini:negele:iilm} the lambda parameter\footnote{Actually, the authors fixed the mean instanton size. But it is trivial to relate the latter to the lambda parameter.} is fixed by computing the meson and nucleon masses, through current correlators of the interpolating fields and their asymptotic spatial decay, and by comparing them with the available lattice data. This study established that the IILM is compatible with the predictions of chiral perturbation theory. We will take this for granted in what follows. 

In order to fix $\Lambda$, we could still use the topological susceptibility as it is routinely measured on the lattice. However, the topological susceptibility depends strongly on the quark masses, see \reffig{fig:adiabatic:gc:Q2}. We can get rid of the mass dependence by using chiral perturbation theory and computing the chiral condensate $\langle \bar{q}{q} \rangle$\footnote{To reiterate, we implicitly rely on the fact that the IILM is describing well the chiral properties of QCD, as has been checked in numerous studies, the most convincing being \cite{cristofotetti:foccioli:traini:negele:iilm}.}. The chiral condensate has been studied within chiral perturbation theory and, more recently, it has been precisely determined on the lattice \cite{chiu:hsieh:tseng:topological:susceptibility,Chiu:Aoki:JLQCD:TWQCD:topological:susceptibility:overlap,Chiu:Aoki:JLQCD:TWQCD:topological:susceptibility:fixed_topology}. We will take it to be $\langle \bar{q}q \rangle_0^{\overline{\mathrm{MS}}}(\mu=2 \units{GeV})=250 \units{MeV}$.

To extract the chiral condensate from the IILM, we will use the procedure adopted in \cite{chiu:hsieh:tseng:topological:susceptibility,Chiu:Aoki:JLQCD:TWQCD:topological:susceptibility:overlap,Chiu:Aoki:JLQCD:TWQCD:topological:susceptibility:fixed_topology}: we compute the topological susceptibility for different sets of quark masses and extrapolate to the chiral limit, $m_i \to 0$. The condensate can then be determined by chiral perturbation theory \cite{leutwyler:smilga:spectrum:dirac},
\begin{eqnarray}
 \chi &=& m_\mathrm{eff} \langle \bar{q}q \rangle_0 + O(m^2), \label{eq:chiral:perturbation:chi:qq}\\
 m_\mathrm{eff} &=& \left(\sum_n^{N_f} \frac{1}{m_n}\right)^{-1}\,. \nonumber
\end{eqnarray}

The chiral condensate has an anomalous dimension and, therefore, depends on the scale. Furthermore, the IILM is set up with a $\mathrm{PV}$ regulator, whereas the quoted result is computed in dimensional regularisation. It is well known that within an unphysical renormalisation scheme such as $\overline{\mathrm{MS}}$\footnote{'t Hooft's computation of the one-loop instanton measure, using Pauli-Villars regularisation, is also unphysical because, instead of poles, logarithms of the regulator mass are subtracted.} the results depend on the regulator (for unphysical quantities like masses, coupling constants and amplitudes). We therefore need to compute the finite counterterms that relate the $\mathrm{PV}$ to the $\overline{\mathrm{MS}}$ regularised results. Deferring the details to appendix \ref{app:renormalization:scheme:change}, we find that $\langle \bar{q}q \rangle_0^{\mathrm{PV}}(\mu=2 \units{GeV}) \approx 244 \units{MeV}$. This is a one-loop result. The two-loop correction can be estimated very roughly to be on the $10\%$ level as is typical for computations around the scale of $\mu=2 \units{GeV}$\footnote{Strictly speaking, we should use the two-loop result because the simulations in the IILM have been obtained using the two-loop improved instanton measure. However, Pauli-Villars regularisation is not straightforward for non-Abelian gauge theories beyond the one-loop level, and we do not have the expertise to embark on this endeavour. In any case, the difference should still be on the $10\%$ level.}.

We will define the scale of the IILM by $\mu_\Lambda = \Lambda/\bar{\rho}$, as suggested in \cite{cristofotetti:foccioli:traini:negele:iilm}, and determine $\Lambda$ from the self-consistency equation
\begin{equation}
 \langle \bar{q}q \rangle_0^{\mathrm{PV}}(\mu_\Lambda)=\Lambda^3 \langle \bar{q}q \rangle_0^{\mathrm{IILM}}\,, \label{eq:lambda:self_consistency}
\end{equation}
where we run the chiral condensate $\langle \bar{q}q \rangle_0^{\mathrm{PV}}(\mu)$ at one loop. To that order, there is no difference between schemes and we can use the $\overline{\mathrm{MS}}$ results, e.g.\ \cite{vermaseren:larin:ritbergen:4loop:anomalous:mass:dimension}.

To get an estimate of the quark mass ratio dependence, we have used two different sets of quark masses, one inspired by the chiral perturbation theory and the other by the quark masses extracted from the lattice \cite{mason:trottier:horgan:davies:lepage:quark:masses:lattice}. The two sets have the following ratios
\begin{equation}
\frac{m_i}{m_j}=\left\{
\begin{array}{c@{\;:\;}c@{\;:\;}cc}
 1 & 1.83 & 36.7 & (M_1) \\
 1 & 2.32 & 45.0 & (M_2)
\end{array}
\right. \,.\label{eq:mass:ratios}
\end{equation}

\begin{figure}[tbp]
\begin{center}
 \includegraphics[width=\figwidth,clip=true,trim=0mm 0mm 15mm 10mm]{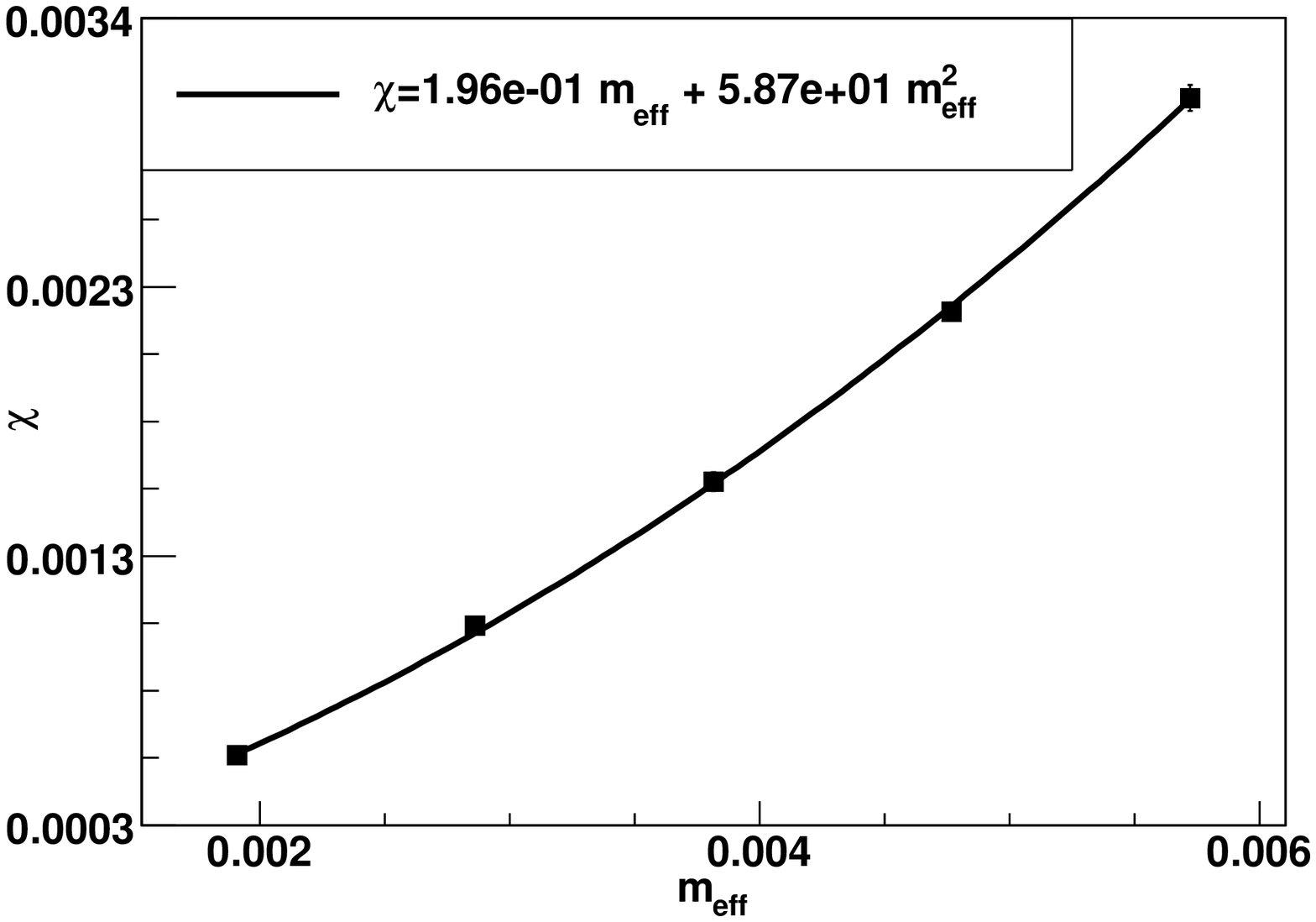}
 \includegraphics[width=\figwidth,clip=true,trim=0mm 0mm 15mm 10mm]{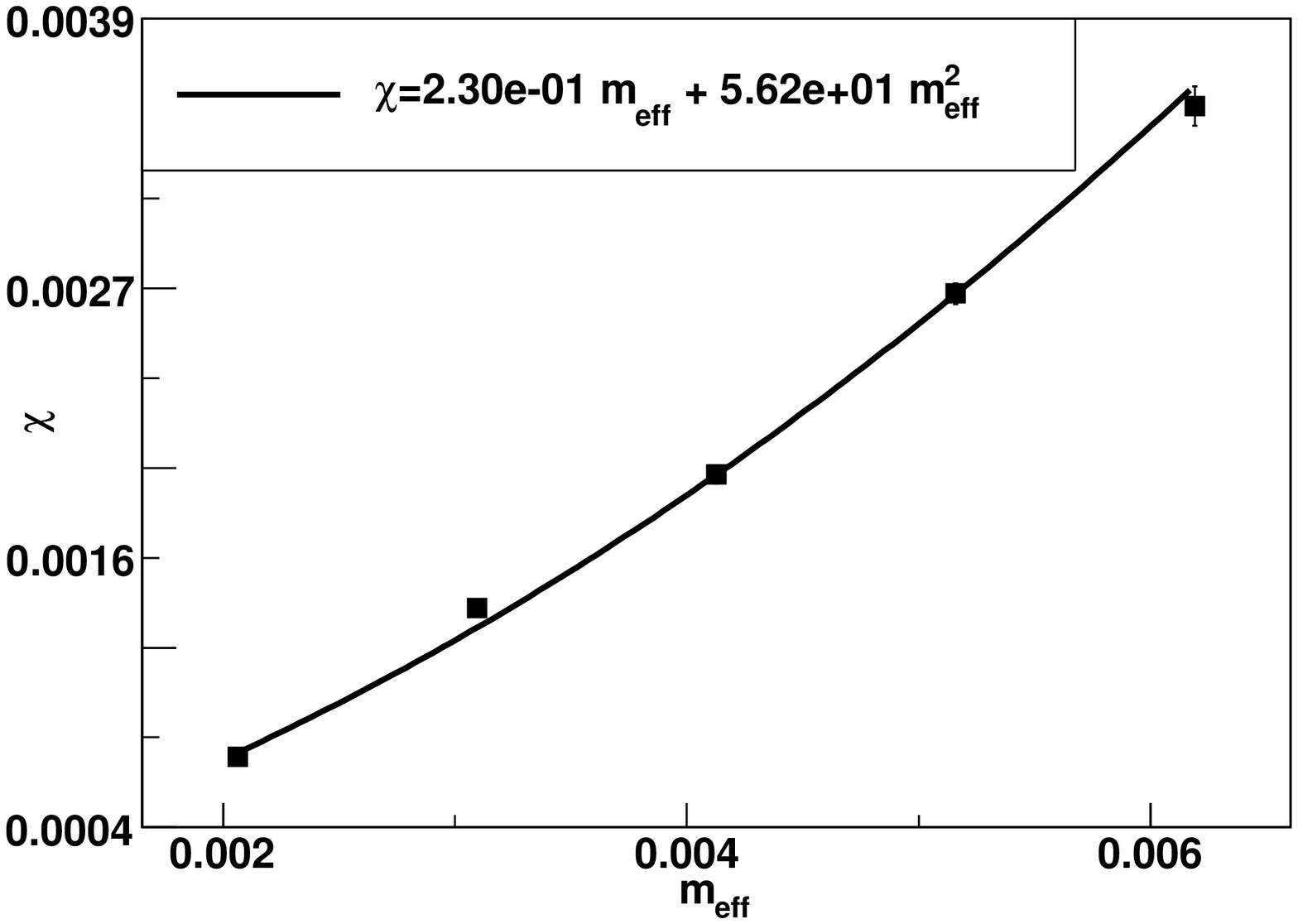}
\end{center}
 \caption{Computing the topological susceptibility allows for the extraction of the chiral condensate by using chiral perturbation theory, (\ref{eq:chiral:perturbation:chi:qq}). The rationale is the same as used in recent lattice studies to extract the chiral condensate \cite{chiu:hsieh:tseng:topological:susceptibility,Chiu:Aoki:JLQCD:TWQCD:topological:susceptibility:overlap,Chiu:Aoki:JLQCD:TWQCD:topological:susceptibility:fixed_topology}. In order to get a rough estimate on the systematic error introduced by the chiral limit, two sets of masses have been used. The upper plot corresponds to the set $M_1$ and the lower plot to $M_2$, as given in (\ref{eq:mass:ratios}). The chiral condensate for both mass ratios agrees on the $1\sigma$ level, and we conclude that it depends only  weakly on the quark mass ratios.}\label{fig:Chi:QQ}
\end{figure}

For each set we perform 5 simulations with ever smaller absolute masses, see \reffig{fig:Chi:QQ}. This data is fitted to (\ref{eq:chiral:perturbation:chi:qq}) to extract the chiral condensate. The results for the two sets agree on the $1\sigma$ level, and we can argue that the chiral condensate depends only weakly on the quark mass ratios, given that the latter vary by roughly $25\%$, see (\ref{eq:mass:ratios}). This is as it should be since the exact chiral condensate does not depend on the quark masses at all. From an operational point of view, the robustness against quark mass ratios\footnote{i.e.\ taking the limit from different directions in quark mass space.} makes the chiral condensate a good quantity to set $\Lambda$.

Solving (\ref{eq:lambda:self_consistency}) we find that the lambda parameter is given by
\begin{equation}
\Lambda_i=\left\{
\begin{array}{cc}
401(5)(40)(15) \units{MeV}\\
389(6)(40)(15) \units{MeV}
\end{array}
\right. \label{eq:lambda}\,,
\end{equation}
where the errors follow from the fit, $\langle \bar{q}q \rangle_0^{\mathrm{PV}}$ and the systematic on $\chi$. This leads to an overall error of $44 \units{MeV}$, or roughly $11\%$, and is strongly dominated by the one-loop result for $\langle \bar{q}q \rangle_0^{\mathrm{PV}}$. Since we run at one loop in the self-consistency equation (\ref{eq:lambda:self_consistency}), such a large error is certainly realistic, if not underestimated.

We found that running at two-loop\footnote{We use $\beta$-functions and anomalous dimensions from the $\overline{\mathrm{MS}}$ scheme, \cite{vermaseren:larin:ritbergen:4loop:anomalous:mass:dimension}, since we do not know them for $\mathrm{PV}$ regularisation. However, both regularisations are thought to give roughly similar results, for instance $\langle \bar{q}q \rangle_0^{\mathrm{PV}}$ and $\langle \bar{q}q \rangle_0^{\overline{\mathrm{MS}}}$ agree on the $3\%$ level at one-loop and $\mu=2\,\mathrm{GeV}$, and for the purpose of estimating errors in the one-loop running this procedure should be fine.} gives results consistent with (\ref{eq:lambda}). Using the prescription of \cite{cristofotetti:foccioli:traini:negele:iilm}, the scale and the mean instanton size for the IILM is\footnote{Remember that in the unquenched case the instanton size is fairly independent of the quark masses.}
\begin{align}
& \mu_{\Lambda_i}=\left\{
\begin{array}{cc}
598(65) \units{MeV}\\
580(64) \units{MeV}
\end{array}
\right.\,,
& \rho_{\Lambda_i}=\left\{
\begin{array}{cc}
0.33(3) \units{fm}\\
0.34(4) \units{fm}
\end{array}
\right. \label{eq:scale:rho}\,.
\end{align}
This is in very good agreement with the precision study \cite{cristofotetti:foccioli:traini:negele:iilm}. Given that both works use chiral properties for the calibrations, the nice overlap is probably not totally unexpected.

Note that (\ref{eq:lambda}) is a prediction for the lambda parameter with 3 active quark flavours. To compare our result with experimental data we run down the coupling constant $\alpha_s^{\overline{\mathrm{MS}}} = 0.117(2)$ \cite{pdg:2002} from $M_Z$ to $\mu_\Lambda$ and convert it to a lambda parameter. This is a rather big difference in scales and it is appropriate to use two-loop running, although not entirely consistent when we compare it to the one-loop result (\ref{eq:lambda}). To deal with threshold effects, we use the \textit{Mathematica} package \textit{RunDec}, \cite{chetyrkin:kuehn:steinhauser:rundec}. The conversion between the $\overline{\mathrm{MS}}$ and $\mathrm{PV}$ lambda parameters is given by \cite{hasenfratz:hasenfratz:scales}, \cite{armoni:shifman:veneziano:quark:condensate}
\begin{equation}
 \Lambda_\mathrm{PV} = \Lambda_{\overline{\mathrm{MS}}} \exp\left(\frac{1}{22 - 4 N_f/3}\right)\,.
\end{equation}
This leads to $\Lambda^{(3)}_\mathrm{PV}=325(40)$ and the IILM result agrees on the $1\sigma$ level. Trusting the perturbative running down to the rather low scale $\mu_\Lambda$ is a leap of faith. However, earlier studies have seen good agreement between IILM and lattice predictions for physical quantities, such as meson masses, and so the agreement between the lambda parameters might not just be a fluke.

To determine the physical quark masses, we will use (\ref{eq:chiral:perturbation:chi:qq}) rewritten in terms of the pion mass
\begin{equation}
 \chi = m^2_\pi f^2_\pi \frac{m_u m_d}{(m_u+m_d)^2}+O\left(\frac{1}{m_s}\right) = \left\{
\begin{array}{c}
 (77.4 \units{MeV})^4 \\
 (75.9 \units{MeV})^4 
\end{array}
\right. \,. \label{eq:chiral:perturbation:chi:pion}
\end{equation}
We used $m_\pi=135 \units{MeV}$ and $f_\pi=93 \units{MeV}$. Together with the fits, \reffig{fig:Chi:QQ}, we can compute the corresponding quark masses. We convert them into $\overline{\mathrm{MS}}$ masses at $2\units{GeV}$, run at one-loop, in order to compare them more easily with other sources. Our results are
\begin{align}
m_i^{\mathrm{PV}}(\mu = 0.6\units{GeV}) &= \left\{
\begin{array}{ccc}
 2.2(2) & 4.0(4) & 80(9) \\
 1.9(2) & 4.4(4) & 87(10)
\end{array}
(\units{MeV})
\right. \,,\\
m_i^{\overline{\mathrm{MS}}}(\mu  = 2\units{GeV}) &= \left\{
\begin{array}{ccc}
 1.9(2) & 3.4(5) & 69(11) \\
 1.7(2) & 3.8(5) & 74(11)
\end{array}
(\units{MeV})
\right. \,.
\end{align}
The errors include an estimate from the 2-loop running. These masses compare well with the particle data group masses \cite{pdg:quark:masses}, i.e.\ $m_u=1.5-3.3 \units{MeV}$, $m_d=3.5-6.0 \units{MeV}$ and $m_s=70-130 \units{MeV}$, and to the lattice masses \cite{mason:trottier:horgan:davies:lepage:quark:masses:lattice}, i.e.\ $m_u=1.9(2) \units{MeV}$, $m_d=4.4(3) \units{MeV}$ and $m_s=87(6) \units{MeV}$.

Very large volume simulations are expensive even in the IILM. We have seen in section \ref{sec:ansaetze} that the instanton density becomes independent of quark masses in the chiral limit. Therefore, in the physical region of parameter space that we are considering the volume is directly proportional to the number of instantons in the box. The computation of the fermionic determinant is the most costly part of the simulations. In a naive implementation it would scale as $O(V^3)$. However, in section \ref{sec:fermion_determinant} we were able to drastically reduce this cost to $O(V^2)$ by implementing fast update algorithms for the Cholesky decomposition. The absolute scale for the volumes is set by considerations regarding finite size effects. To have these under control the lightest propagating particle, in our case the pion, should fit into the box. For masses beyond the chiral limit the instanton density increases, see again section \ref{sec:ansaetze}. However, the pion mass increases too and it turns out that for the same level of control over finite size effects the system size can be smaller, i.e.\ larger quark masses are computationally cheaper.

We have studied the thermodynamic limit on four volumes, in the range $2 \lesssim Lm_{\pi} \lesssim 3$. Even though the data has displayed a nice scaling with the volume, it is important to check whether the thermodynamic limit was consistent. To this end we will run large volume simulations, $Lm_{\pi} \in [2.11,3.7]$, for the particular set of physical masses inspired by chiral perturbation theory: in dimensionless units, $m_u=0.00546$, $m_d=0.01001$ and $m_s=0.2002$. This will allow us to estimate the systematic error introduced by performing the thermodynamic on the set of smaller simulation boxes.

\begin{figure}[tbp]
\begin{center}
\includegraphics[width=\figwidth,clip=true,trim=0mm 0mm 15mm 10mm]{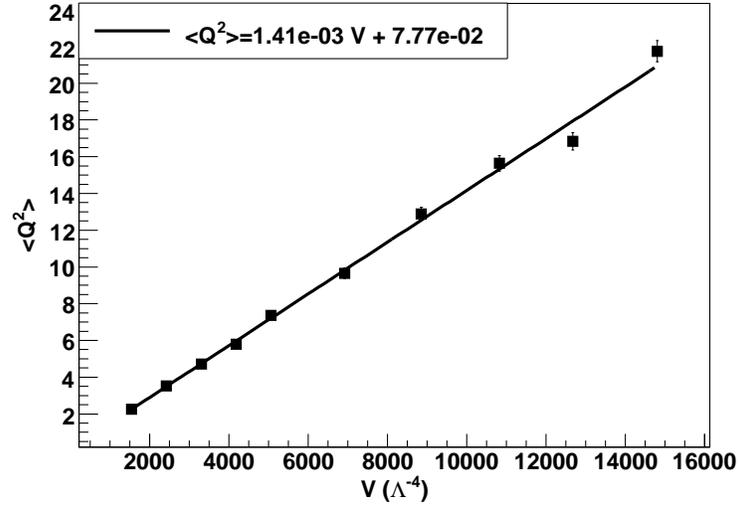}
\end{center}
 \caption{The fluctuations of the topological charge $\langle Q^2 \rangle$ show a nice linear dependence with the volume box $V$, as it should be for an extensive quantity. Applying the thermodynamic limit to the 4 smallest volumes yields a topological susceptibility that agrees on the $1\sigma$ level with the corresponding result using all available volumes. The mean instanton number ranges from $\langle N \rangle \approx 200$ to $\langle N \rangle \approx 1600$.}\label{fig:unquenched:chi}
\end{figure}

\begin{figure}[tbp]
\begin{center}
\includegraphics[width=\figwidth,clip=true,trim=0mm 0mm 15mm 10mm]{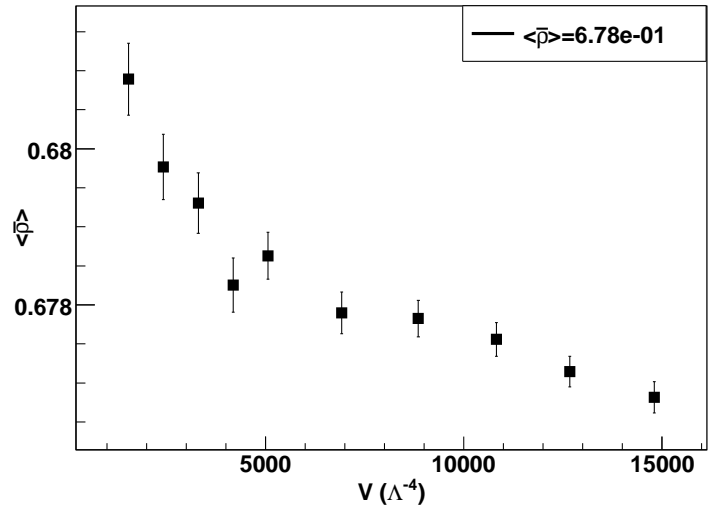}
\end{center}
 \caption{Even for the largest volumes the mean instanton size $\bar{\rho}$ is still decreasing and does not seem to converge to a constant. We are rather lucky that, although the effect is clearly systematic, the variation of $\bar{\rho}$ is small in absolute terms.} \label{fig:unquenched:rho}
\end{figure}

The thermodynamic limit on the topological susceptibility turns out to be rather insensitive, see \reffig{fig:unquenched:chi}. However, the mean instanton size does not converge to a constant even for the largest volumes, see \reffig{fig:unquenched:rho}. It is a rather lucky fact that the mean instanton size does not vary much in absolute terms. The slope is clearly decreasing and we might estimate the convergence to occur somewhere in the range $\bar{\rho} \in [0.68,0.66]$. A fit to $\bar{\rho} = \bar{\rho}_{\infty} + \alpha V^{-0.25}$ gives $\bar{\rho}_{\infty} = 0.6720(5) \, \Lambda^{-1} = 0.33(3) \units{fm}$, in good agreement with the phenomenological value. The instanton density turns out to be $n = 1.7(7) \units{fm}^{-4}$, and like the topological susceptibility displays a nice thermodynamic limit.

\section{Conclusion}

With the discovery of the new non-trivial holonomy calorons \cite{kraan:baal:caloron:I,kraan:baal:caloron:II}, \cite{lee:lu:caloron}, there is renewed interest in studying the role of non-trivial field configurations in QCD, especially their role in the confinement/deconfinement phase transition. A lattice based approach \cite{gerhold:ilgenfritz:mueller_preussker:kvbll:gas:confinement} is well suited for the pure gauge sector because it easily incorporates many-body instanton interactions in the classical action. However, the introduction of fermions will be plagued by the same problems that full lattice computations face. Most notably, the computations will become very costly.

A different approach, pioneered by Shuryak, Diakonov and Petrov, is to formulate the instanton liquid in the continuum, see for instance \cite{schaefer:shuryak:iilm}. This approach suffers from the fact that it is not straightforward to include many-body interactions. Incidentally, only two-body effects are taken into account. The strongly localised profiles of instantons and the, a posteriori, fact that the instanton ensemble is rather dilute make this a viable working premise. The big advantage of the continuum formulation is the ease with which quarks can be incorporated, see also \cite{wagner:fermions:pseudoparticle:approach}. From this perspective, both the lattice and the continuum models complement each other rather well. Meanwhile, different publications have investigated the confining nature of other backgrounds, such as regular gauge instantons and merons \cite{lenz:negele:thies:confinement:meron,negele:lenz:thies:confinement:instanton:meron}, \cite{wagner:confining} at zero temperature.

So far the continuum models used explicit analytic formulas for the interactions. They have been obtained through asymptotic considerations and fits to numerical evaluations of the classical action. We noted that these formulas do not possess a thermodynamic limit at finite temperature. More importantly perhaps, the more complex moduli-space of the non-trivial holonomy calorons probably demand a more systematic approach. In this paper we have set up a framework which we believe is numerically well-defined, can be extended to more complicated backgrounds and does not suffer from the parametrisation bias introduced implicitly through analytical formulas motivated by symmetry arguments and fits. The price to pay is a larger numerical overhead because evaluation of the interactions through look-up tables and asymptotic matching formulas is computationally more expensive than through simple fitting formulas.

We have found that the analytic formulas of \cite{schaefer:shuryak:iilm} agree very well with our interactions at zero temperature. Especially for the case of equal instanton sizes, where strong symmetry arguments support the analytic formulas, the agreement can be seen as a validation of the numerics. In general, however, the interactions of both schemes differ; the differences are especially pronounced for the quark overlaps because in this paper we use the full ratio ansatz whereas a sum ansatz is used in \cite{schaefer:shuryak:iilm}. Shifting the point of view, we considered the formulas of \cite{schaefer:shuryak:iilm} to be another valid scheme; together with the streamline ansatz we studied the dependence of bulk properties of the IILM on the choice of these three rather different interactions and found that this introduces a systematic effect which depends on the quantity under consideration, but was generally rather large, up to $20\%$.

The IILM has been shown to be compatible with the chiral properties of QCD, see for instance \cite{cristofotetti:foccioli:traini:negele:iilm}. A key chiral property, the topological susceptibility, has not been studied extensively within the IILM, see however \cite{shuryak:verbaarschot:screening}. One reason might be that the IILM was so far set up for simulations in the canonical ensemble whereas the topological susceptibility is most naturally studied in the grand canonical ensemble. We have enlarged the Monte Carlo moves to incorporate insertion/deletion steps in order to simulate an open ensemble. Apart from technical problems related to book-keeping issues this is rather straightforward.

A major incentive for this work was to investigate the regime of physical quark masses. In order to deal with such light quarks, rather large volumes need to be considered. We dealt with this issue by, first, reducing the complexity of the algorithm from $O(N^3)$ to $O(N^2)$; this is achieved by rewriting the updates in a form suitable for fast matrix modifications. Secondly, we study the thermodynamic limit and monitor some bulk quantities to guarantee a consistent large volume extrapolation.

The topological susceptibility is easy to compute in the IILM and has been studied extensively on the lattice. It represents a natural candidate to fix units in the IILM. For quenched simulations we found rather good agreement between the IILM and the lattice in this way. We are, however, mainly interested in the unquenched case. Instead of using the topological susceptibility directly, we decided to use the chiral condensate, extracted through the topological susceptibility and chiral perturbation theory, to set units. The reason we decided against a direct use of the topological susceptibility is that it depends strongly on the quark masses. We found that the chiral condensate has a very weak dependence on the chiral limit, and use it to set the units in the unquenched sector. We achieve good agreement with previous work and also with experimental data on the strong coupling constant $\alpha_s$. Using chiral perturbation theory, we are able to determine physical quark masses. These turn out to compare well with experimental bounds and lattice simulations.

Finally, we investigated the uncertainties introduced by the large volume extrapolations, and found that our procedure, of bounding the volume in such a way that the quark masses are smaller than the smallest Dirac eigenvalues, allows for a systematic thermodynamic limit.

In a further publication we will use these input parameters to study the IILM at finite temperature.

\section*{Acknowledgements}

We are very grateful for many informative discussions with P. Faccioli and E.P.S. Shellard. Simulations were performed on the COSMOS supercomputer (an Altix 4700) which is funded by STFC, HEFCE and SGI. This work was supported by STFC grant PPA/S/S2004/03793 and an Isaac Newton Trust European Research Studentship.

\appendix
\section{Gluonic Interactions}
\label{app:interaction:gluonic}

The ratio ansatz for an instanton--anti-instanton pair is defined by 
\begin{equation}
 A^a_\mu = - \frac{\bar{\eta}^a_{\mu\nu}\partial_\nu \Pi_1(x,\{x_1,\rho_1\}) + O^{ab} \eta^b_{\mu\nu} \partial_\nu \Pi_2(x,\{x_2,\rho_2\})}{1+\Pi_1(x,\{x_1,\rho_1\})+\Pi_2(x,\{x_2,\rho_2\})}\,,
\end{equation}
where $\Pi(x,\{y,\rho\}) = \frac{\rho^2}{r^2}$ and $O=O_1^t O_2$, with $O_i$ the respective colour embeddings. A global colour rotation has been performed to bring the gauge potential into this form, which is irrelevant since the action is gauge invariant. Instanton--instanton and anti-instanton--anti-instanton pairs differ by having either only $\bar{\eta}$ or $\eta$ in the above formula. A brute force computation then gives 

\begin{multline}
 F^a_{\mu\nu} F^a_{\mu\nu} = I + (\Tr O^tO + (\bar{\eta} O \eta)_{\mu\nu\mu\nu}) J + (\bar{\eta} O \eta)_{\rho\mu\rho\nu} I_{\mu\nu} \\
+ (\bar{\eta} O \eta)_{\mu\rho\nu\sigma} I_{\mu\rho\nu\sigma} + (\eta O^tO \eta)_{\mu\rho\nu\sigma} J_{\mu\rho\nu\sigma} + (\bar{\eta} O \eta)_{\alpha\mu\alpha\rho} (\bar{\eta} O \eta)_{\beta\nu\beta\sigma} K_{\mu\rho\nu\sigma}\,.
\end{multline}

The different terms have the following form
\ifthenelse{\equal{\Qclass}{revtex4}}{
\begin{multline}
I = \frac{4}{(1+\Pi_1+\Pi_2)^2} \left[ (\partial_\mu \partial_\nu \Pi_1) (\partial_\mu \partial_\nu \Pi_1) + (\partial_\mu \partial_\nu \Pi_2) (\partial_\mu \partial_\nu \Pi_2) \right]\\
\shoveleft{- \frac{8}{(1+\Pi_1+\Pi_2)^3} \left[ (\partial_\mu \partial_\nu \Pi_1) (\partial_\mu \Pi_1)(\partial_\nu \Pi_2) + (\partial_\mu \partial_\nu \Pi_2) (\partial_\mu \Pi_2)(\partial_\nu \Pi_1) \right.}\\
\shoveright{\left. + 2(\partial_\mu \partial_\nu \Pi_1) (\partial_\mu \Pi_1)(\partial_\nu \Pi_1) + 2(\partial_\mu \partial_\nu \Pi_2) (\partial_\mu \Pi_2)(\partial_\nu \Pi_2) \right]}\\
\shoveleft{+ \frac{4}{(1+\Pi_1+\Pi_2)^4} \left[3(\partial_\mu \Pi_1 \partial_\mu \Pi_1)(\partial_\mu \Pi_1 \partial_\mu \Pi_2) +  3(\partial_\mu \Pi_2 \partial_\mu \Pi_2)(\partial_\mu \Pi_2 \partial_\mu \Pi_1)\right.}\\
\left .+ 3(\partial_\mu \Pi_1 \partial_\mu \Pi_1)^2 + 3(\partial_\mu \Pi_2 \partial_\mu \Pi_2)^2 + 2(\partial_\mu \Pi_1 \partial_\mu \Pi_1)(\partial_\mu \Pi_2 \partial_\mu \Pi_2) + (\partial_\mu \Pi_1 \partial_\mu \Pi_2)^2  \right]\,.
\end{multline}
}{}
\ifthenelse{\equal{\Qclass}{elsarticle}}{
\begin{multline}
I = \frac{4}{(1+\Pi_1+\Pi_2)^2} \left[ (\partial_\mu \partial_\nu \Pi_1) (\partial_\mu \partial_\nu \Pi_1) + (\partial_\mu \partial_\nu \Pi_2) (\partial_\mu \partial_\nu \Pi_2) \right]\\
\shoveleft{- \frac{8}{(1+\Pi_1+\Pi_2)^3} \left[ (\partial_\mu \partial_\nu \Pi_1) (\partial_\mu \Pi_1)(\partial_\nu \Pi_2) + (\partial_\mu \partial_\nu \Pi_2) (\partial_\mu \Pi_2)(\partial_\nu \Pi_1) \right.}\\
\shoveright{\left. + 2(\partial_\mu \partial_\nu \Pi_1) (\partial_\mu \Pi_1)(\partial_\nu \Pi_1) + 2(\partial_\mu \partial_\nu \Pi_2) (\partial_\mu \Pi_2)(\partial_\nu \Pi_2) \right]}\\
\shoveleft{+ \frac{4}{(1+\Pi_1+\Pi_2)^4} \left[3(\partial_\mu \Pi_1 \partial_\mu \Pi_1)(\partial_\mu \Pi_1 \partial_\mu \Pi_2) +  3(\partial_\mu \Pi_2 \partial_\mu \Pi_2)(\partial_\mu \Pi_2 \partial_\mu \Pi_1)\right.}\\
\left .+ 3(\partial_\mu \Pi_1 \partial_\mu \Pi_1)^2 + 3(\partial_\mu \Pi_2 \partial_\mu \Pi_2)^2 + 2(\partial_\mu \Pi_1 \partial_\mu \Pi_1)(\partial_\mu \Pi_2 \partial_\mu \Pi_2) \right.\\
\left. + (\partial_\mu \Pi_1 \partial_\mu \Pi_2)^2  \right]\,.
\end{multline}
}{}

\begin{multline}
\shoveright{J = \frac{2}{(1+\Pi_1+\Pi_2)^4} (\partial_\mu \Pi_1 \partial_\mu \Pi_1)(\partial_\mu \Pi_2 \partial_\mu \Pi_2)\,.}
\end{multline}

\ifthenelse{\equal{\Qclass}{revtex4}}{
\begin{multline}
I_{\mu\nu} = \frac{4}{(1+\Pi_1+\Pi_2)^2} (\partial_\mu \partial_\sigma \Pi_1) (\partial_\mu \partial_\sigma \Pi_2)\\
\shoveleft{+ \frac{4}{(1+\Pi_1+\Pi_2)^3} \left[ (\partial_\mu \partial_\nu \Pi_1)(\partial_\sigma \Pi_2 \partial_\sigma \Pi_2) + (\partial\mu \partial_\nu \Pi_2)(\partial_\sigma \Pi_1 \partial_\sigma \Pi_1) - 2(\partial_\mu \partial_\sigma \Pi_1)(\partial_\nu \Pi_2)(\partial_\sigma \Pi_2) \right.}\\
\shoveright{\left. - 2(\partial_\mu \Pi_1)(\partial_\sigma \Pi_1)(\partial_\nu \partial_\sigma \Pi_2) - 2(\partial_\mu \partial_\sigma \Pi_1)(\partial_\sigma \Pi_1)(\partial_\nu \Pi_2) - 2(\partial_\mu \Pi_1)(\partial_\nu \partial_\sigma \Pi_2)(\partial_\sigma \Pi_2) \right]}\\
\shoveleft{+ \frac{4}{(1+\Pi_1+\Pi_2)^4} \left[ - (\partial_\mu \Pi_1)(\partial_\nu \Pi_1)(\partial_\sigma \Pi_2 \partial_\sigma \Pi_2) - (\partial_\mu \Pi_2)(\partial_\nu \Pi_2)(\partial_\sigma \Pi_1 \partial_\sigma \Pi_1) \right.}\\
\shoveright{\left. + 3(\partial_\mu \Pi_1)(\partial_\nu \Pi_2)(\partial_\sigma \Pi_1 \partial_\sigma \Pi_1) + 3(\partial_\mu \Pi_1)(\partial_\nu \Pi_2)(\partial_\sigma \Pi_2 \partial_\sigma \Pi_2)\right.}\\
\left.  + 3(\partial_\mu \Pi_1)(\partial_\nu \Pi_2)(\partial_\sigma \Pi_1 \partial_\sigma \Pi_2) \right]\,.
\end{multline}
}{}
\ifthenelse{\equal{\Qclass}{elsarticle}}{
\begin{multline}
I_{\mu\nu} = \frac{4}{(1+\Pi_1+\Pi_2)^2} (\partial_\mu \partial_\sigma \Pi_1) (\partial_\mu \partial_\sigma \Pi_2)\\
\shoveleft{+ \frac{4}{(1+\Pi_1+\Pi_2)^3} \left[ (\partial_\mu \partial_\nu \Pi_1)(\partial_\sigma \Pi_2 \partial_\sigma \Pi_2) + (\partial\mu \partial_\nu \Pi_2)(\partial_\sigma \Pi_1 \partial_\sigma \Pi_1)\right.}\\
\left. - 2(\partial_\mu \partial_\sigma \Pi_1)(\partial_\nu \Pi_2)(\partial_\sigma \Pi_2) - 2(\partial_\mu \Pi_1)(\partial_\sigma \Pi_1)(\partial_\nu \partial_\sigma \Pi_2) \right.\\
\shoveright{\left.  - 2(\partial_\mu \partial_\sigma \Pi_1)(\partial_\sigma \Pi_1)(\partial_\nu \Pi_2) - 2(\partial_\mu \Pi_1)(\partial_\nu \partial_\sigma \Pi_2)(\partial_\sigma \Pi_2) \right]}\\
\shoveleft{+ \frac{4}{(1+\Pi_1+\Pi_2)^4} \left[ - (\partial_\mu \Pi_1)(\partial_\nu \Pi_1)(\partial_\sigma \Pi_2 \partial_\sigma \Pi_2) - (\partial_\mu \Pi_2)(\partial_\nu \Pi_2)(\partial_\sigma \Pi_1 \partial_\sigma \Pi_1) \right.}\\
\shoveright{\left. + 3(\partial_\mu \Pi_1)(\partial_\nu \Pi_2)(\partial_\sigma \Pi_1 \partial_\sigma \Pi_1) + 3(\partial_\mu \Pi_1)(\partial_\nu \Pi_2)(\partial_\sigma \Pi_2 \partial_\sigma \Pi_2)\right.}\\
\left.  + 3(\partial_\mu \Pi_1)(\partial_\nu \Pi_2)(\partial_\sigma \Pi_1 \partial_\sigma \Pi_2) \right]\,.
\end{multline}
}{}

\begin{multline}
I_{\mu\rho\nu\sigma} = \frac{4}{(1+\Pi_1+\Pi_2)^2} (\partial_\mu \partial_\nu \Pi_1)(\partial_\rho \partial_\sigma \Pi_2)\\
\shoveleft{+ \frac{8}{(1+\Pi_1+\Pi_2)^3} \left[ (\partial_\mu \Pi_2)(\partial_\rho \partial_\nu \Pi_1)(\partial_\sigma \Pi_2) + (\partial_\mu \Pi_1)(\partial_\rho \partial_\nu \Pi_2)(\partial_\sigma \Pi_1) \right]}\\
\shoveleft{+ \frac{8}{(1+\Pi_1+\Pi_2)^4} (\partial_\mu \Pi_1)(\partial_\rho \Pi_2)(\partial_\nu \Pi_1)(\partial_\sigma \Pi_2)\,.}\hfill
\end{multline}

\begin{multline}
\shoveright{J_{\mu\rho\nu\sigma} = \frac{2}{(1+\Pi_1+\Pi_2)^4}(\partial_\mu \Pi_1)(\partial_\rho \Pi_2)(\partial_\nu \Pi_1)(\partial_\sigma \Pi_2)\,.}
\end{multline}

\begin{multline}
\shoveright{K_{\mu\rho\nu\sigma} = \frac{2}{(1+\Pi_1+\Pi_2)^4}(\partial_\mu \Pi_1)(\partial_\rho \Pi_2)(\partial_\nu \Pi_1)(\partial_\sigma \Pi_2)\,.}
\end{multline}

\subsection{Exact Interactions}
\label{app:interaction:gluonic:exact}

When computing the look-up tables, we use global translations and rotations in $\mathbb{R}^4$ to place one instanton at the origin and the partner at $y'_4=R=\sqrt{R_\mu R_\mu}=|y^{I_1}-y^{I_2}|$, where $y^i$ are the instanton centres. The rotation will reemerge in contractions of $R_\mu$ with the colour structure, as we will now see. The relation between the position vector $R_\mu$ and $R_\mu' \equiv (0,0,0,R)$ is given by the following rotation matrix
\begin{eqnarray}
 R_{\mu}' &=&  \mathcal{O}^t_{\mu\nu} R_\nu\,,\nonumber\\
 \mathcal{O}_{\mu 4} &=& \frac{R_\mu}{R}\,,\label{eq:space:colour}
\end{eqnarray}
and the other components of the rotation matrix are irrelevant.

Note that, with the choice of $R_\mu'$, the integrands are $O(3)$ symmetric in the subspace orthogonal to the $4$-direction. Denoting the arguments of the 't Hooft potentials $\Pi(x,\{y,\rho\})$ by $x_\mu$ and $\tilde{x}_\mu \equiv x_\mu-R_\mu$, we can extract extract the $R_\mu$ from the integrands with help of the following formulas, which we order according to the tensor structure of the $x_\mu$-dependence on the integrand.

\begin{equation}
 \int x_\mu = \mathcal{O}_{\mu4} \int x_4'\,.
\end{equation}

\begin{equation}
 \int x_\mu x_\nu = \delta_{\mu \nu} \int x'^2_1+ \mathcal{O}_{\mu 4} \mathcal{O}_{\nu 4} \int (x'^2_4-x'^2_1)\,.
\end{equation}

\begin{eqnarray}
 \int x_\mu x_\nu x_\kappa &=&  (\delta_{\mu\nu} \mathcal{O}_{\kappa 4} + \delta_{\kappa\mu} \mathcal{O}_{\nu 4} + \delta_{\nu\kappa} \mathcal{O}_{\mu 4})\int x'^2_1 x'_4\\
 &+& \mathcal{O}_{\mu 4} \mathcal{O}_{\nu 4} \mathcal{O}_{\kappa 4} \int (x'^3_4-3x'^2_1 x'_4)\,.
\end{eqnarray}

\begin{eqnarray}
 \int x_\mu x_\nu x_\kappa x_\delta &=&  (\delta_{\mu\nu}\delta_{\kappa\delta} + \delta_{\mu\kappa}\delta_{\nu\delta} + \delta_{\mu\delta}\delta_{\kappa\nu})\int x'^2_1 x'^2_2\\
 &+& (\delta_{\mu\nu}\mathcal{O}_{\kappa 4} \mathcal{O}_{\delta 4} + \mathrm{perm.} ) \int (x'^2_4 x'^2_1-x'^2_1 x'^2_2)\\
&+& \mathcal{O}_{\mu 4} \mathcal{O}_{\nu 4} \mathcal{O}_{\kappa 4} \mathcal{O}_{\delta 4} \int (x'^4_4-6x'^2_1 x'^2_4 + 3 x'^2_1 x'^2_2)\,.
\end{eqnarray}

Terms with $\tilde{x}$ can be constructed from these. Incidentally, splitting the different integrands according to the above formulas is the most stable procedure numerically. Taking into account the antisymmetry of the 't Hooft symbols, we end up with the following integrands.

\begin{multline}
I = \frac{4}{(1+\Pi_1+\Pi_2)^2} \left[ (\Pi''_1)^2 + 3(\Pi'_1/r)^2 + (\Pi''_2)^2 +3(\Pi'_2/\tilde{r})^2\right]\\
\shoveleft{- \frac{8}{(1+\Pi_1+\Pi_2)^3} \left[ 2\Pi''_1(\Pi'_1)^2 + 2\Pi''_2(\Pi'_2)^2 + \frac{x\tilde{x}}{r\tilde{r}} \left( \Pi''_1 \Pi'_1 \Pi'_2 + \Pi'_1 \Pi''_2 \Pi'_2 \right) \right]}\\
\shoveleft{+ \frac{4}{(1+\Pi_1+\Pi_2)^4} \left[12 (\Pi'_1)^4 + 12(\Pi'_2)^4 + 8 (\Pi'_1)^2 (\Pi'_2)^2 + 4 (\frac{x\tilde{x}}{r\tilde{r}} \Pi'_1 \Pi'_2)^2\right.}\\
\left. + 12(\Pi'_1)^2(\frac{x\tilde{x}}{r\tilde{r}} \Pi'_1 \Pi'_2 ) + 12(\frac{x\tilde{x}}{r\tilde{r}} \Pi'_1 \Pi'_2 ) (\Pi'_2)^2 \right]\,.
\end{multline}

Note that to achieve good numerical precision, we need to subtract the one-instanton integrands from the above before performing the numerical integration. 

\begin{multline}
\shoveright{J = \frac{2}{(1+\Pi_1+\Pi_2)^4} (\Pi'_1)^2(\Pi'_2)^2\,.}
\end{multline}

\begin{multline}
\shoveright{I_{\mu\nu} = \delta_{\mu\nu} \tilde{I}_{\mu\mu} + \frac{R_\mu R_\nu}{R^2} \tilde{I}_{\mu\nu}\,.}
\end{multline}

\ifthenelse{\equal{\Qclass}{revtex4}}{
\begin{multline}
 \tilde{I}_{\mu\mu} = \frac{4}{(1+\Pi_1+\Pi_2)^2} \left[ \frac{x'^2_1}{r^2} ( \Pi''_1- (\Pi'_1/r))(\Pi'_2/\tilde{r}) + \frac{x'^2_1}{\tilde{r}^2} (\Pi'_1/r)( \Pi''_2-(\Pi'_2/\tilde{r})) \right.\\
 \left. + (\Pi'_1/r)(\Pi'_2/\tilde{r}) + \frac{x'^2_1}{r\tilde{r}}\frac{x\tilde{x}}{r\tilde{r}}(\Pi''_1 \Pi''_2 -\Pi''_1 (\Pi'_2/\tilde{r})-(\Pi'_1/r)\Pi''_2 + (\Pi'_1/r)(\Pi'_2/\tilde{r})) \right]\\
\shoveleft{+ \frac{1}{(1+\Pi_1+\Pi_2)^3}\left[ 4 ((\Pi'_1/r)(\Pi'_2)^2 + (\Pi'_1)^2(\Pi'_2/\tilde{r})) \right.}\\
\left. + \frac{x'^2_1}{r^2}( 4(\Pi''_1-(\Pi'_1/r))(\Pi'_2)^2-8 (\Pi'_1)^2(\Pi'_2/\tilde{r})) \right.\\
\left. +\frac{x'^2_1}{\tilde{r}^2} (4(\Pi'_1)^2 ( \Pi''_2-(\Pi'_2/\tilde{r})) - 8 (\Pi'_1/r)(\Pi'_2)^2) \right.\\
\shoveright{\left.  + \frac{x'^2_1}{r\tilde{r}}(-8\frac{x\tilde{x}}{r\tilde{r}} ((\Pi''_1-(\Pi'_1/r))(\Pi'_2)^2+(\Pi'_1)^2(\Pi''_2-(\Pi'_2/\tilde{r}))) -8 \Pi''_1 \Pi'_1 \Pi'_2 - 8 \Pi'_1 \Pi''_2 \Pi'_2 \right]}\\
\shoveleft{+\frac{1}{(1+\Pi_1+\Pi_2)^4} \left[ -4 \frac{x'^2_1}{r^2} (\Pi'_1)^2 (\Pi'_2)^2 -4 \frac{x'^2_1}{\tilde{r}^2} (\Pi'_1)^2 (\Pi'_2)^2 \right.}\\
\left. + 12 \frac{x'^2_1}{r\tilde{r}} \Pi'_1 \Pi'_2 ( (\Pi'_1)^2 + (\Pi'_2)^2 + \frac{x\tilde{x}}{r\tilde{r}} \Pi'_1 \Pi'_2) \right].
\end{multline}
}{}
\ifthenelse{\equal{\Qclass}{elsarticle}}{
\begin{multline}
 \tilde{I}_{\mu\mu} = \frac{4}{(1+\Pi_1+\Pi_2)^2} \left[ \frac{x'^2_1}{r^2} ( \Pi''_1- (\Pi'_1/r))(\Pi'_2/\tilde{r}) + \frac{x'^2_1}{\tilde{r}^2} (\Pi'_1/r)( \Pi''_2-(\Pi'_2/\tilde{r})) \right.\\
 \left. + (\Pi'_1/r)(\Pi'_2/\tilde{r}) + \frac{x'^2_1}{r\tilde{r}}\frac{x\tilde{x}}{r\tilde{r}}(\Pi''_1 \Pi''_2 -\Pi''_1 (\Pi'_2/\tilde{r})-(\Pi'_1/r)\Pi''_2 + (\Pi'_1/r)(\Pi'_2/\tilde{r})) \right]\\
\shoveleft{+ \frac{1}{(1+\Pi_1+\Pi_2)^3}\left[ 4 ((\Pi'_1/r)(\Pi'_2)^2 + (\Pi'_1)^2(\Pi'_2/\tilde{r})) \right.}\\
\left. + \frac{x'^2_1}{r^2}( 4(\Pi''_1-(\Pi'_1/r))(\Pi'_2)^2-8 (\Pi'_1)^2(\Pi'_2/\tilde{r})) \right.\\
\left. +\frac{x'^2_1}{\tilde{r}^2} (4(\Pi'_1)^2 ( \Pi''_2-(\Pi'_2/\tilde{r})) - 8 (\Pi'_1/r)(\Pi'_2)^2) \right.\\
\left.  + \frac{x'^2_1}{r\tilde{r}}(-8\frac{x\tilde{x}}{r\tilde{r}} ((\Pi''_1-(\Pi'_1/r))(\Pi'_2)^2+(\Pi'_1)^2(\Pi''_2-(\Pi'_2/\tilde{r}))) \right.\\
\shoveright{ \left. -8 \Pi''_1 \Pi'_1 \Pi'_2 - 8 \Pi'_1 \Pi''_2 \Pi'_2 \right]}\\
\shoveleft{+\frac{1}{(1+\Pi_1+\Pi_2)^4} \left[ -4 \frac{x'^2_1}{r^2} (\Pi'_1)^2 (\Pi'_2)^2 -4 \frac{x'^2_1}{\tilde{r}^2} (\Pi'_1)^2 (\Pi'_2)^2 \right.}\\
\left. + 12 \frac{x'^2_1}{r\tilde{r}} \Pi'_1 \Pi'_2 ( (\Pi'_1)^2 + (\Pi'_2)^2 + \frac{x\tilde{x}}{r\tilde{r}} \Pi'_1 \Pi'_2) \right].
\end{multline}
}{}

\begin{multline}
 \tilde{I}_{\mu\nu} = \frac{4}{(1+\Pi_1+\Pi_2)^2} \left[ \frac{x'^2_4-x'^2_1}{r^2} ( \Pi''_1- (\Pi'_1/r))(\Pi'_2/\tilde{r}) \right.\\
\left.  + \frac{(x'_4-R)^2-x'^2_1}{\tilde{r}^2} (\Pi'_1/r)( \Pi''_2-(\Pi'_2/\tilde{r})) \right.\\
 \left.  + \frac{x'_4(x'_4-R)-x'^2_1}{r\tilde{r}}\frac{x\tilde{x}}{r\tilde{r}}(\Pi''_1 \Pi''_2 -\Pi''_1 (\Pi'_2/\tilde{r})-(\Pi'_1/r)\Pi''_2 + (\Pi'_1/r)(\Pi'_2/\tilde{r})) \right]\\
\shoveleft{+ \frac{1}{(1+\Pi_1+\Pi_2)^3}\left[ \frac{x'^2_4-x'^2_1}{r^2}( 4(\Pi''_1-(\Pi'_1/r))(\Pi'_2)^2-8 (\Pi'_1)^2(\Pi'_2/\tilde{r})) \right.}\\
\left. +\frac{(x'_4-R)^2-x'^2_1}{\tilde{r}^2} (4(\Pi'_1)^2 ( \Pi''_2-(\Pi'_2/\tilde{r})) - 8(\Pi'_1/r)(\Pi'_2)^2) \right.\\
\left. + \frac{x'_4(x'_4-R)-x'^2_1}{r\tilde{r}}(-8\frac{x\tilde{x}}{r\tilde{r}} ((\Pi''_1-(\Pi'_1/r))(\Pi'_2)^2+(\Pi'_1)^2(\Pi''_2-(\Pi'_2/\tilde{r}))) \right.\\
\shoveright{\left. -8 \Pi''_1 \Pi'_1 \Pi'_2 - 8 \Pi'_1 \Pi''_2 \Pi'_2 \right]}\\
\shoveleft{+\frac{1}{(1+\Pi_1+\Pi_2)^4} \left[ -4 \frac{x'^2_4-x'^2_1}{r^2} (\Pi'_1)^2 (\Pi'_2)^2 -4 \frac{(x'_4-R)^2-x'^2_1}{\tilde{r}^2} (\Pi'_1)^2 (\Pi'_2)^2 \right.}\\
\left. + 12 \frac{x'_4(x'_4-R)-x'^2_1}{r\tilde{r}} \Pi'_1 \Pi'_2 ( (\Pi'_1)^2 + (\Pi'_2)^2 + \frac{x\tilde{x}}{r\tilde{r}} \Pi'_1 \Pi'_2) \right]\,.
\end{multline}

\begin{multline}
\shoveright{I_{\mu\rho\nu\sigma} = \delta_{\mu \nu} \delta_{\rho \sigma} \tilde{I}_{\mu \nu \mu \nu} + \delta_{\mu \nu} \frac{R_\rho R_\sigma}{R^2} \tilde{I}_{\mu \rho \mu \sigma}\,.}
\end{multline}

\begin{multline}
\shoveright{\tilde{I}_{\mu \nu \mu \nu} = 0 \quad (\mathrm{analytically})\,.}
\end{multline}

\begin{multline}
\tilde{I}_{\mu \rho \mu \sigma} = \frac{4}{(1+\Pi_1+\Pi_2)^2} \left[ \frac{x'^2_4-x'^2_1}{r^2} ( \Pi''_1-(\Pi'_1/r)) (\Pi'_2/\tilde{r}) \right.\\
\hfill \left. + \frac{(x'_4-R)^2-x'^2_1}{\tilde{r}^2} (\Pi'_1/r)(\Pi''_2-(\Pi'_2/\tilde{r})) \frac{x'^2_1 R^2}{(r\tilde{r})^2}(\Pi''_1-(\Pi'_1/r))(\Pi''_2-(\Pi'_2/\tilde{r})) \right]\\
\shoveleft{ - \frac{8}{(1+\Pi_1+\Pi_2)^3} \left[ \frac{x'^2_4-x'^2_1}{r^2} (\Pi'_1)^2(\Pi'_2/\tilde{r}) + \frac{(x'_4-R)^2-x'^2_1}{\tilde{r}^2} (\Pi'_1/r)(\Pi'_2)^2 \right.}\\
\hfill \left. + \frac{x'^2_1 R^2}{(r\tilde{r})^2} ( (\Pi''_1-(\Pi'_1/r))(\Pi'_2)^2 + (\Pi'_1)^2 (\Pi''_2-(\Pi'_2/\tilde{r}))) \right]\\
+ \frac{8}{(1+\Pi_1+\Pi_2)^4}\left[ \frac{x'^2_1 R^2}{(r\tilde{r})^2} (\Pi'_1)^2 (\Pi'_2)^2 \right]\,.\hfill
\end{multline}

\begin{multline}
\shoveright{\tilde{J}_{\mu\rho\nu\sigma} = \delta_{\mu \nu} \frac{R_\rho R_\sigma}{R^2} \frac{2}{(1+\Pi_1+\Pi_2)^4} \frac{x'^2_1 R^2}{(r\tilde{r})^2} (\Pi'_1)^2 (\Pi'_2)^2\,.}
\end{multline}

\begin{multline}
\tilde{K}_{\mu\rho\nu\sigma} = \left[ (\delta_{\mu \nu} \delta_{\rho \sigma} + \delta_{\mu \rho} \delta_{\nu \sigma} + \delta_{\mu \sigma} \delta_{\rho \nu }) x'^2_1 x'^2_2 \right.\\
\left. + \delta_{\mu \nu} \frac{R_\rho R_\sigma}{R^2} ( (x'_4-R)^2 x'^2_1- x'^2_1 x'^2_2) + \delta_{\rho \sigma} \frac{R_\mu R_\nu}{R^2} ( x'^2_4 x'^2_1 - x'^2_1 x'^2_2)\right.\\
+ \left. (\delta_{\mu \rho} \frac{R_\nu R_\sigma}{R^2} + \delta_{\mu \sigma} \frac{R_\nu R_\rho}{R^2} + \delta_{\nu \rho} \frac{R_\mu R_\sigma}{R^2} + \delta_{\nu \sigma} \frac{R_\mu R_\rho}{R^2} ) (x'_4(x'_4-R)x'^2_1 - x'^2_1 x'^2_2) \right.\\
\left. + \frac{R_\mu R_\nu R_\rho R_\sigma}{R^4} (x'^2_4(x'_4-R)^2 + 3x'^2_1 x'^2_2 - x'^2_4 x'^2_1 - (x'_4-R)^2 x'^2_1 \right. \\
\left. - 4x'_4(x'_4-R)x'^2_1)\right] \frac{2}{(1+\Pi_1+\Pi_2)^4} \frac{1}{(r\tilde{r})^2} (\Pi'_1)^2 (\Pi'_2)^2\,.
\end{multline}

\subsection{Asymptotic Interactions}
\label{app:interaction:gluonic:asymptotic}

As explained in the main text, the small separation asymptotic formulas get contributions which have the same functional form as those for the large separation asymptotics; the difference lies in the integration limit. We will therefore start with the large separation formulas and leave the integrals explicit. 

\subsubsection{Large Separation}
\label{app:interaction:gluonic:large}

The upper integration limit $z$ follows from variable substitution and has the the following for integration over $I_1$, with $I_2$ held fixed,
\begin{equation}
 z^2_1=\frac{1+\Pi_2}{\rho^2_1} r^2\,.
\end{equation}

Apart from the dependence of $z$ on $\Pi$, the rational form of the 't Hooft potential allows for a complete factoring out of $\Pi$ under the above mentioned variable substitution. For the large separation formulas it is understood that $z^2 \to \infty$ because the initial integration variable $r$ extends to infinity.

The integral over $I$ contains terms that do not mix the 't Hooft potential $\Pi_1$ and $\Pi_2$ except for the denominators. At zeroth order in our expansion, these terms can be transformed to exactly match the single instanton contributions by exploiting scale invariance. Remembering that we actually subtract the one-instanton contributions to get the interactions, we can neglect these terms altogether. We then end up with the following formulas.

\begin{multline}
\shoveright{\int I = 72 \pi^2 \rho^2 \frac{\partial_\mu \Pi \partial_\mu \Pi}{(1+\Pi)^3} \int^z \frac{s^5 ds}{(s^2+1)^4} + \mathrm{sym}\,.}
\end{multline}

\begin{multline}
 \shoveright{\int J = 16 \pi^2 \rho^2 \frac{\partial_\mu \Pi \partial_\mu \Pi}{(1+\Pi)^3} \int^z \frac{s^5 ds}{(s^2+1)^4} + \mathrm{sym}\,.}
\end{multline}

\begin{multline}
 \int I_{\mu\nu} = 16 \pi^2 \rho^2 \frac{\partial_\mu \partial_\nu \Pi}{(1+\Pi)^2} \int^z \frac{s^3 ds}{(s^2+1)^3}\\
- \left( 8 \pi^2 \rho^2 \delta_{\mu\nu} \frac{\partial_\sigma \Pi \partial_\sigma \Pi}{(1+\Pi)^3} + 8\pi^2 \rho^2 \frac{(\partial_\mu \Pi)(\partial_\nu \Pi)}{(1+\Pi)^3}\right) \int^z \frac{s^5 ds}{(s^2+1)^4} + \mathrm{sym}\,.
\end{multline}

At zeroth order, partial integration and the antisymmetry of the 't Hooft symbols can be used to simplify
\begin{equation}
 I_{\mu\rho\nu\sigma} \to \frac{8}{(1+\Pi_1+\Pi_2)^4} (\partial_\mu \Pi_1) (\partial_\rho \Pi_2) (\partial_\nu \Pi_1) (\partial_\sigma \Pi_2)\,,
\end{equation}
with asymptotic behaviour
\begin{multline}
 \shoveright{\int I_{\mu\rho\nu\sigma} = 16 \pi^2 \rho^2 \delta_{\mu\nu}\frac{(\partial_\rho \Pi)(\partial_\sigma \Pi)}{(1+\Pi)^3} \int^z \frac{s^5 ds}{(s^2+1)^4} + \mathrm{sym}\,.}
\end{multline}

\begin{multline}
 \shoveright{\int J_{\mu\rho\nu\sigma} = 4 \pi^2 \rho^2 \delta_{\mu\nu}\frac{(\partial_\rho \Pi)(\partial_\sigma \Pi)}{(1+\Pi)^3} \int^z \frac{s^5 ds}{(s^2+1)^4} + \mathrm{sym}\,.}
\end{multline}

For $K_{\mu\rho\nu\sigma}$ no 't Hooft symbols can be used to exchange the index pairs $(\mu,\nu) \leftrightarrow (\rho,\sigma)$, and so we cannot simplify with a symmetry argument anymore.

\begin{multline}
 \int K_{\mu\rho\nu\sigma} = 4 \pi^2 \rho^2_1 \delta_{\mu\nu}\frac{(\partial_\rho \Pi_2)(\partial_\sigma \Pi_2)}{(1+\Pi_2)^3} \int^{z_1} \frac{s^5 ds}{(s^2+1)^4}\\
 + 4 \pi^2 \rho^2_2 \delta_{\rho\sigma}\frac{(\partial_\mu \Pi_1)(\partial_\nu \Pi_1)}{(1+\Pi_1)^3} \int^{z_2} \frac{s^5 ds}{(s^2+1)^4}\,.
\end{multline}

\subsubsection{Small Separation}
\label{app:interaction:gluonic:small}

As explained in the main text, the small separation asymptotic formulas get contributions from the large asymptotics. Also, in this case we have performed a global translation so that the instantons sit at $\pm R_\mu /2$. Therefore, in the large separation formulas we need to put $z^2 = \frac{1+\Pi}{\rho^2} (R/2)^2$.

We now turn to the proper small separation asymptotic formulas that encode the repulsion through the gauge singularity. We will again introduce an explicit upper limit for the integrals; abusing notation we will use the same letter as before, but here the meaning becomes
\begin{equation}
 z^2 = \frac{R^2}{\rho^2_1+\rho^2_2}, \quad z^2_i = \frac{R^2}{\rho^2_i}\,.
\end{equation}

To derive these formulas, we approximate the arguments $x_\mu \pm R_\mu/2 \to x_\mu$. We have, therefore, explicitly restored $O(4)$ symmetry, which can be exploited to set several integrals to zero. Eventually, we arrive at

\ifthenelse{\equal{\Qclass}{revtex4}}{
\begin{multline}
 \int I = 384 \pi^2 \left[ \frac{\rho^4_1 + \rho^4_2}{(\rho^2_1 + \rho^2_2)^2} \int_z \frac{ds}{s(s^2+1)^2} -\left( \frac{\rho^2_1 \rho^2_2}{(\rho^2_1 + \rho^2_2)^2} + 2 \frac{\rho^6_1 + \rho^6_2}{(\rho^2_1 + \rho^2_2)^3}  \right) \int_z \frac{ds}{s(s^2+1)^3} \right.\\
\left. +\frac{\rho^8_1 + \rho^8_2 + \rho^4_1 \rho^4_2 + \rho^6_1 \rho^2_2 + \rho^2_1 \rho^6_2}{(\rho^2_1 + \rho^2_2)^4} \int_z \frac{ds}{s(s^2+1)^4} -\int_{z_1} \frac{s^4 ds}{s(s^2+1)^4} -\int_{z_2} \frac{s^4 ds}{s(s^2+1)^4} \right]\,.
\end{multline}
}{}
\ifthenelse{\equal{\Qclass}{elsarticle}}{
\begin{multline}
 \int I = 384 \pi^2 \left[ \frac{\rho^4_1 + \rho^4_2}{(\rho^2_1 + \rho^2_2)^2} \int_z \frac{ds}{s(s^2+1)^2} \right.\\
 \left. -\left( \frac{\rho^2_1 \rho^2_2}{(\rho^2_1 + \rho^2_2)^2} + 2 \frac{\rho^6_1 + \rho^6_2}{(\rho^2_1 + \rho^2_2)^3}  \right) \int_z \frac{ds}{s(s^2+1)^3} \right.\\
\left. +\frac{\rho^8_1 + \rho^8_2 + \rho^4_1 \rho^4_2 + \rho^6_1 \rho^2_2 + \rho^2_1 \rho^6_2}{(\rho^2_1 + \rho^2_2)^4} \int_z \frac{ds}{s(s^2+1)^4}\right.\\
\left. -\int_{z_1} \frac{s^4 ds}{s(s^2+1)^4} -\int_{z_2} \frac{s^4 ds}{s(s^2+1)^4} \right]\,.
\end{multline}
}{}

\begin{multline}
\shoveright{ \int J = 64 \pi^2 \frac{\rho^4_1 \rho^4_2}{(\rho^2_1 + \rho^2_2)^4} \int_z \frac{ds}{s(s^2+1)^4}\,.}
\end{multline}

\begin{multline}
 \int I_{\mu\nu} = \delta_{\mu\nu} \left[ 96 \pi^2 \frac{\rho^2_1 \rho^2_2}{(\rho^2_1 + \rho^2_2)^2} \int_z \frac{ds}{s(s^2+1)^2} - 192 \pi^2 \frac{\rho^2_1 \rho^2_2}{(\rho^2_1 + \rho^2_2)^2} \int_z \frac{ds}{s(s^2+1)^3} \right.\\
\left. + 32 \pi^2 \frac{\rho^4_1 \rho^4_2 + 3\rho^6_1 \rho^2_2 + 3\rho^2_1 \rho^6_2}{(\rho^2_1 + \rho^2_2)^4} \int_z \frac{ds}{s(s^2+1)^4} \right]\,.
\end{multline}

\begin{multline}
 \int I_{\mu\rho\nu\sigma} = \delta_{\mu\nu} \delta_{\rho\sigma} \left[ -32 \pi^2 \frac{\rho^2_1 \rho^2_2}{(\rho^2_1 + \rho^2_2)^2} \int_z \frac{ds}{s(s^2+1)^2} \right.\\
\left. +32 \pi^2 \frac{\rho^2_1 \rho^2_2}{(\rho^2_1 + \rho^2_2)^2} \int_z \frac{ds}{s(s^2+1)^3} \right]\,.
\end{multline}

\begin{multline}
 \shoveright{\int J_{\mu\rho\nu\sigma} = 0\,.}
\end{multline}

\begin{multline}
 \int K_{\mu\rho\nu\sigma} = \frac{8}{3} \pi^2 ( \delta_{\mu\nu} \delta_{\rho\sigma} + \delta_{\mu\rho} \delta_{\nu\sigma} + \delta_{\mu\sigma} \delta_{\nu\rho})\frac{\rho^4_1 \rho^4_2}{(\rho^2_1 + \rho^2_2)^4} \int_z \frac{ds}{s(s^2+1)^4}\,.
\end{multline}

\section{Fermionic Interactions}
\label{app:interaction:quark}

The Dirac overlap matrix elements are given by
\begin{equation}
T_{IA} = \int d^4x \frac{1}{4\pi^2\rho_I \rho_A} \frac{1}{2}\Tr (U \tau^{+}_{\beta}) I_\beta\,.
\end{equation}
Note that $\frac{1}{2}\Tr (U \tau^{+}_{\beta}) \equiv i u_\beta$ is the colour four-vector used for instance in \cite{schaefer:shuryak:iilm}. After some straightforward algebra, we find that $I_\beta$ has the following form
\begin{multline}
I_\beta = \frac{-1}{(1+\Pi_I+\Pi_A)(1+\Pi_I)^{3/2}(1+\Pi_A)^{3/2}}\\
\left(\frac{\Pi_A}{1+\Pi_I} (\partial_\mu \Pi_I \partial_\mu \Pi_I) \partial_\beta \Pi_A + (\partial_\mu \Pi_A \partial_\mu \Pi_A) \partial_\beta \Pi_I \right)\,.
\end{multline}

\subsection{Exact Interactions}
\label{app:interaction:quark:exact}

Using the same rotations (\ref{eq:space:colour}) as for the gluonic interactions to marry the space-time with the colour indices, we get

\begin{multline}
I_\beta = \frac{R_\beta}{R} \frac{-3}{(1+\Pi_I+\Pi_A)(1+\Pi_I)^{3/2}(1+\Pi_A)^{3/2}}\\
\left\{ \frac{x'_4}{r} \Pi_I' (\Pi_A')^2 + \frac{x'_4-R}{\tilde{r}} (\Pi_I')^2 \Pi_A' \frac{\Pi_A}{1+\Pi_I} \right\}\,.
\end{multline}

\subsection{Asymptotic Interactions}

\subsubsection{Large Separation}
\label{app:interaction:quark:large}

In order to get rather simple formulas, we make the following additional simplification
\begin{equation}
 1+\Pi_I+\Pi_A \to 
\left\{
\begin{array}{c@{\; : \quad}l}
 1+\Pi_I & \mathrm{Integration\; over\; \Pi_I}\\
 1+\Pi_A & \mathrm{Integration\; over\; \Pi_A}
\end{array}
\right. \,.
\end{equation}

Given these further assumption, we can proceed as for the gluonic interactions. Finally, caution needs to be taken in the case of an anti-instanton because it sits at $-R_\mu$ so that $\partial_\beta \Pi_A$ generates an extra minus sign.

\ifthenelse{\equal{\Qclass}{revtex4}}{
\begin{multline}
\int I_\beta = 8 \pi^2 \rho_I^2 \frac{\Pi_A \partial_\beta \Pi_A}{(1+\Pi_A)^{3/2}} \int^{z_I} \frac{s^4 ds}{(s^2+1)^{7/2}} - 8 \pi^2 \rho_A^2 \frac{\partial_\beta \Pi_I}{(1+\Pi_I)^{3/2}} \int^{z_A} \frac{s^4 ds}{(s^2+1)^{5/2}}\,.
\end{multline}
}{}
\ifthenelse{\equal{\Qclass}{elsarticle}}{
\begin{multline}
\int I_\beta = 8 \pi^2 \rho_I^2 \frac{\Pi_A \partial_\beta \Pi_A}{(1+\Pi_A)^{3/2}} \int^{z_I} \frac{s^4 ds}{(s^2+1)^{7/2}} \\
- 8 \pi^2 \rho_A^2 \frac{\partial_\beta \Pi_I}{(1+\Pi_I)^{3/2}} \int^{z_A} \frac{s^4 ds}{(s^2+1)^{5/2}}\,.
\end{multline}
}{}

\subsubsection{Small Separation}
\label{app:interaction:quark:small}

At zeroth order, i.e.\ $x_\mu \pm R_\mu /2 \to x_\mu$, the contribution to $I_\beta$ vanishes because of $O(4)$ symmetry. It turns out that the large separation asymptotics falls off too quickly as $R \to 0$. However, this is not important because in this regime the gluonic interaction is dominant.

\section{Cholesky decomposition update}
\label{app:cholesky:update}

In this appendix we look in detail at how the structure suitable for the Cholesky decomposition update comes about. We will also see that insertion can be performed faster whereas deletions will be the most costly.

\subsection{Canonical Moves}

Upon updating instanton $I$, we have that $T \to T + \Delta T$, with $(\Delta T)_{ij} = \delta_{iI} \xi^{*}_j$. This induces the following changes

\begin{eqnarray}
 (T^\dagger T)_{ij} &\to& (T^\dagger T)_{ij} + T^\dagger_{iI} \xi^{*}_j + \xi_i T_{Ij} + \xi_i \xi^{*}_j\,,\\
 \psi_i &\equiv& T^{*}_{Ii}\,,\\
 \phi_i &\equiv& \xi_i + \psi_i\,,\\
 T^\dagger T &\to& T^\dagger T + \phi \phi^\dagger - \psi \psi^\dagger\,.
\end{eqnarray}

\begin{eqnarray}
 (T T^\dagger)_{ij} &\to& (T T^\dagger)_{ij} + \delta_{iI} \xi^{*}_k T^\dagger_{kj} + T_{ik} \xi_k \delta_{Ij} + \delta_{iI}\delta_{jI}\,,\\
 \psi_i &\equiv& \frac{1}{|\xi|}(T\xi)_i\,,\\
 \phi_i &\equiv& \delta_{Ii}|\xi| + \psi_i\,,\\
 T^\dagger T &\to& T^\dagger T + \phi \phi^\dagger - \psi \psi^\dagger\,.
\end{eqnarray}

Changes in an anti-instanton will have analogous formulas.

\subsection{Insertion}

We focus on inserting an instanton. Insertion of an anti-instanton is then similar. Since in the code we always add an instanton at the end of the arrays, an insertion corresponds to adding a row to $T$.
\begin{eqnarray}
 T &\to& \left( \begin{array}{c} T \\ \xi^\dagger \end{array} \right)\,,\\
 T T^\dagger &\to& \left( \begin{array}{c} T \\ \xi^\dagger \end{array} \right) \left( \begin{array}{cc} T^\dagger & \xi \end{array} \right) = \left( \begin{array}{cc} TT^\dagger & T\xi \\ \xi^\dagger T^\dagger & \xi^\dagger \xi \end{array} \right)\,.
\end{eqnarray}

On the level of the Cholesky decomposition
\begin{eqnarray}
 L &\to& = \left( \begin{array}{cc} L & 0 \\ \chi^\dagger & 1 \end{array} \right)\,,\\
 D &\to& = \left( \begin{array}{cc} D & 0 \\ 0 & d \end{array} \right)\,,\\
 LDL^\dagger &\to& = \left( \begin{array}{cc} LDL^\dagger & LD\chi \\ \chi^\dagger D L^\dagger & \chi^\dagger D \chi +d \end{array} \right)\,.
\end{eqnarray}

Remembering that the insertion also adds a mass term in the diagonal, we have to solve the following system
\begin{equation}
 \left\{
\begin{array}{c@{\;=\;}l}
 LD\chi & T\xi \\
 d & \xi^\dagger \xi + m^2 - \chi^\dagger D \chi
\end{array}
 \right. \,,
\end{equation}
which can be solved in $O(N^2)$ by using backsubstitution. The case for $T^\dagger T$ is simply given by
\begin{equation}
 T^\dagger T \to \left( \begin{array}{cc} T^\dagger & \xi \end{array} \right) \left( \begin{array}{c} T \\ \xi^\dagger \end{array} \right) = T^\dagger T + \xi \xi^\dagger\,,
\end{equation}
which is a rank-1 update.

\subsection{Deletion}

We focus again on an instanton. Deletion will be a two step process. We first delete the last instanton and then swap it with that instanton that has been selected for deletion. The swapping is similar to a canonical move, where now $\xi$ is given by the difference between the last instanton and the selected instanton.

The proper deletion part is given by
\begin{eqnarray}
 T T^\dagger &\to& = \left( \begin{array}{cc} TT^\dagger & 0 \\ 0 & 0 \end{array} \right)\,,\\
 L &\to& = \left( \begin{array}{cc} L & 0 \\ 0 & 1 \end{array} \right)\,,\\
 D &\to& = \left( \begin{array}{cc} D & 0 \\ 0 & 0 \end{array} \right)\,.
\end{eqnarray}

The $T^\dagger T$ part is again simply related to a rank-1 update because, upon rearranging the result from the insertion part, we get
\begin{equation}
T^\dagger T \to T^\dagger T - \xi \xi^\dagger\,.
\end{equation}

\section{$\overline{\mathrm{MS}}$ to $\mathrm{PV}$}
\label{app:renormalization:scheme:change}

Operators with anomalous dimensions run, and for mass independent renormalisation prescriptions they depend on the scheme. The IILM makes predictions within a subtraction scheme that uses Pauli-Villars regularisation. However, the lattice results have been quoted in $\overline{\mathrm{MS}}$, and so we need to work out the relation between the two.

It is not hard to convince ourselves that the quark masses run inversely to the chiral condensate: note that chiral perturbation theory, (\ref{eq:chiral:perturbation:chi:pion}), relates the topological susceptibility to the pion mass and decay constant, which are physical quantities; it also relates the chiral condensate and the quark masses to the topological susceptibility through (\ref{eq:chiral:perturbation:chi:qq}), and therefore the renormalisation scheme dependence must exactly cancel among the two.

\begin{figure}[tbp]
\begin{center}
\begin{fmffile}{MassRenormalization}
\begin{fmfgraph*}(200,100)
\fmfleft{i}
\fmfright{o}
\fmf{quark,tension=3,label=$p$}{i,v1}
\fmf{quark,right,tension=1.5}{v1,v2}
\fmf{gluon,left,tension=1.5}{v1,v2}
\fmf{quark,tension=3,label=$p$}{v2,o}
\fmfdot{v1,v2}
\end{fmfgraph*}
\end{fmffile}
\end{center}
\caption{Feynman diagram needed to compute the difference between the $\overline{\mathrm{MS}}$ and $\mathrm{PV}$ scheme at one-loop.}
\label{feyn:quark:mass:renormalization}
\end{figure}
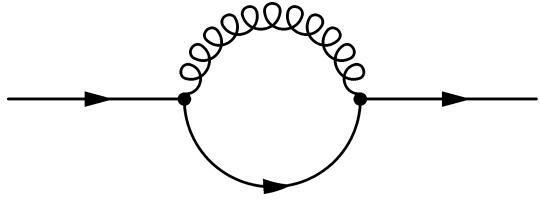

Eventually, we will also relate the quark masses of the two schemes, and so here we will focus on mass renormalisation. We will only work at one-loop\footnote{Maintaining manifest gauge-invariance in Yang-Mills theories using Pauli-Villars regularisation is not straightforward beyond one-loop.}, i.e.\ we will need to evaluate (\reffig{feyn:quark:mass:renormalization}) in both schemes.

This is a textbook computation, \cite{peskin:schroeder:qft}. After subtracting off the divergences, we end up with
\begin{equation}
\Sigma_{\mathrm{PV}} = \frac{\alpha_s}{2\pi} C(3) \left\{-2m +\frac{1}{4} \slashed{p}  + \int dx (2m-(1-x) \slashed{p} ) \ln \frac{\mu^2}{x m^2 - x(1-x)  p^2} \right\}\,,
\end{equation}
and
\begin{equation}
\Sigma_{\overline{\mathrm{MS}}} = \frac{\alpha_s}{2\pi} C(3) \left\{-m +\frac{1}{2} \slashed{p} + \int dx (2m-(1-x) \slashed{p} ) \ln \frac{\mu^2}{x m^2 - x(1-x)  p^2} \right\}\,.
\end{equation}

Relating both through the pole mass and using that, in our case $C(3)=4/3$, we get
\begin{equation}
 m_{\overline{\mathrm{MS}}}=m_{\mathrm{PV}}(1-\frac{\alpha_s}{2\pi} \frac{5}{3})\,.
\end{equation}


\begin{thebibliography}{10}

\bibitem{armoni:shifman:veneziano:quark:condensate}
A.~Armoni, M.~Shifman, and G.~Veneziano.
\newblock {QCD Quark Condensate from SUSY and the Orientifold Large-N
  Expansion}.
\newblock {\em Phys. Lett. B}, 579:384--390, 2004.

\bibitem{bpst:instanton}
A.~A. Belavin, A.~M. Polyakov, A.~S. Schwartz, and Yu.~S. Tyupkin.
\newblock {Pseudoparticle solutions of the Yang-Mills equations}.
\newblock {\em Phys. Lett. B}, 59:85--87, 1975.

\bibitem{bruckmann:gattringer:ilgenfritz:al:filtering}
F.~Bruckmann, C.~Gattringer, E.M. Ilgenfritz, M.~M\"{u}ller-Preussker,
  A.~Schafer, and S.~Solbrig.
\newblock {Quantitative comparison of filtering methods in lattice QCD}.
\newblock {\em Eur. Phys. J. A}, 33:333--338, 2007.

\bibitem{chetyrkin:kuehn:steinhauser:rundec}
K.~Chetyrkin, J.~K\"uhn, and M.~Steinhauser.
\newblock {RunDec: a Mathematica package for running and decoupling of the
  strong coupling and quark masses.}
\newblock {\em Comput. Phys. Commun.}, 133:43--65, 2000.

\bibitem{Chiu:Aoki:JLQCD:TWQCD:topological:susceptibility:fixed_topology}
T.~Chiu, S.~Aoki, H.~Fukaya, S.~Hashimoto, T.~Hsieh, T.~Kaneko, H.~Matsufuru,
  J.~Noaki, K.~Ogawa, T.~Onogi, and N.~Yamada.
\newblock {Topological susceptibility in 2-flavor lattice QCD with fixed
  topology}.
\newblock {\em PoS {L}attice 2007}, 068, 2007.

\bibitem{Chiu:Aoki:JLQCD:TWQCD:topological:susceptibility:overlap}
T.~Chiu, S.~Aoki, S.~Hashimoto, T.~Hsieh, T.~Kaneko, H.~Matsufuru, J.~Noaki,
  T.~Onogi, and N.~Yamada.
\newblock {Topological susceptibility in (2+1)-flavor lattice QCD with overlap
  fermion}.
\newblock {\em PoS {L}attice 2008}, 072, 2008.

\bibitem{chiu:hsieh:tseng:topological:susceptibility}
T.~Chiu, T.~Hsieh, and P.~Tseng.
\newblock {Topological susceptibility in (2+1) flavors lattice QCD with
  domain-wall fermions}.
\newblock {\em Phys. Lett. B}, 671:135--138, 2008.

\bibitem{chu:grandy:huang:negele:instanton:lattice:evidence}
M.C. Chu, J.~Grandy, S.~Huang, and J.~W. Negele.
\newblock {Evidence for the Role of Instantons in Hadron Structure from Lattice
  QCD}.
\newblock {\em Phys.Rev. D}, 49:6039--6050, 1994.

\bibitem{cristofotetti:foccioli:traini:negele:iilm}
M.~Cristoforetti, P.~Faccioli, M.C. Traini, and J.W. Negele.
\newblock {Exploring the Chiral Regime of QCD in the Interacting Instanton
  Liquid Model}.
\newblock {\em Phys.Rev. D}, 75:034008, 2007.

\bibitem{diakonov:polyakov:weiss:instanton:vacuum}
D.~Diakonov, M.~Polyakov, and C.~Weiss.
\newblock {Hadronic matrix elements of gluon operators in the instanton
  vacuum}.
\newblock {\em Nucl. Phys.}, B461:539--580, 1996.

\bibitem{dunne:hur:lee:min:instanton:determinant:mass}
G.V. Dunne, J.~Hur, C.~Lee, and H.~Min.
\newblock {Calculation of QCD Instanton Determinant with Arbitrary Mass}.
\newblock {\em Phys.Rev. D}, 71:085019, 2005.

\bibitem{durr:fodor:hoebling:kurth:su3:topological:susceptibility}
S.~Durr, Z.~Fodor, C.~Hoelbling, and T.~Kurth.
\newblock {Precision study of the SU(3) topological susceptibility in the
  continuum}.
\newblock {\em JHEP}, 04:055, 2007.

\bibitem{diakonov:instanton:variational}
D.I. Dyakonov and V.Yu. Petrov.
\newblock {Instanton-based vacuum from the Feynman variational principle}.
\newblock {\em Nucl. Phys. B}, 245:259--292, 1984.

\bibitem{diakonov:instanton:quarks}
D.I. Dyakonov and V.Yu. Petrov.
\newblock {A theory of light quarks in the instanton vacuum}.
\newblock {\em Nucl. Phys. B}, 272:457--489, 1986.

\bibitem{pdg:quark:masses}
C.~Amsler et~al.
\newblock {\em Phys. Lett. B}, 667:1, 2008.

\bibitem{pdg:2002}
K.~Hagiwara et~al.
\newblock {Quantum Chromodynamics}.
\newblock {\em Phys. Rev. D}, 66:010001--1, 2002.

\bibitem{faccioli:strong:CP:theta}
P.~Faccioli.
\newblock {Strong CP breaking and quark-antiquark repulsion in QCD, at finite
  $\theta$}.
\newblock {\em Phys. Rev. D}, 71:091502, 2005.

\bibitem{faccioli:guadagnoli:simula:neutron:dipole:moment}
P.~Faccioli, D.~Guadagnoli, and S.~Simula.
\newblock {The Neutron Electric Dipole Moment in the Instanton Vacuum: Quenched
  Versus Unquenched Simulations}.
\newblock {\em Phys.Rev.D}, 70:074017, 2004.

\bibitem{frenkel:smit:molecular:simulation}
D.~Frankel and B~Smit.
\newblock {\em {Molecular Simulation: From Algorithm to Applications}}.
\newblock Academic {P}ress, second edition edition, 2002.

\bibitem{garcia:osborn:iilm:chiral}
A.~M. Garc\'{i}a-Garc\'{i}a and J.~C. Osborn.
\newblock {Chiral phase transition and Anderson localization in the Instanton
  Liquid Model for QCD}.
\newblock {\em Nucl. Phys. A}, 770:141--161, 2006.

\bibitem{gerhold:ilgenfritz:mueller_preussker:kvbll:gas:confinement}
P.~Gerhold, E.M. Ilgenfritz, and M.~M\"{u}ller-Preussker.
\newblock {An $SU(2)$ KvBLL caloron gas model and confinement}.
\newblock {\em Nucl. Phys. B}, 760:1--37, 2007.

\bibitem{gill:golub:murray:saunders:modify:factorization}
P.~Gill, G.~Golub, W.~Murray, and M.~Saunders.
\newblock Methods for {M}odifying {M}atrix {F}actorizations.
\newblock {\em Mathematics of Computations}, 28:505--535, 1974.

\bibitem{goeckeler:horsley:irving:pleiter:rakow:schierholz:stueben:lambda:para%
meter}
M.~G\"ockeler, R.~Horsley, A.~Irving, D.~Pleiter, P.~Rakow, G.~Schierholz, and
  H.~St\"uben.
\newblock {A Determination of the Lambda Parameter from Full Lattice QCD}.
\newblock {\em Phys. Rev. D}, 73:014513, 2006.

\bibitem{grossman:dirac:zeromode}
B.~Grossman.
\newblock {Zero energy solutions if the Dirac equation in an N-pseudoparticle
  field}.
\newblock {\em Phys. Lett. A}, 61:86--88, 1977.

\bibitem{harrington:shepard:caloron}
B.J. Harrington and H.K. Shepard.
\newblock {Periodic Euclidean solutions and the finite-temperature Yang-Mills
  gas}.
\newblock {\em Phys. Rev. D}, 17:2122--2125, 1978.

\bibitem{hasenfratz:hasenfratz:scales}
A.~Hasenfratz and P.~Hasenfratz.
\newblock {The scales of euclidean and hamiltonian lattice QCD}.
\newblock {\em Nucl. Phys. B}, 193:210--220, 1981.

\bibitem{kleinert:path_integrals}
H.~Kleinert.
\newblock {\em {Path Integral in Quantum Mechanics, Statistics, Polymer
  Physics,and Financial Markets}}.
\newblock World Scientific, 2006.

\bibitem{kraan:baal:caloron:I}
T.C. Kraan and P.~van Baal.
\newblock {Exact $T$-duality between Calorons and Taub-NUT spaces}.
\newblock {\em Phys.Lett. B}, 428:268--276, 1998.

\bibitem{kraan:baal:caloron:monopole}
T.C. Kraan and P.~van Baal.
\newblock {Monopole Constituents inside $SU(N)$ Calorons}.
\newblock {\em Phys.Lett. B}, 435:389--395, 1998.

\bibitem{kraan:baal:caloron:II}
T.C. Kraan and P.~van Baal.
\newblock {Periodic Instantons with non-trivial Holonomy}.
\newblock {\em Nucl.Phys. B}, 533:627--659, 1998.

\bibitem{lee:lu:caloron}
K.~Lee and C.~Lu.
\newblock {$SU(2)$ Calorons and Magnetic Monopoles}.
\newblock {\em Phys.Rev. D}, 58:025011, 1998.

\bibitem{lenz:negele:thies:confinement:meron}
F.~Lenz, J.W. Negele, and M.~Thies.
\newblock {Confinement from merons}.
\newblock {\em Phys. Rev. D}, 69:074009, 2004.

\bibitem{leutwyler:smilga:spectrum:dirac}
H.~Leutwyler and A.~Smilga.
\newblock {Spectrum of Dirac operator and role of winding number in QCD}.
\newblock {\em Phys. Rev. D}, 46:5607--5632, 1992.

\bibitem{mason:trottier:horgan:davies:lepage:quark:masses:lattice}
Q.~Mason, H.~Trottier, R.~Horgan, C.~Davies, and G.~Lepage.
\newblock {High-precision determination of the light-quark masses from
  realistic lattice QCD}.
\newblock {\em Phys. Rev. D}, 73:114501, 2006.

\bibitem{munster:kamp:instanton:gas}
G.~Munster and C.~Kamp.
\newblock {Distribution of instanton sizes in a simplified instanton gas
  model}.
\newblock {\em Eur. Phys. J.}, C17:447--454, 2000.

\bibitem{negele:lenz:thies:confinement:instanton:meron}
J.W. Negele, F.~Lenz, and M.~Thies.
\newblock {Confinement from Instantons or Merons}.
\newblock {\em Nucl. Phys. Proc. Suppl.}, 140:629, 2005.

\bibitem{nowak:rho:zahed:chiral_nuclear_dynamics}
M.A. Nowak, M.~Rho, and I.~Zahed.
\newblock {\em {Chiral Nuclear Dynamics}}.
\newblock World Scientific, 1996.

\bibitem{peskin:schroeder:qft}
M.~Peskin and D.~Schroeder.
\newblock {\em {An Introduction to Quantum Field Theory}}.
\newblock Perseus {B}ooks, 1995.

\bibitem{schaefer:shuryak:iilm}
T.~Sch\"{a}fer and E.V. Shuryak.
\newblock {Interacting instanton liquid in QCD at zero and finite
  temperatures}.
\newblock {\em Phys.Rev. D}, 53:65226542, 1996.

\bibitem{shuryak:instantons:liquid:I}
E.V. Shuryak.
\newblock {Toward the quantitative theory of the instanton liquid (I).
  Phenomenology and the method of collective coordinates}.
\newblock {\em Nucl. Phys. B}, 302:559--579, 1988.

\bibitem{shuryak:instantons:liquid:II}
E.V. Shuryak.
\newblock {Toward the quantitative theory of the instanton liquid (II). The
  $SU(2)$ gluodynamics.}
\newblock {\em Nucl. Phys. B}, 302:574--598, 1988.

\bibitem{shuryak:instantons:liquid:III}
E.V. Shuryak.
\newblock {Toward the quantitative theory of the instanton liquid (III).
  Instantons and light fermions.}
\newblock {\em Nucl. Phys. B}, 302:599--620, 1988.

\bibitem{shuryak:instantons:qcd:I}
E.V. Shuryak.
\newblock {Instantons in QCD (I). Properties of the "instanton liquid".}
\newblock {\em Nucl. Phys. B}, 319:521--540, 1989.

\bibitem{shuryak:instantons:qcd:II}
E.V. Shuryak.
\newblock {Instantons in QCD (II). Correlators of pseudoscalar and scalar
  currents.}
\newblock {\em Nucl. Phys. B}, 319:541--569, 1989.

\bibitem{shuryak:verbaarschot:interactions:finite:T}
E.V. Shuryak and J.J.M. Verbaarschot.
\newblock {QCD Instantons at finite temperature}.
\newblock {\em Nucl. Phys. B}, 364:255--282, 1991.

\bibitem{shuryak:verbaarschot:screening}
E.V. Shuryak and J.J.M. Verbaarschot.
\newblock {Screening of the topological charge in a correlated instanton
  vacuum}.
\newblock {\em Phys. Rev. D}, 52:295--306, 1995.

\bibitem{thooft:instanton:fluctuations}
G.~'t~Hooft.
\newblock {Computation of the quantum effects due to a four-dimensional
  pseudoparticle}.
\newblock {\em Phys.Rev. D}, 14:3432--3448, 1976.

\bibitem{verbaarschot:streamline}
J.J.M. Verbaarschot.
\newblock {Streamlines and conformal invariance in Yang-Mills theories}.
\newblock {\em Nucl. Phys. B}, 362:33--53, 1991.

\bibitem{vermaseren:larin:ritbergen:4loop:anomalous:mass:dimension}
J.~Vermaseren, S.~Larin, and T.~Ritbergen.
\newblock {The 4-loop quark mass anomalous dimension and the invariant quark
  mass}.
\newblock {\em Phys. Lett. B}, 405:327--333, 1997.

\bibitem{wagner:confining}
M.~Wagner.
\newblock {Classes of confining gauge field configurations}.
\newblock {\em Phys.Rev. D}, 75:016004, 2007.

\bibitem{wagner:fermions:pseudoparticle:approach}
M.~Wagner.
\newblock {Fermions in the pseudoparticle approach}.
\newblock {\em Phys. Rev.}, D76:076002, 2007.

\bibitem{wang:lu:wang:lattice:instanton:vacuum:dominance}
Z.Q. Wang, X.F. Lu, and F.~Wang.
\newblock {Dilute liquid of instanton and its topological charge dominate the
  QCD vacuum}.
\newblock {\em AIP Conf.Proc}, 865:242--247, 2006.

\bibitem{wantz:iilm:2}
Olivier Wantz.
\newblock {The topological susceptibility from grand canonical simulations in
  the interacting instanton liquid model: strongly associating fluids and
  biased Monte Carlo}.
\newblock {\em Nucl. Phys. B}, 829:91--109, 2010.

\bibitem{wantz:iilm:3}
Olivier Wantz and E.~P.~S. Shellard.
\newblock {The topological susceptibility from grand canonical simulations in
  the interacting instanton liquid model: chiral phase transition and axion
  mass}.
\newblock {\em Nucl. Phys. B}, 829:110--160, 2010.

\end{thebibliography}

\end{document}